\documentclass[12pt]{article} 
\pdfoutput=1

\usepackage{amsmath,amssymb,amsfonts}
\usepackage{color}
\definecolor{darkblue}{rgb}{0.1,0.1,.7}
\usepackage[colorlinks, linkcolor=darkblue, citecolor=darkblue, urlcolor=darkblue, linktocpage]{hyperref} 
\usepackage[square, comma, compress,numbers]{natbib}
\usepackage[]{amsmath}
\usepackage{amsthm}
\usepackage[]{graphicx}
\usepackage[]{latexsym}
\usepackage{geometry}
\usepackage{amscd}
\usepackage[all,cmtip]{xy}
\usepackage{mathrsfs}
\usepackage{bbold}
\usepackage{subfigure}
\usepackage[margin=10pt,font=small,labelfont=bf]{caption}
\usepackage{simplewick}
\usepackage{changepage}
\usepackage[]{algorithm2e}
\usepackage{booktabs,multirow}
\usepackage[font={small}]{caption}

\usepackage{newfloat}
\DeclareFloatingEnvironment[
    listname={Algorithm},
    name=Algorithm,
    placement=tbhp,
]{algorithmfloat}

\definecolor{darkblue}{rgb}{0.,0.,0.4}
\definecolor{darkred}{rgb}{0.5,0.,0.}
\definecolor{BlueViolet}{RGB}{138,43,226}
\definecolor{SkyBlue}{RGB}{30,144,255}
\definecolor{DarkGreen}{RGB}{0,100,0}

\theoremstyle{definition}
\newtheorem{numerics}{Numerics Setup}

\newtheorem{analytics}{Analytics Setup}
\newtheorem{hybrid}{Hybrid Setup}

\newcommand\s{\sigma}
\newcommand\e{\epsilon}
\newcommand\<{\langle}
\renewcommand\>{\rangle}

\newcommand\ds{\Delta_{\s}}
\newcommand\de{\Delta_{\e}}
\newcommand\Ds{\Delta_{\s}}
\newcommand\De{\Delta_{\e}}
\newcommand\D{\Delta}
\newcommand\f{f}
\newcommand\fT{\f_T}
\newcommand\fsse{\f_{\s\s\e}}
\newcommand\feee{\f_{\e\e\e}}
\newcommand\fssT{\f_{\s\s T}}
\newcommand\feeT{\f_{\e\e T}}
\newcommand\calO{\mathcal{O}}

\newcommand\itie{\textit{i.e. }}

\newcommand\zb{\bar{z}}
\def\hb {\bar{h}}

\newcommand\Vp{\vec{V}^{(+)}}
\newcommand\Vm{\vec{V}^{(-)}}
\newcommand\Vth{\vec{V}^{(\theta)}}

\def\cO {{\cal O}}
\def\cOp{\mathcal{O}'}
\def\cN {{\cal N}} 
\def\bbZ {{\mathbb{Z}}} 
\def\dsZ {{\mathbb{Z}}} 

\newcommand{\be}{\begin{eqnarray}}
\newcommand{\ee}{\end{eqnarray}}
\newcommand{\bea}{\begin{eqnarray}}
\newcommand{\eea}{\end{eqnarray}}

\begin{document}

\vspace*{-.6in} \thispagestyle{empty}
\vspace{.2in} {\Large
\begin{center}
{\bf 
The Hybrid Bootstrap
}
\end{center}
}
\vspace{.2in}
\begin{center}
{\bf 
Ning Su
\\
\vspace{.2in} 
{\it Department of Physics, University of Pisa, I-56127 Pisa, Italy}
}
\end{center}

\vspace{.2in}

\begin{abstract}
Finding a method to combine the numerical bootstrap with the analytic lightcone bootstrap is an important goal to advance the conformal bootstrap program. We propose a hybrid bootstrap method to do just that. The numerical and analytic bootstrap approaches are sensitive to different regions of the spectrum and complement each other. When they are effectively combined, the hybrid bootstrap enjoys the best of both worlds and the prediction for the actual CFT can be significantly improved. In this work, we discuss the general strategy to perform such a hybrid bootstrap, and we make a partial implementation of the strategy for 3D Ising CFT $\{\s,\e\}$ system. Even at relatively low derivative order $\Lambda=19$, the hybrid bootstrap predicts very precise values for the scaling dimension $\ds,\de$ that are within the previous $\Lambda=43$ rigorous error bars.
\end{abstract}

\newpage

\tableofcontents
\newpage

\section{Introduction}\label{sec:introduction}

The conformal bootstrap program has achieved much progress, both numerically and analytically, in the past few years. On the numerical side, critical exponents of various CFTs have been determined precisely with rigorous error bars\footnote{See \cite{Poland:2018epd} for a thorough review. See \cite{Kos:2016ysd,Rong:2018okz,Chester:2019ifh,Chester:2020iyt,Chester:2021aun} for some recent computations, where rigorous error bars are determined in various CFTs.}. On the analytic side, tools have been developed to solve bootstrap equations in the lightcone limit $z \ll 1-\zb \ll 1$\cite{Alday:2007mf,Komargodski:2012ek,Fitzpatrick:2012yx,Alday:2015eya,Alday:2015ewa,Alday:2016njk,Simmons-Duffin:2016wlq,Albayrak:2019gnz,Caron-Huot:2017vep,Liu:2020tpf}\footnote{See also \cite{Henriksson:2020jwk} for a review of the analytic bootstrap.}. The result is that the conformal data of the leading twist operators can be accurately approximated using the data of low lying operators. 

The numerical and analytic approaches are sensitive to different regions. In the numerical bootstrap, we project the conformal block function
$G_{\Delta ,\ell}\left(z,\bar{z}\right)$
to a finite dimension vector by expanding it around the crossing symmetric point $z=\zb=1/2$ as
$G^{(m,n)}=\partial_z^m\partial _{\bar{z}}^nG_{\Delta ,\ell}\left(z,\bar{z}\right)| {}_{z=\bar{z}=1/2}$.
However, in the lightcone limit, we know
$G_{\Delta ,\ell}\left(z,\bar{z}\right) \propto z^{\Delta /2}\log (\zb)$.
It would be difficult to capture the $\text{log}(\zb)$ divergence using the Taylor expansion at $\zb=1/2$. On the other hand, the analytic bootstrap predicts the large spin operators have to exist in order to match the $\text{log}(\zb)$ divergence in the lightcone limit. This is just the shortcomings of the numerical bootstrap, because these large spin operators are not visible to the numerical methods : using the extreme functional method (EFM)\cite{El-Showk:2012vjm}, the numerically determined spectrum only has accurate information for the small spin operators even if we start with a large spin set in the setup. 

To better improve the prediction from the numerical bootstrap, we should make the numerics converge not just around $z=\zb=1/2$, but also more uniformly in other regions of $z,\zb$. To achieve that, a natural idea is to inject the information from the analytic lightcone bootstrap, which is valid around the lightcone limit $z \ll 1-\zb \ll 1$, into the numerical bootstrap. In this work, we propose a hybrid bootstrap method to realize this idea\footnote{Alternative approaches might be to follow \cite{CastedoEcheverri:2016fxt} or \cite{Paulos:2019fkw}.}. A rough idea about the hybrid method was actually already suggested in \cite{Simmons-Duffin:2016wlq}. But in the past, there were two difficulties in the implementation of the hybrid bootstrap. One is that the lightcone analytics is parameterized by a few variables that one has to scan over, but numerically scanning over a higher dimensional parameter space was forbiddingly expensive\footnote{The scan problem was partially solved in \cite{Chester:2019ifh}, where the surface cutting algorithm was invented to handle the space of the OPE coefficient. But this algorithm is not suitable for the generic parameter space that involves the scaling dimensions.}. The second difficulty was that the lightcone analytics is not exact, which contradicts the previous paradigm of the numerical bootstrap : the bootstrap numerics works in the ``feasibility mode", \textit{i.e.} we make an assumption about the CFT spectrum, then the numerics rigorously rule out the spectrum if it contradicts with the bootstrap equation. In this feasibility mode, the bootstrap numerics can't distinguish a ``less wrong parameter" from ``a more wrong parameter" --- they are all just infeasible parameters. Therefore it was not obvious how to incorporate non-rigorous information in a controllable way. Fortunately, recently the navigator function \cite{Reehorst:2021ykw} was developed, where the binary information “feasible/infeasible” was replaced by a continuous measure of success. With this technique, both difficulties can be effectively handled. In this paper, we will investigate the hybrid bootstrap method based on the navigator function.

\begin{figure}[!htpb]
\centering
\includegraphics[width = 0.8\textwidth]{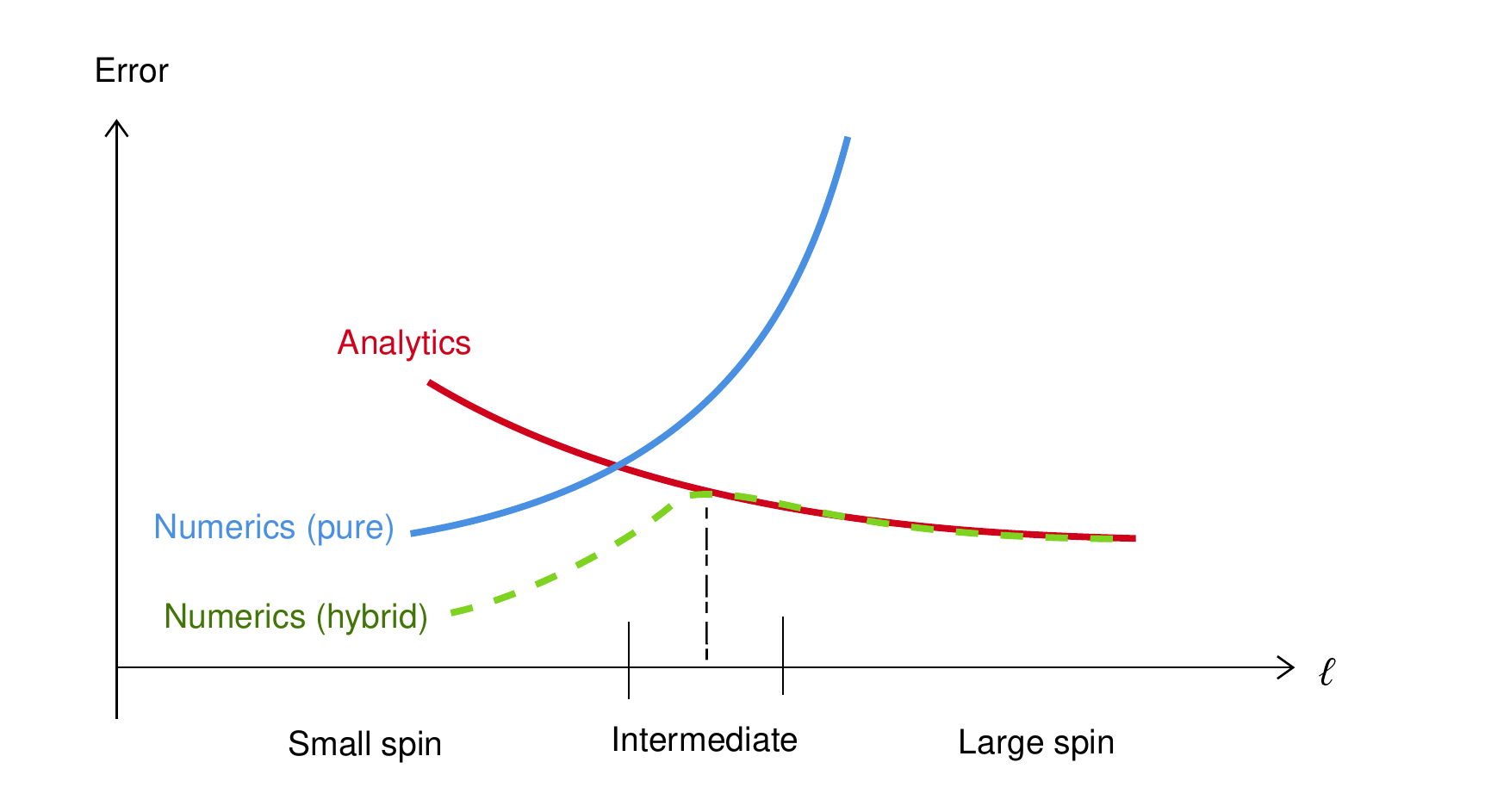}
\caption{\label{fig:NAcurve_hybrid_vs_pure}In the hybrid numerics, the accuracy of the small spin is improved, while the large spin operators matches with the analytics.} 
\end{figure}

A summary of the content and the structure of this paper is the following. In Section \ref{sec:review}, we briefly review the numerical bootstrap using the navigator technique, and lightcone analytic bootstrap in the language of the inversion formula. We compute a few ingredients which are needed for the later sections. In Section \ref{sec:main_strategy} and \ref{sec:alternative_approaches}, we discuss the general strategy of the hybrid bootstrap. We also show a few alternative approaches and explain why we choose our main approach. The key idea is that the numerical bootstrap is accurate for low twist operators with small spins, while the analytic bootstrap work better for the large spins, therefore we should trust each approach in the region where it has an advantage and glue the spectrum in some intermediate range of the spin. The gluing process effectively single out an unique point in the theory space whose leading twist operators match with the analytics as much as possible. At that point, the errors for a wide range of spins are improved, as the schematic illustration in Figure \ref{fig:NAcurve_hybrid_vs_pure}). There are many technical aspects when we actually try to implement this strategy. For example, how to decide the intermediate region and how to match the spectrum. From Section \ref{sec:fullsetup} to \ref{sec:liebig}, we demonstrate how to use the hybrid bootstrap in the example of 3D Ising $\{\s,\e\}$ system and address those technical issues. We will see the hybrid bootstrap yields very accurate predictions for the Ising CFT data, even at a low derivative order.

\section{Review of the numerical bootstrap and analytic bootstrap} \label{sec:review}

\subsection{Numerical bootstrap} \label{sec:numerics}

The bootstrap equation of the 3D Ising $\{\s,\e\}$ system is
\be
\label{eq:crossingequationwithv}
 \sum_{\cO^+} \begin{pmatrix}\f_{\s\s\cO} & \f_{\e\e\cO}\end{pmatrix} \vec{V}^{(+)}_{\D,\ell}\begin{pmatrix} \f_{\s\s\cO} \\ \f_{\e\e\cO} \end{pmatrix}+ \sum_{\cO^-} \f_{\s\e\cO}^2 \vec{V}^{(-)}_{\D,\ell} & = & 0 ,
\ee
where $\vec{V}^{(+/-)}_{\D,\ell}$ is the crossing vectors defined in (3.12) of \cite{Kos:2014bka}.

To use the navigator method, we take the parameter space to be $(\ds,\de,\theta)$ with $\theta=\text{tan}^{-1}(\f_{\e\e\e}/\f_{\s\s\e})$, and on each point of the parameter space, we define a semi-definite program (SDP) as:
\begin{align}
	&objective=\vec{\alpha}.(\Vp_{\Delta=0,\ell=0}), \nonumber\\
	&conditions:\nonumber\\
	&\vec{\alpha}\cdot(\Vth)\ge 0, \nonumber\\
	&\vec{\alpha}\cdot(\Vp_{\D,\ell=0})\ge 0 \text{ for } \D \ge 3,\nonumber\\
	&\vec{\alpha}\cdot(\Vm_{\D,\ell=0})\ge 0 \text{ for } \D \ge 3,\nonumber\\
	&\vec{\alpha}\cdot(\Vp_{\D,\ell})\ge 0 \text{ for } \D \ge \D_{unitary} \text{ and } \ell=2, 4 ... ,\nonumber\\
	&\vec{\alpha}\cdot(\Vm_{\D,\ell})\ge 0 \text{ for } \D \ge \D_{unitary} \text{ and } \ell=1, 2, ... ,
	\label{cond:nvg3par}
\end{align}
where
\be\label{eq:Vtheta}
 \Vth=\Vp_{\de,0} + \Vm_{\ds,0} \otimes \left(\begin{array}{cc}1 &0 \\0 &0\end{array}\right).
\ee
and $\alpha$ is a linear functional acting on the crossing vectors.
For the normalization of the SDP, we use the $\Sigma$-navigator normalization \cite{Reehorst:2021ykw}. Specifically the normalization is defined by summing the crossing vectors in each channel at 20 equally spaced $\D_i$ between the first gap and the gap+6 (for above SDP, the gap is $\D_{unitary}$ for $\ell>0$ and 3 for $\ell=0$).

\begin{figure}[!htpb]
\centering
\includegraphics[width = 0.45\textwidth]{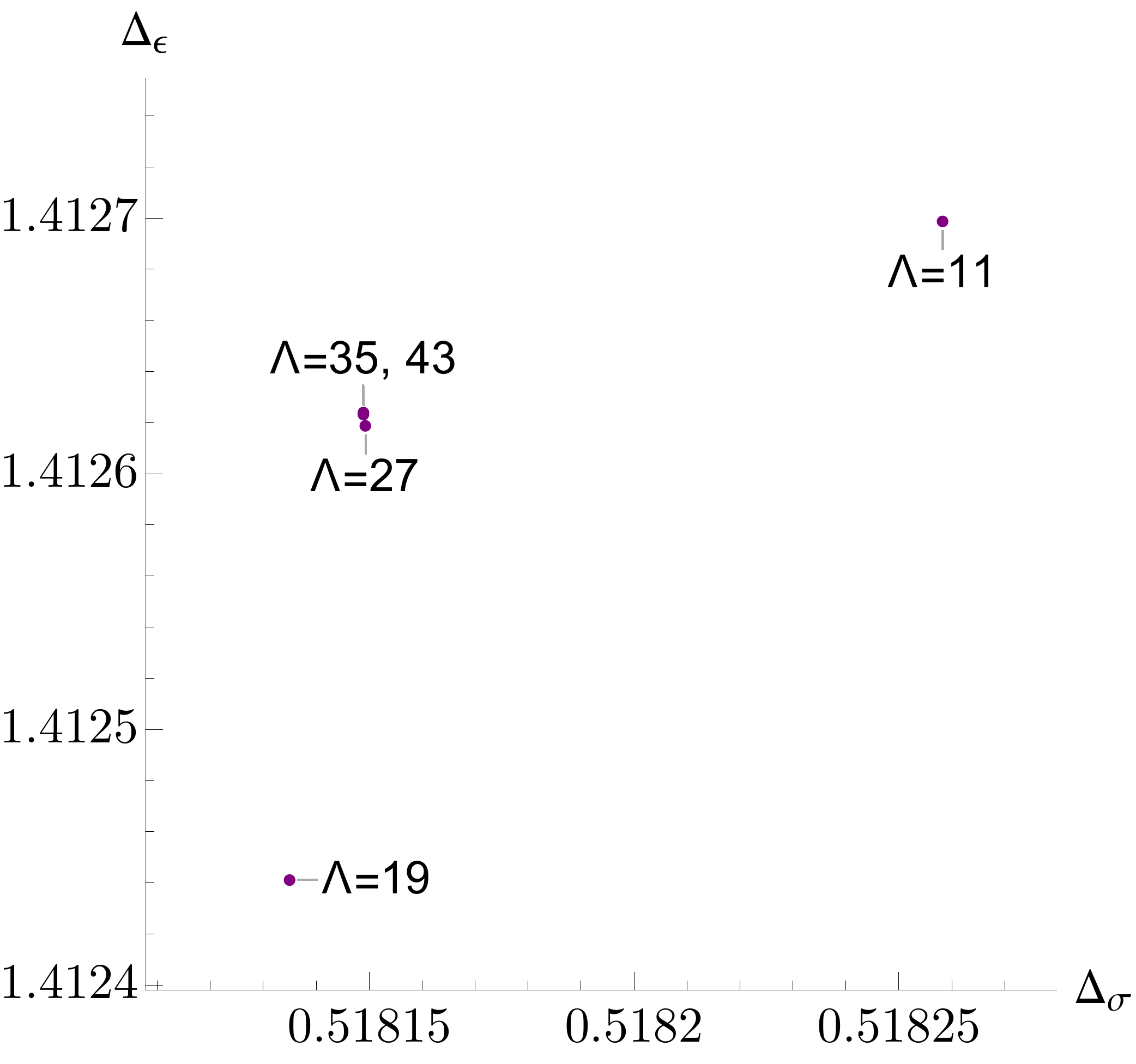}
\quad
\includegraphics[width = 0.45\textwidth]{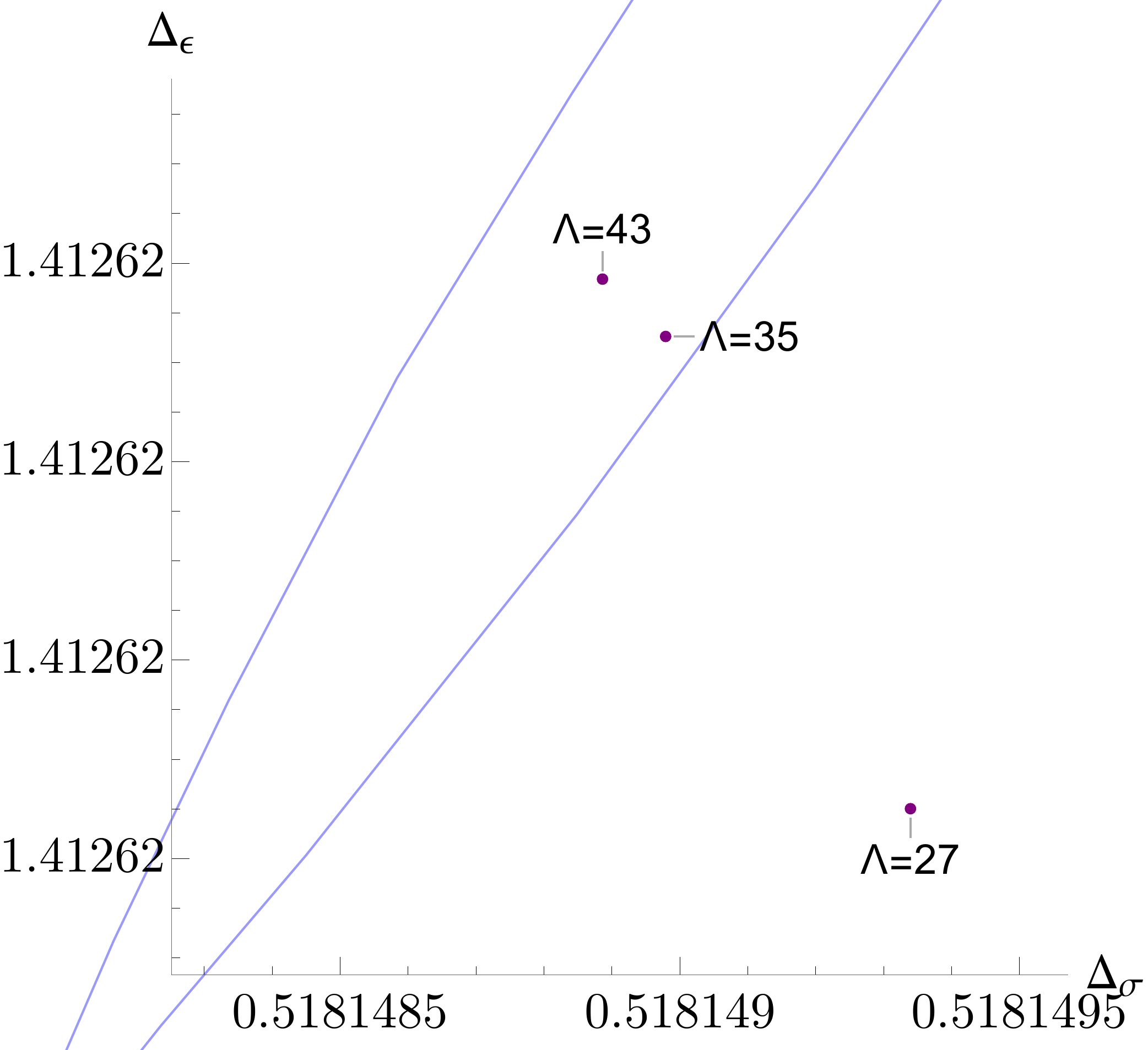}
\caption{\label{fig:3par}The locations of the minimum navigator at $\Lambda=11, 19, 27, 35, 43$ from the computation of the Numerics Setup \ref{numerics:1}. The light blue curve on the right figure is the boundary of the $\Lambda=43$ Ising island of \cite{Kos:2016ysd}. 
}
\end{figure}

The SDP can be solved using the \texttt{sdpb} program \cite{Simmons-Duffin:2015qma} and the optimal value of the objective defines a navigator function $\cN(\ds,\de,\theta)$. The positive/negative sign of the navigator function indicates the parameter $(\ds,\de,\theta)$ is infeasible/feasible. We can use Algorithm 2 of \cite{Reehorst:2021ykw} to maximize/minimize certain parameters in the feasible region or use Algorithm 1 of \cite{Reehorst:2021ykw} to minimize the navigator value $\cN(\ds,\de,\theta)$. After the maximizing/minimizing procedure, we obtain an unique point in the parameter space. In this paper, we may refer to this unique point as the optimal point, or simply the result of the optimization. As shown in \cite{Reehorst:2021ykw}, the result of the navigator minimization gives a good prediction of the actual CFT. Throughout this paper, we refer this computation as 
\begin{numerics}\label{numerics:1}
Minimize the navigator function $\cN(\ds,\de,\theta)$ subject to the SDP condition of (\ref{cond:nvg3par}).
\end{numerics}
In Figure ~\ref{fig:3par}, we present the $(\ds,\de)$ from the navigator minimization for $\Lambda=11, 19, 27, 35, 43$. The technical details of the computation (such as the choice of various parameters) are presented in the Appendix \ref{app:parameters}, while the full result of all computations in this paper are presented in the Appendix \ref{app:results}.
The $\Lambda=43$ result gives the best prediction of the actual CFT:
\be\label{data:pd43}
\ds\approx0.518148884,\qquad \de\approx1.41262383,\qquad \qquad \qquad\nonumber\\
\f_{\e\e\e}\approx1.05185442,\qquad \f_{\s\s\e}\approx1.53243407,\qquad \fT\approx1.25885570
\ee
where $\ds,\de,\f_{\e\e\e}/\f_{\s\s\e}$ is obtained from the minimal navigator point while the rest of the data\footnote{The convention of the OPE coefficient and the block in this paper is defined in the following way. We refer the block convention of the \textit{n}-th row of the Table 1 of \cite{Poland:2018epd} as ``Convention \textit{n}". We say the OPE coefficient is in Convention \textit{n} if the OPE coefficient is defined by $G(u,v)=\sum_\cO \f^2 g_{\D_{\cO},\ell_\cO}(u,v)$ where $G$ is the four-point function and $g$ is the block in Convention \textit{n}. The block and OPE coefficient are in the Convention 1 throughout this paper except Section \ref{sec:analytics}. For Section \ref{sec:analytics}, the block and OPE coefficient is in the Convention 6, because this convention is more convenient for the analytic computation. When using the result of Section \ref{sec:analytics} in the numerics, one has to first convert the convention accordingly.} is extracted using the extremal functional method (EFM)\cite{El-Showk:2012vjm}. The $\fT$ is defined by $\fT=\fssT/\ds=\feeT/\de$ using the Ward identity\footnote{The data from our numerics has a small difference between $\fssT/\ds,\feeT/\de$. What we present in (\ref{data:pd43}) is the mean value.}.

\begin{figure}[!htpb]
\centering
\includegraphics[width = 0.9\textwidth]{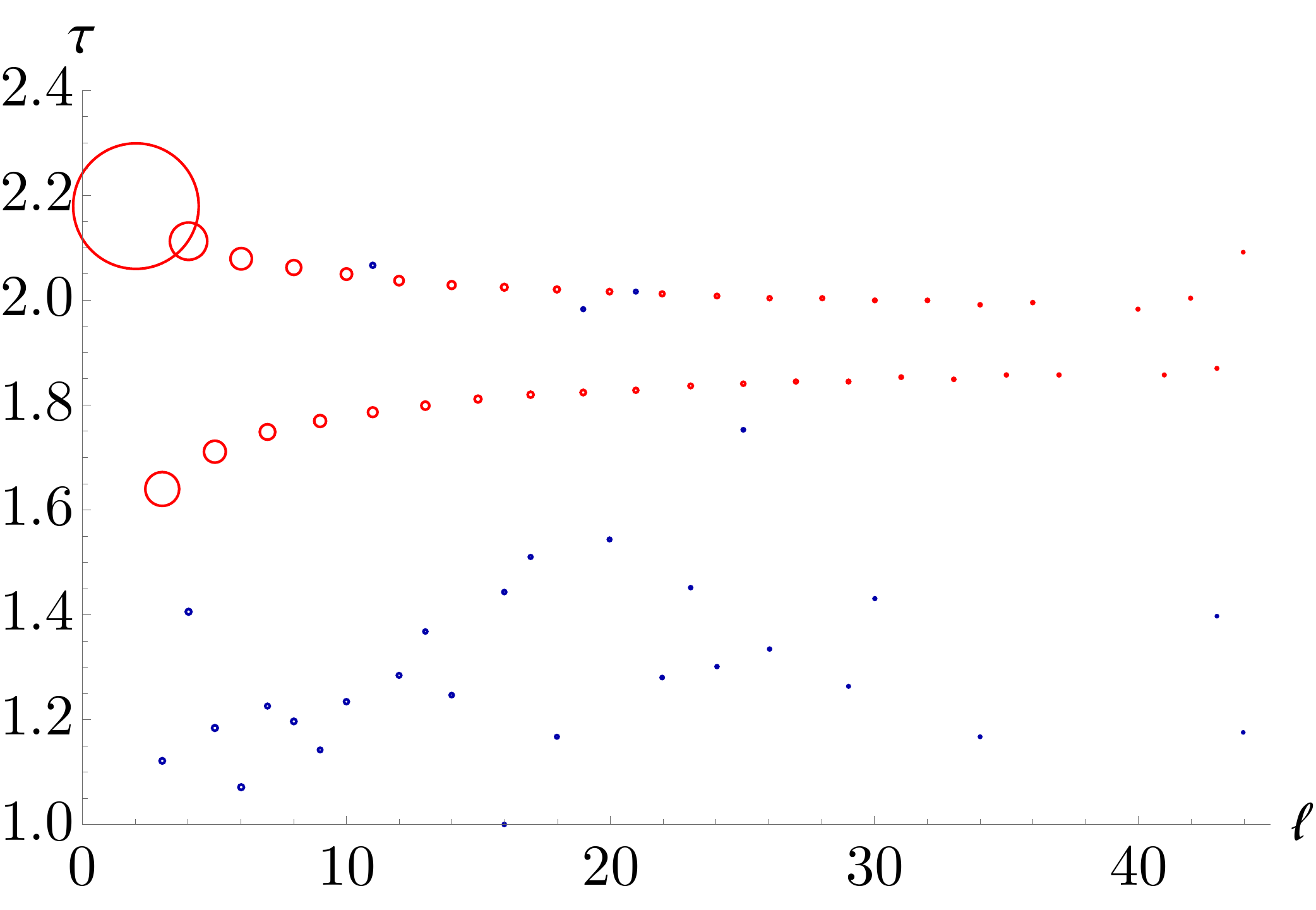}
\caption{\label{fig:spectrum_se0}The spectrum in $\bbZ_2$-odd channel with twist $\tau<2$. The red circles are the operators in the $[\s\e]_0$ family, while the blue circles are the rest. The size of the circle is proportional to $1/|\text{log}(\lambda_{\s\e [\s\e]_0})|$.}
\end{figure}

The spectrum from the EFM at the minimal navigator point has a good quality. In the $\dsZ_2$ even channel, we observed there are fake operators with dimensions exactly at the unitarity bound, due to the sharing effect \cite{Simmons-Duffin:2016wlq,Liu:2020tpf}. The dimension and OPE coefficient of our data for $[\s\s]_0$ is obtained by a cleanup procedure $\Delta _{\text{eff}}=(\sum _{\mathcal{O}}\f _{\mathcal{O}}^2\Delta _{\mathcal{O}})/(\sum _{\mathcal{O}}\f _{\mathcal{O}}^2)$ and $\f _{\text{eff}}=\sqrt{\sum _{\mathcal{O}}\f _{\mathcal{O}}^2}$, where $\sum_\cO$ is a summation over all operators with twist smaller than 1.1 in each spin. After the cleanup, both the dimension and the OPE coefficient matches much better with the analytics in Section \ref{sec:analytics}. On the other hand, the situation in the $\dsZ_2$-odd channel is more complicated. We present the raw data for operators in $\bbZ_2$-odd channel with twist $\tau<2$ in Figure~\ref{fig:spectrum_se0}. 
There are some operators below $[\s\e]_0$ with OPE coefficient roughly two orders of magnitude smaller than the OPE coefficient of $[\s\e]_0$. Whether those operators are physical operators or fake operators due to the sharing effect, is not clear. In fact, the $[\s\e]_0$ shouldn't be the lowest twist family in the $\dsZ_2$ odd channel. The lowest should be $[\s T]_0$, because $\tau_\s>\tau_T$\footnote{We thank Johan Henriksson for pointing out this fact and sharing some results of their ongoing work\cite{johan:2022}}. So the blue circles in Figure~\ref{fig:spectrum_se0} could be related to physical operators\footnote{However according to \cite{Henriksson:2022rnm}, $[\s T]$ shouldn't exist for $\ell<6$ in the $4-\epsilon$ expansion.}. Another observation is that if we process the raw data in the same way as the $\dsZ_2$ even channel, the result is further away from the analytics in Section \ref{sec:analytics}. Therefore we choose to clean up the sharing effect for $[\s\s]_0$, while for $[\s\e]_0$ we use the raw data. In later sections, when we refer to ``$\Lambda=43$ EFM spectrum", we always mean the data obtained in this way.

\subsection{Analytical bootstrap} \label{sec:analytics}

In \cite{Simmons-Duffin:2016wlq}, it was demonstrated how to solve the Ising bootstrap equation in the lightcone limit $z \ll 1-\zb \ll 1$. The conclusion is that the conformal data of the leading twist $[\s\s]_{0,\ell}, [\s\e]_{0,\ell}$ at large spin can be effectively approximated using only five parameters $\{\ds,\de,\feee,\fsse,\fT\}$, where $\fT=\fssT/\ds=\feeT/\de$. Alternatively, one can use the inversion formula \cite{Caron-Huot:2017vep, Simmons-Duffin:2017nub} to do similar computations \cite{Liu:2020tpf}. The difference is that, certain non-perturbative term that is relevant to small $\ell$ is omitted in the lightcone bootstrap\cite{Albayrak:2019gnz}, while the inversion formula approach has the full contribution of an isolated operator, thus is more accurate for small $\ell$. On the other hand, the result of \cite{Simmons-Duffin:2016wlq} is given in term of simple $\Gamma$ function and can be evaluated efficiently, while the formalism of \cite{Liu:2020tpf} contains hypergeometric function ${}_4F_3$ or ${}_3F_2$ and is computationally very expensive.

In this section we briefly review some of the results of \cite{Simmons-Duffin:2016wlq,Liu:2020tpf}. We then derive an improved version for $[\s\s]_{0,\ell}$, which will be used later. Our goal is actually slightly different from \cite{Caron-Huot:2020ouj,Simmons-Duffin:2016wlq,Liu:2020tpf,Atanasov:2022bpi}. For the purpose of hybrid bootstrap, the analytics has to simultaneously meet (1), computational efficiency; (2), accuracy at large spin; (3), dependence only on a few parameters (at most $\{\ds,\de,\feee,\fsse,\fT\}$). On the other hand, the accuracy at small spins doesn't matter too much for us. The reason will be clear later.

Let's consider the correlator $\<\s\s\s\s\>$. Following the formalism and notations of \cite{Liu:2020tpf}, we define a generating function that capturing all the information in the $\s \times \s$ OPE:
\be
C(z,\bar{h})=\sum _{\mathcal{O}\in \sigma \times \sigma }\lambda _{\sigma \sigma \mathcal{O}}^2(\bar{h}_{\mathcal{O}})z^{h_{\mathcal{O}}}
\ee
where $h=\frac{\Delta -\ell }{2},\bar{h}=\frac{\Delta +\ell }{2}$ and the $\lambda$ is related to the OPE coefficient that we use in the numerical bootstrap by
$f_{\sigma \sigma \mathcal{O}}^2(\bar{h}_{\mathcal{O}})=(1-\frac{\partial h_{\mathcal{O}}(\bar{h})}{\partial \bar{h}}){}^{-1}\lambda _{\sigma \sigma \mathcal{O}}^2(\bar{h}_{\mathcal{O}}).$

The inversion formula prediction that the contribution of a single operator $\mathcal{O}'$ to the t-channel generating function $C_{\mathcal{O}'}^t\left(z,\bar{h}\right)$ at small $z$ is
\be\label{equ:cOp}
C_{\mathcal{O}'}^t(z,\bar{h})\approx 2\sin (\pi  (h_{\mathcal{O}'}-2h_{\sigma })){}^2\sum _{p=0}^{\infty }\sum _{q=-p}^p\mathcal{A}_{p,q}^{h_{32},h_{14}}(h_{\mathcal{O}'},\bar{h}_{\mathcal{O}'})\frac{z^{h_1+h_2}k_{\bar{h}_{\mathcal{O}'}+q}^{h_{32},h_{14}}(1-z)}{(1-z)^{h_2+h_3}}\kappa _{2\bar{h}}\Omega _{\bar{h},h_{\mathcal{O}'+p},h_2+h_3}^{h_1,h_2,h_3,h_4}.
\ee
where the function $\mathcal{A}, k, \kappa, \Omega$ is defined in \cite{Liu:2020tpf}.
At large $\bar{h}=h_5$ the $\kappa \Omega$ in ~(\ref{equ:cOp}) can expand to asymptotic series \footnote{This is essentially (G.1) of \cite{Liu:2020tpf}, except we flipped the $\Gamma$ function for easier numerical implementation.} :
\begin{align}\label{equ:kappaOmega}
(\kappa \Omega )_{h_5,h_6,h_2+h_3}^{h_1,h_2,h_3,h_4}=&\sum _{m=0}^{\infty }\frac{\Gamma (m-h_2-h_3+h_6+1) \Gamma (m-h_1-h_4+h_6+1)}{\pi ^2}\times \\ \nonumber
&\frac{\Gamma (-h_1+h_2+h_5) \Gamma (-h_3+h_4+h_5) \Gamma (-m+h_1+h_3+h_5-h_6-1)}{\Gamma (2 h_5-1) \Gamma (m-h_1-h_3+h_5+h_6+1)}\times \\\nonumber
&\frac{(-1)^m \Gamma (2 h_6) \Gamma (m+h_2-h_3+h_6) \Gamma (m-h_1+h_4+h_6)}{2 m! \Gamma (h_2-h_3+h_6) \Gamma (-h_1+h_4+h_6) \Gamma (m+2 h_6)}.
\end{align}

If we have information of a set of operators $\{\mathcal{O}'\}$, we can use them to approximate the t-channel generating function as $C^t(z,\bar{h})\approx \sum _{\mathcal{O}'}\lambda _{\sigma \sigma \mathcal{O}'}^2C_{\mathcal{O}'}^t(z,\bar{h})$ and the full generating function $C(z,\bar{h})=2C^t(z,\bar{h})$.
We can extract the dimension and OPE coefficient of the leading twist family $[\s\s]_0$ by matching our approximated generating function $C(z,\bar{h})$ with the ansatz $C(z,\bar{h})=C_{[\sigma \sigma ]_0}(\bar{h})z^{2h_{\sigma }+\delta h_{[\sigma \sigma ]_0}(\bar{h})}+\text{...}=C_{[\sigma \sigma ]_0}(\bar{h})z^{2h_{\sigma }}(1+\log (z)\delta h_{[\sigma \sigma ]_0}+\text{...})+\text{...}$. This matching is of course can't be exact and we will use the recipe in \cite{Simmons-Duffin:2016wlq,Albayrak:2019gnz} : if our approximated generating function has the form of $C(z,\bar{h})=z^{2h_{\sigma }}(A(\bar{h})+B(\bar{h}) \log (z))$, we extract the OPE coefficient and dimension as
\be\label{equ:h&f}
\delta h_{[\sigma \sigma ]_{0,\ell }}(\bar{h})=\frac{B(\bar{h})}{A(\bar{h})+\log (z_0)B(\bar{h})},\qquad f_{\sigma \sigma [\sigma \sigma ]_{0,\ell }}^2(\bar{h})=A(\bar{h})
\ee
for a choice of $z_0\in(0,1]$. Now we solve the equation
\be\label{equ:solve_h_to_Delta}
\Delta _{[\sigma \sigma ]_{0,\ell }}=2\Delta _{\sigma }+\ell +2\delta h_{[\sigma \sigma ]_{0,\ell }}(\bar{h})
\ee
for $\Delta _{[\sigma \sigma ]_{0,\ell }},\bar{h}_{[\sigma \sigma ]_{0,\ell }}$ and plug back to (\ref{equ:h&f}) to get $f_{\sigma \sigma [\sigma \sigma ]_{0,\ell }}$.

\begin{figure}[!htpb]%
\centering
\includegraphics[width = 0.47\textwidth]{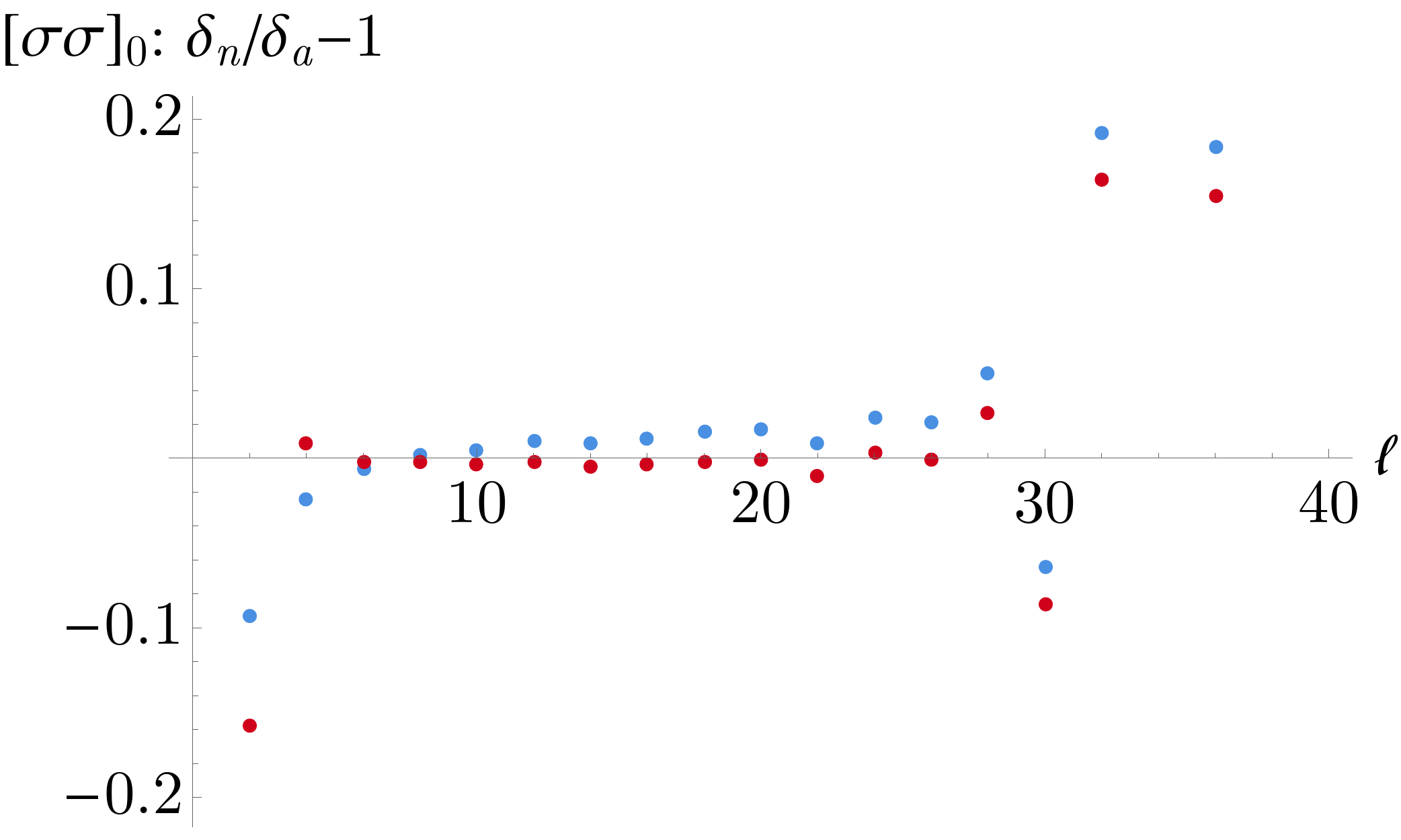}
\includegraphics[width = 0.47\textwidth]{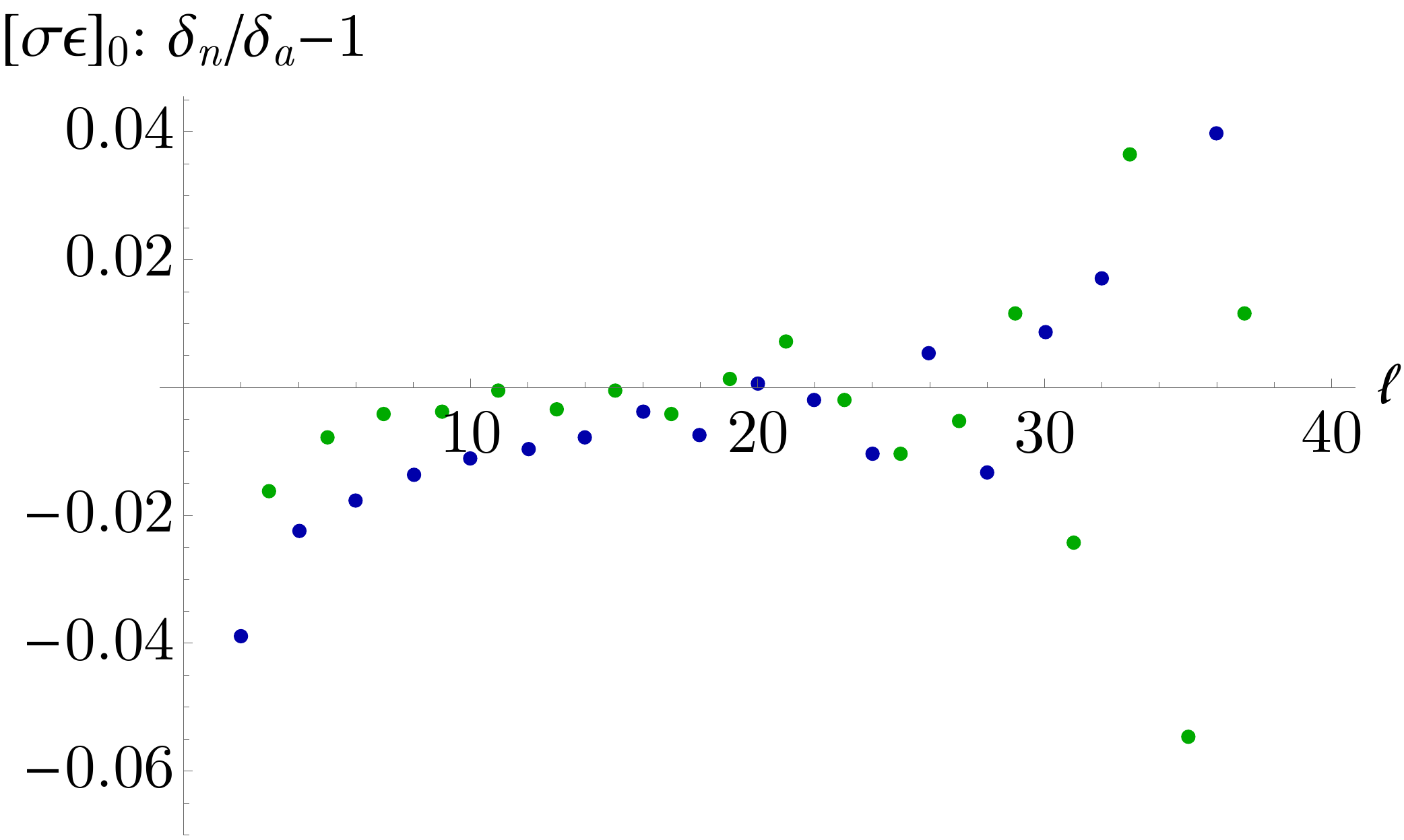}
\caption{\label{fig:analyticerr}Comparing the lightcone analytics with the $\Lambda=43$ EFM spectrum. Horizontal axes : spin. Vertical axes : $\delta^{(analytic)}/\delta^{(numerics)}-1$, where $\delta$ is the correction to the GFF dimension. On the left: Red dots are from Analytics Setup \ref{analytics:1}; Blue dots are from Analytics Setup \ref{analytics:2}. On the right : green/blue dots are for odd/even spin, both from Analytics Setup \ref{analytics:2}. All the analytics are computed using the values in (\ref{data:pd43}).}
\end{figure}

If we take the operator set $\{\mathcal{O}'\}$ to be $\{\e,T\}$ and $p=q=m=0$ in (\ref{equ:cOp}), (\ref{equ:kappaOmega}), $z_0=1$ in (\ref{equ:h&f}), we recover the lightcone bootstrap result (6.1) of \cite{Simmons-Duffin:2016wlq}. We show the percentage difference of this result to $\Lambda=43$ result in Figure~\ref{fig:analyticerr} (the blue dots). There is an almost constant $1.8\%$ discrepancy at larger spins, which doesn't meet our goal. Now let's try to understand why.

Consider a generic set of operators $\{\mathcal{O}'\}$. Expand $\lambda _{\sigma \sigma [\sigma \sigma ]_0}^2, \delta h(\bar{h})$ in large $\hb$ limit and keep only the leading term for each operator $\cOp$:
\begin{align}\label{equ:large_hb}
&\delta h(\bar{h})\approx -\sum _{\mathcal{O}' }\lambda _{\sigma \sigma \mathcal{O}'}^2\frac{\Gamma (2 h_{\sigma })^2 \Gamma (2 \bar{h}_{\mathcal{O}'}) }{\Gamma (2h_{\sigma }-h_{\mathcal{O}'}) ^2\Gamma (\bar{h}_{\mathcal{O}'})^2} \bar{h}^{-2 h_{\mathcal{O}'}}\\\nonumber
&\lambda _{\sigma \sigma [\sigma \sigma ]_0}^2/\lambda _{\text{GFF}}^2\approx (1-\sum _{\mathcal{O}' }\lambda _{\sigma \sigma \mathcal{O}'}^2\frac{ \Gamma (2 \bar{h}_{\mathcal{O}'}) }{\Gamma (\bar{h}_{\mathcal{O}'})^2}\frac{\Gamma (2 h_{\sigma })^2}{\Gamma (2h_{\sigma }-h_{\mathcal{O}'})^2}2 H_{\bar{h}_{\mathcal{O}'}-1}\bar{h}^{-2 h_{\mathcal{O}'}})\\\nonumber
&\lambda _{\text{GFF}}^2\approx \frac{\sqrt{\pi }}{\Gamma (2 h_{\sigma })^2}2^{3-2 \bar{h}} \bar{h}^{-\frac{3}{2}+4h_{\sigma }}
\end{align}
where $H$ is the harmonic number. We see that at large spin, the leading contributions come from operators with small twists. In $\dsZ_2$-even channel, $\{\e,T,[\s\s]_{0,\ell}\}$ are the small twist operators with $\tau<1.5$. Therefore potentially they can all affect the large spin result, so we should sum up the contribution from $[\s\s]_{0,\ell}$. Following the double twist improvement (DTI) procedure of \cite{Albayrak:2019gnz}), a single operator in $\mathcal{O}'\in [\sigma \sigma ]_{0,\ell }$ contribute to $C^t$ as
\begin{align}\label{equ:DTI}
C^t(z,\bar{h})_{[\sigma \sigma ]_{0,\ell }}\approx \frac{\sqrt{\pi }}{\Gamma (2 h_{\sigma })^2}2^{4-2 \bar{h}_{\mathcal{O}'}} \bar{h}_{\mathcal{O}'}^{-\frac{3}{2}+4h_{\sigma }}(\lambda _{\sigma \sigma [\sigma \sigma ]_0}^2/\lambda _{\text{GFF}}^2)(\delta h_{\mathcal{O}'})^2\\\nonumber
\sum _{p=0}^{\infty }\sum _{q=-p}^p\hat{\mathcal{A}}_{p,q}^{0,0}(2h_{\sigma })\frac{z^{2 h_{\sigma }}k_{\bar{h}_{\mathcal{O}'}+q}^{0,0}(1-z)}{(1-z)^{2 h_{\sigma }}}\kappa \Omega _{\bar{h},2h_{\sigma }+p,2 h_{\sigma }}^{h_{\sigma }}
\end{align}
Now we can expand $(\lambda _{\sigma \sigma [\sigma \sigma ]_0}^2/\lambda _{\text{GFF}}^2)(\delta h_{\mathcal{O}'})^2$ to a linear combination of $1/\hb^\tau_i$. For each term, we use (4.41) of \cite{Simmons-Duffin:2016wlq} to approximately sum up the contribution from $\ell=\ell_0, \ell_0+2, ...\infty$. For example, if we pick up a term proportional to $1/\bar{h}^{2 h_{\mathcal{O}'_k}}$ from $(\lambda _{\sigma \sigma [\sigma \sigma ]_0}^2/\lambda _{\text{GFF}}^2)$ and a term proportional to $(1/\bar{h}^{2 h_{\mathcal{O}'_i}})(1/\bar{h}^{2 h_{\mathcal{O}'_j}})$ from $(\delta h_{\mathcal{O}'})^2$, the summation\footnote{Interestingly, this summation also contributes to the term $z^{\tau_{tot}}$, which seems to be corresponding to a triple trace operator $[\mathcal{O}_i\mathcal{O}_j\mathcal{O}_k]$. Similarly for $[\s\e]_0$, $\delta h_{[\s\e]_0}$ is a linear combination of $\hb^{-h_\s},1/\hb^{-h_T}, ...$ at large spin. Taking $\hb^{-(h_\s+h_T)}$ term in $\delta h_{[\s\e]_0}^2$, with the help of (4.47) of \cite{Simmons-Duffin:2016wlq} we can sum up the double twist $[\s\e]_0$ contribution to the generating function of $\<\s\e\s\e\>$. This summation contribute to $z^{h_\s+h_T}$ of the generating function. This seems to suggests that $[\s T]_0$ in principle should appear in the bootstrap of the $\{\s,\e\}$ system.} contributes to the $z^{2 h_\sigma }$ as 
%
%
\begin{align}
C^t(z,\bar{h})_{[\sigma \sigma ]_{0,\ell =\ell _0\text{...}\infty }}\approx 4 \pi ^2 D_{\text{tot}} \frac{\Gamma (\Delta _{\sigma }-\frac{\tau _{\text{tot}}}{2})^2}{\Gamma (\Delta _{\sigma })^2}\sum _{p=0}^{\infty }\sum _{q=-p}^p2^{2q}\hat{\mathcal{A}}_{p,q}^{0,0}(2h_{\sigma })z^{2h_{\sigma }}\kappa \Omega _{\bar{h},\Delta _{\sigma }+p,\Delta _{\sigma }}^{h_{\sigma }}\\\nonumber
\left(\alpha _0^{\text{even}}\left[S_{-\Delta _{\sigma }+\frac{\tau _{\text{tot}}}{2}}\right](q+\Delta _{\sigma }+\ell _0)\log (z)+\beta _0^{\text{even}}\left[S_{-\Delta _{\sigma }+\frac{\tau _{\text{tot}}}{2}}\right](q+\Delta _{\sigma }+\ell _0)\right)
\end{align}
where $\tau _{\text{tot}}=\tau _i+\tau _j+\tau _k, D_{\text{tot}}=D_i D_j D_k 2H_{\bar{h}_{\mathcal{O}'_k}-1}$ with $D_{\mathcal{O}'}=-\lambda _{\sigma \sigma \mathcal{O}'}^2\frac{\Gamma (2 h_{\sigma })^2\Gamma (2 \bar{h}_{\mathcal{O}'}) }{\Gamma (2h_{\sigma }-h_{\mathcal{O}'}) ^2\Gamma (\bar{h}_{\mathcal{O}'})^2}$, and $\alpha _0^{\text{even}}, \beta _0^{\text{even}}$ are defined in (4.41) of \cite{Simmons-Duffin:2016wlq}.

Now we can compute the individual contribution of $\{\e,T\}$ using (\ref{equ:cOp}) and sum up the double twist contribution using (\ref{equ:DTI}) to get an approximate generating function. To further improve it, we may iterate the inversion formula a few times, \itie we take the spectrum extracted from the approximate generating function and plug it into the RHS of the inversion formula\footnote{Similar procedure has been used in \cite{Simmons-Duffin:2016wlq,Atanasov:2022bpi}.}. We also have to make a balance between our goals : in various infinite summations, we choose cut-off parameters small enough to make the computation fast, but also large enough so that the result is accurate. Specifically, our choice is the following.
\begin{analytics}\label{analytics:1}
We take cut-off parameters $p_\text{max}=1$ in (\ref{equ:cOp}), $m_\text{max}=1$ in (\ref{equ:kappaOmega}), $K_\text{max}=\ell_\text{max}=7$ in (4.40) of \cite{Simmons-Duffin:2016wlq} and $z_0=0.05$. To solve (\ref{equ:solve_h_to_Delta}), we use GFF value as the initial value for $\Delta$ and iterate (\ref{equ:solve_h_to_Delta}) for 3 times. In first step, we take $\{\e,T\}$ as individual contributions using (\ref{equ:cOp}) and sum up double twist contribution from $\ell_0=4$ using (\ref{equ:DTI}). To do the double twist summation, we take the identity contribution for $\lambda _{\sigma \sigma [\sigma \sigma ]_0}^2/\lambda _{\text{GFF}}^2$ and the $\{\e,T\}$ contributions for $\delta h(\bar{h})$ in (\ref{equ:large_hb}). Using this generating function, we compute the dimension and OPE coefficient of $[\sigma \sigma ]_{0,\ell}$ for spin $4,... , 20$. In the second step. we take the spectrum $[\sigma \sigma ]_{0,\ell=4, ...20}$ from the first step as individual contributions and repeat the process with $\ell_0=22$.
\end{analytics}
The result only depends on $\{\ds,\de,\fsse,\fT\}$. It's also accurate and fast enough for our purpose\footnote{When implemented in Mathematica, on a computer with 64 AMD EPYC 7532 32-Core Processor, computations for $[\sigma \sigma ]_{0,\ell=4,... , 52}$ takes about 7 seconds.}. The result is shown as red dots in Figure~\ref{fig:analyticerr} by comparing with the $\Lambda=43$ spectrum of the Numerics Setup \ref{numerics:1}. Now we understand the $1.8\%$ discrepancy in the red dots : eventually the discrepancy will disappear for very large spins, but for the spin range of the figure, it has to be corrected by the DTI procedure.

In this work, we will mainly use the analytics in \cite{Simmons-Duffin:2016wlq} as our analytic input in the hybrid bootstrap, and occasionally we also use Analytics Setup \ref{analytics:1}. We won't repeat the derivation of \cite{Simmons-Duffin:2016wlq} here, but simply layout the notation:
\begin{analytics}\label{analytics:2}
For twist family $[\s\s]_{0,\ell}$, the analytic predictions of the conformal data~ $\D^{(analytic)}_{[\s\s]_{0,\ell}}=2\ds+\ell+\delta^{(analytic)}_{[\s\s]_{0,\ell}}$, $\f^{(analytic)}_{\s\s[\s\s]_{0,\ell}}$, $\f^{(analytic)}_{\e\e[\s\s]_{0,\ell}}$ are given by (6.1), (6.2), (6.31) of \cite{Simmons-Duffin:2016wlq}. For $[\s\e]_{0,\ell}$, the data $\D^{(analytic)}_{[\s\e]_{0,\ell}}=\ds+\de+\ell+\delta^{(analytic)}_{[\s\e]_{0,\ell}}$, $\f^{(analytic)}_{\s\e[\s\e]_{0,\ell}}$ is given by (6.32), (6.34) of \cite{Simmons-Duffin:2016wlq}.
\end{analytics}

The result of Analytics Setup \ref{analytics:2} is shown in Figure~\ref{fig:analyticerr} by comparing with the $\Lambda=43$ EFM spectrum\footnote{In this paper, when we want to show the spectrum of the leading twist families, we often plot the percentage difference respect to either the analytics or the $\Lambda=43$ EFM spectrum, because otherwise it's hard to see the tiny difference in the spectrum with bare eyes.}. One might notice they match well for the spin roughly from 10 to 25, but mismatch for smaller or larger spins. This is a key observation that inspired the construction of the hybrid method which will be discussed in the next subsection.

\section{The hybrid bootstrap}\label{sec:Ising_hybrid}

\subsection{General strategy}\label{sec:main_strategy}
\begin{figure}[!htpb]
\centering
\includegraphics[width = 0.6\textwidth]{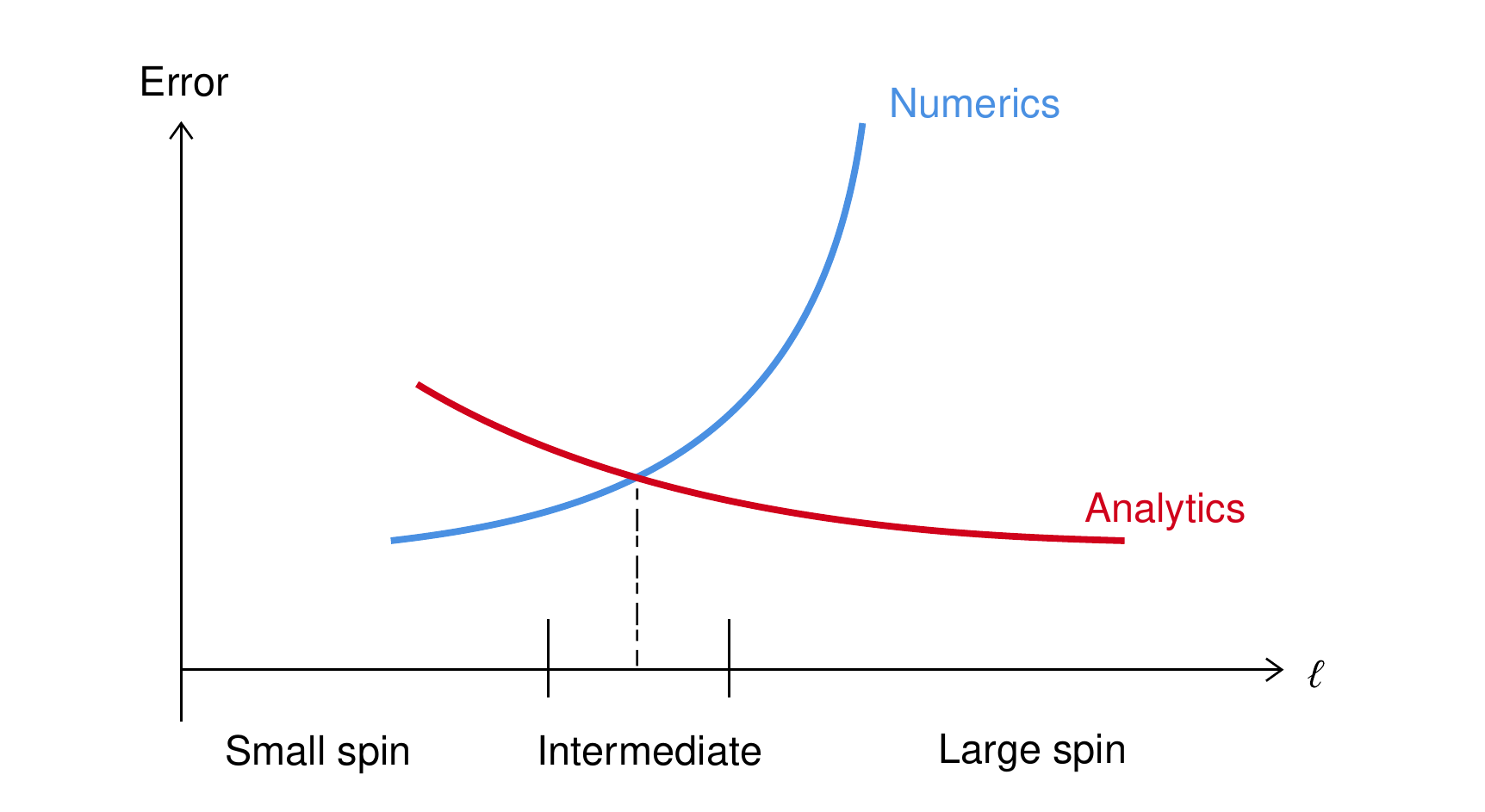}
\caption{\label{fig:NAcurve_schematic}A schematic figure to show the error in analytics v.s. numerics.} 
\end{figure}

Let's consider the leading twist family in a CFT. A general intuition is that the bootstrap numerics is sensitive to operators with small $\ell$. On the other hand, as the spin gets larger, the analytic lightcone bootstrap in Section \ref{sec:analytics} is more accurate. So there must be some intermediate spin range around $L_0$ where the analytics and numerics have similar accuracy. See the schematic Figure~\ref{fig:NAcurve_schematic}. That's why in Figure~\ref{fig:analyticerr}, the analytics and numerics don't precisely agree for small spin and large spin, but they match well for some intermediate values of spin. Since they make up for each other's shortcomings, one might expect to find a hybrid method that takes the best of both sides. To do that, we should trust the analytics for \textit{large spin region}, trust the numerics for \textit{small spin region} and demand they have similar results in the \textit{intermediate region}.

Since our analytic predictions are not exact, we need to find a relatively robust way to use them. A natural idea is that, starting from a certain spin $L_0$, we may demand the operator exist within a \textit{bounding band} around the analytic prediction. The width of the band should be smaller as the spin gets larger, because we expect the analytics is more accurate and trustable. We then demand there is a gap between this band and the next twist family.
\begin{figure}[!ht]
\centering
\includegraphics[width = 0.8\textwidth]{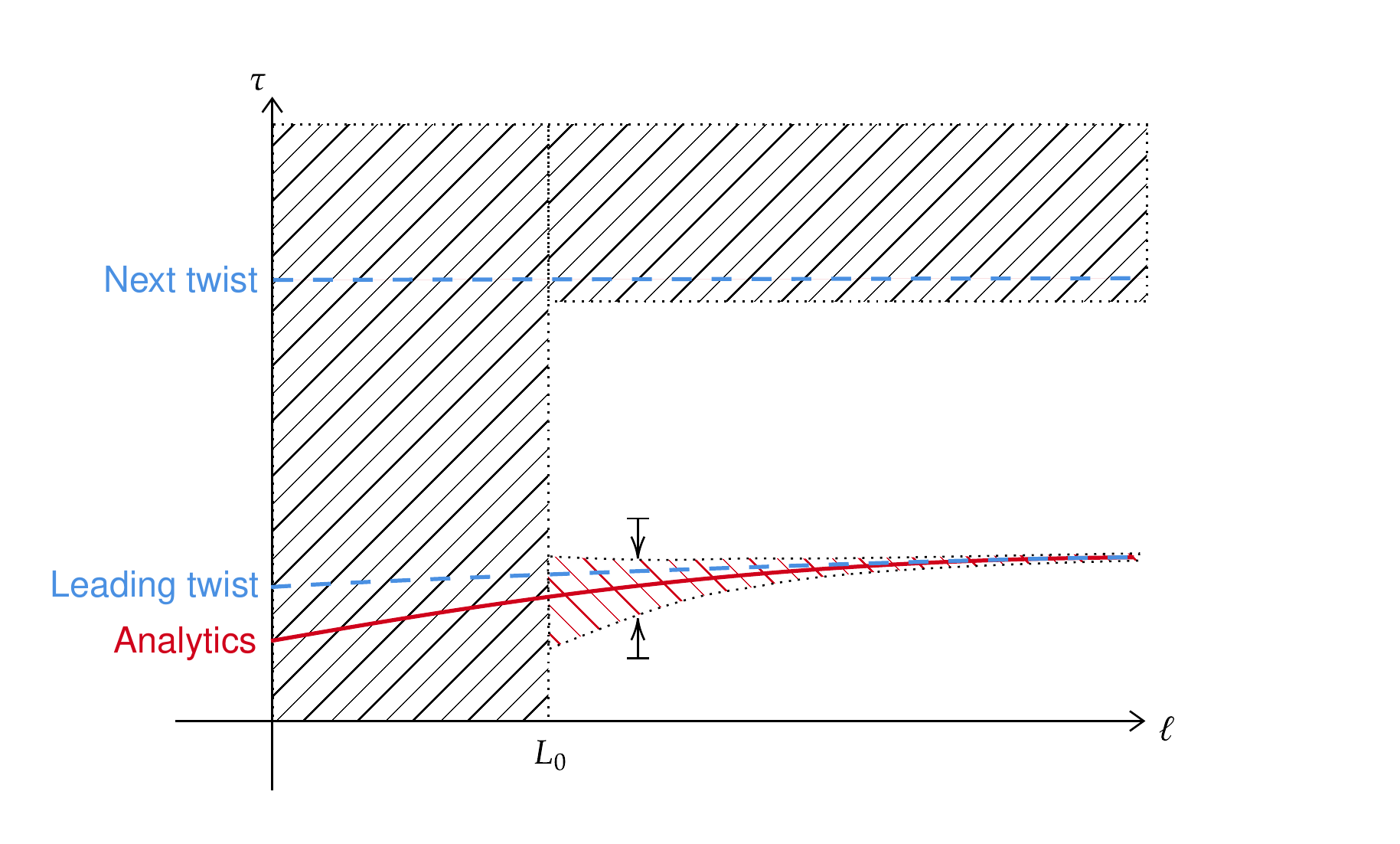}
\caption{\label{fig:hybrid_strategy_schematic}A schematic figure to show the strategy of the hybrid bootstrap. The red curve is the analytic prediction of the leading twist family. The blue dashed line is the leading twist and the next twist family of the actual CFT. We impose positivity on the shaded region and shrink the red shaded region as much as possible in order to glue the analytics with the numerics.} 
\end{figure}
See Figure~\ref{fig:hybrid_strategy_schematic} for a schematic illustration. Technically, those requirement translate to the following positivity requirement:
\be
\vec{\alpha}\cdot(V_{\D,\ell})\ge 0 \text{ for } \D^{(\text{band})}_\text{min} \le \D \le \D^{\text{(band)}}_\text{max} \text{ and } \D \ge \D_{gap},
\ee
where $V$ is the crossing vector. Such conditions can be implemented using the ``interval positivity"\cite{Albayrak:2021xtd}. To demand that the leading twist is as close as possible to the analytics, we may minimize the width of the band until the feasible region of the parameter (such as the dimension of external operators) shrinks to a point.

One may make sense of this process in the following way. The bootstrap equation at finite $\Lambda$ is a truncated bootstrap equation. It admits many solutions, located at different points of the feasible region. By demanding the band is as narrow as possible until the feasible region shrinks to a point, we single out an unique solution whose leading twist matches the most with our analytics.

If we also have the analytic prediction for the OPE coefficient of the leading twist, we may inject that information into numerics as well. To use the information of the OPE coefficient, we have to sum the contribution of the operators with the identity block (\textit{i.e.} $G_{0,0}(u,v) + \f_{\calO}^2 G_{\D_\calO,\ell_\calO}(u,v)+...$). To make the process more robust, when plug in OPE coefficient, we may make it slightly smaller than the analytic prediction, then demand it to be as close as possible to the analytics until the feasible region shrink to a point\footnote{A useful observation is that, from the sharing effect in \cite{Simmons-Duffin:2016wlq}, we see that the numerics care more about the total ``effective" OPE coefficient from a group of nearby operators, but care less about the specific $\D$ of those operators. Therefore assuming our OPE coefficient and $\D$ is not very accurate, it can still be corrected by another operator with a small OPE coefficient in the band.} Since we summed the major portion of the OPE coefficient with the identity, we already demanded the existence of the leading twist family with the spectrum data being compatible with the analytic prediction. 

If we know the information of higher twist families, we can continue this process by imposing more bounding bands for those twist families. We may also encounter a situation where we have major twist families with big OPE coefficients and minor twist families with much smaller OPE coefficients. The 3D Ising $\dsZ_2$-odd channel is such an example: the leading twist $[\s T]_0$ has very small OPE coefficients while the major twist $[\s\e]_0$ is above $[\s T]_0$ but have a much bigger OPE coefficients. In this situation, the above strategy still works for the major twist families as long as $L_0$ is not too small. We can simply ignore the minor twist families, because for spin larger than $L_0$ the numerics are not very sensitive and won't be affected too much by the minor twist families. Of course, if we have analytic predictions for the minor twists, we may plug in those data and sum the contributions of the twist families with the identity block to make the bootstrap more robust.

Now we still need to know concretely how to choose $L_0$ as well as the form of the band. Careful analysis of the error on both the numerics and the analytics could certainly help to determine the $L_0$. But even if we don't have the quantitative version of Figure~\ref{fig:NAcurve_schematic}, there are some simple and model independent rules to determine $L_0$. We will discuss this in detail in Section \ref{sec:how_to_determine_L0}. For the form of the band, in fact any form that shrinks sufficiently fast in large spins is fine. We have some simple and model independent choices that work well.

In Section \ref{sec:Ising_hybrid}, \ref{sec:how_to_determine_L0}, \ref{sec:improve_error}, we will implement above strategy to the Ising $\{\s,\e\}$ system and show concretely how and why it works. But before that, in the next Section \ref{sec:alternative_approaches}, let's comment on a few alternative approaches and their advantages and disadvantages. We will explain why we choose our main approach in this section over the alternatives.

\subsection{Alternative approaches} \label{sec:alternative_approaches}

A natural idea is that why don't we simply plug the analytic data of the leading twist into the bootstrap equation and just minimize the navigator function to get a prediction for the actual CFT? We certainly could do this for the large spin region of Figure~\ref{fig:NAcurve_schematic}. But if we also plug in data for the intermediate spin and even small spin, the feasible region in the parameter space will disappear and the scale of the minimum navigator value (positive) is usually much larger than the scale of the minimum navigator in Section~\ref{sec:numerics}, indicating that we get an even more wrong solution. Now let's assume we only do it for the large spin region. This is in fact equivalent to our main strategy with a large $L_0$ and the band shrinks to a line. The analytics is indeed much more accurate than the numerics in the large spin region. However the numerics is also exponentially insensitive to those large spin data, therefore the constraining power is also loose. It is really the process of ``gluing" the numerics and analytics in the intermediate region generating a strong constraining power. Let's make an analogy to illustrate this argument a bit more. In the loop diagram computation, one may compute $4-\epsilon$ expansion and the $2+\epsilon$ expansion, then use the 2-sided pedé approximation to smoothly interpolate between 2d and 4d. If one only uses $2+\epsilon$ expansion, usually the result won't be very accurate for larger $\epsilon$ due to a branch cut at 4d. Similarly, in the hybrid bootstrap the numerics and the analytics work for different regions and we should smoothly connect them in the intermediate region. Without demanding they match in the intermediate region is just like that in the loop computation, we use $2+\epsilon$ and $4-\epsilon$ expansion for $2<d<3$ and $3<d<4$ respectively but don't demand they smoothly match at $d=3$: then the $4-\epsilon$ expansion won't help with the overall accuracy for $2<d<3$. However this alternative approach could still be very useful for another purpose. In Section \ref{sec:how_to_determine_L0}, we will use it to determine the intermediate region.

\begin{figure}[!htpb]%
\centering
\includegraphics[width = 0.47\textwidth]{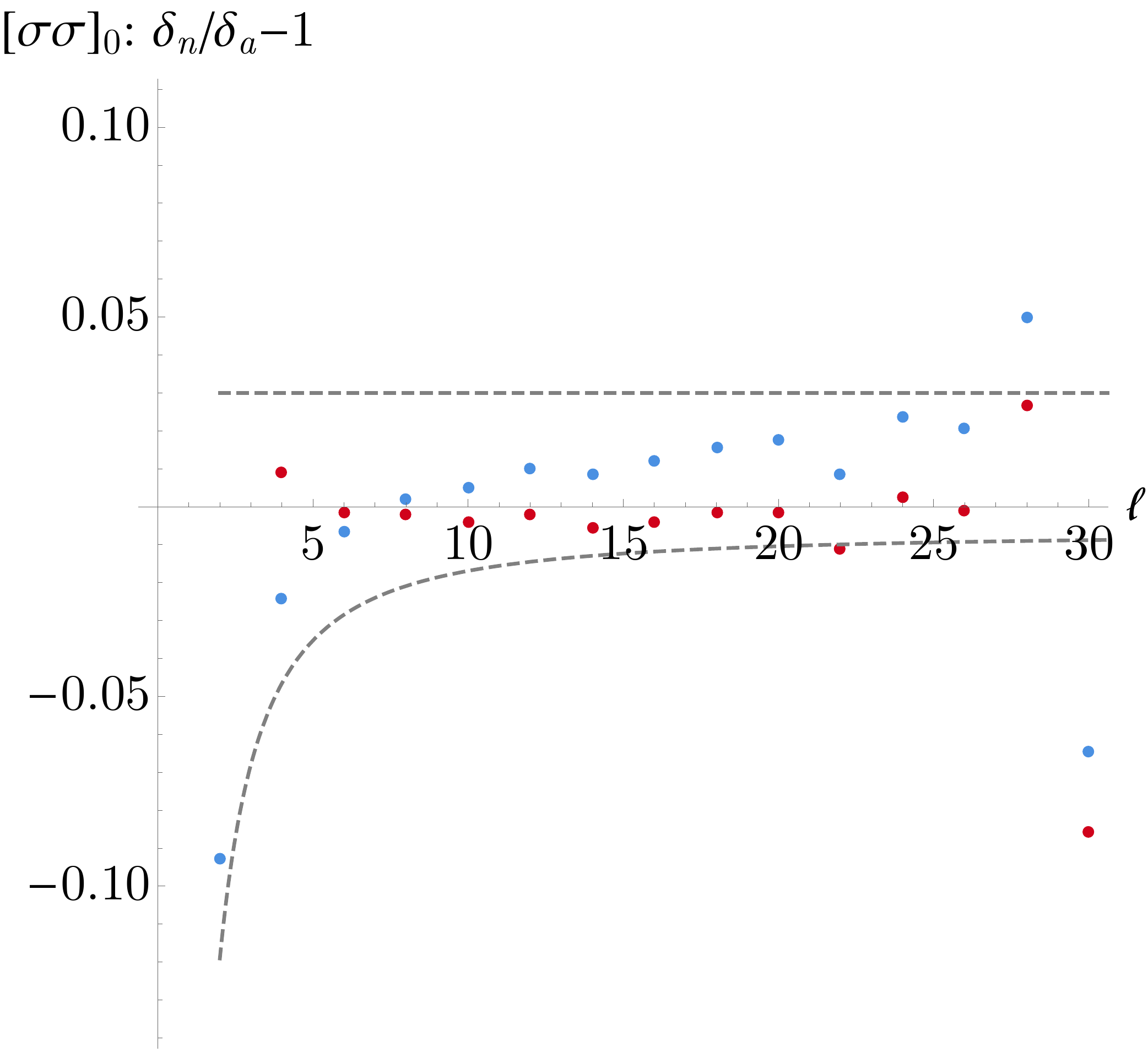}
\includegraphics[width = 0.44\textwidth]{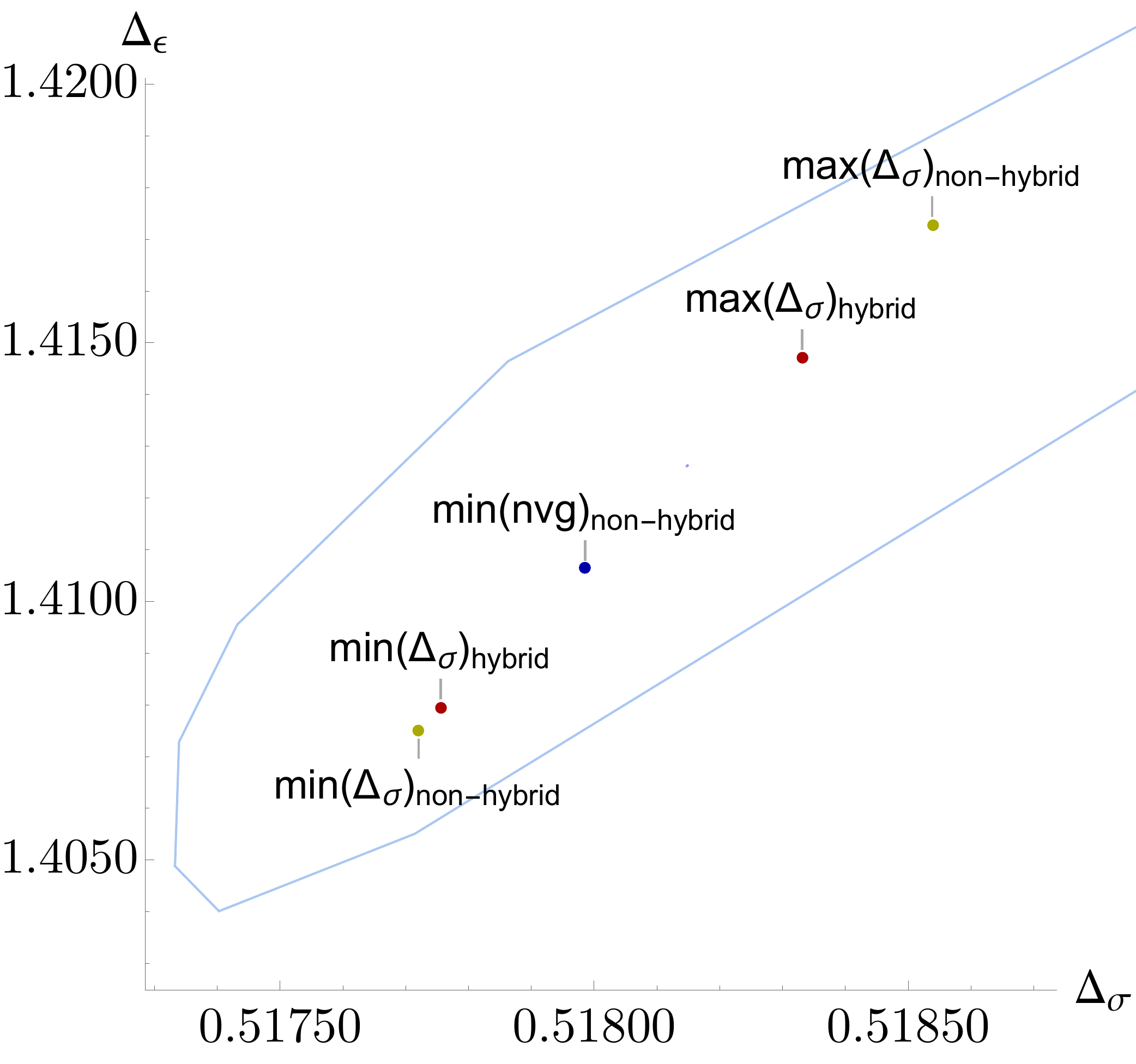}
\caption{\label{fig:soliderr}On the left: The red dots (from Analytics Setup \ref{analytics:1}) and blue dots (from Analytics Setup \ref{analytics:2}) are same as Figure \ref{fig:analyticerr}. The two dashed lines are $\text{Err}^{(upper)}_{[\s\s]_0}=0.03$, $\text{Err}^{(lower)}_{[\s\s]_0}=-0.317285/\ell^{1.5}-0.00687528$. On the right : The optimal points from maximizing/minimizing $\ds$ and minimizing the navigator function in the feasible region at $\Lambda=11$. The ``hybrid" points are from the computation of the Hybrid Setup \ref{hybrid:mild_gaps}. The ``non-hybrid" points are from the computation of the Numerics Setup \ref{numerics:pure}. The blue line is the boundary of the $\Lambda=11$ island from Figure 13 of \cite{Reehorst:2021ykw}.}
\end{figure}

Another alternative approach is that we may make a very mild assumption of the error of our analytics, then maximize/minimize the various parameters in the feasible region. If our assumption is mild enough so that we are totally confident about it's validity, the feasible region we get from the hybrid bootstrap will be almost a rigorous bound. This strategy actually could work. Let's use the Ising $\{\s,\e\}$ system as an example. Based on Figure~\ref{fig:analyticerr}, we can make a conservative assumption on the error of $\D^{(analytic)}_{[\s\s]_0}$ from the Analytics Setup \ref{analytics:2} (blue dot), by comparing it with either the $\Lambda=43$ EFM spectrum, or the Analytics Setup \ref{analytics:1} (red dots). Our assumption is plotted as dashed lines in the left of Figure \ref{fig:soliderr}.

We assume the actual CFT operators $\D_{[\s\s]_0,\ell \ge 4}$ located between the two dashed lines. To be rigorous, we won't put a bound for $[\s\e]_0$ family in the $\dsZ_2$ odd channel, because there are operators (such as $[\s T]_0$) below the $[\s\e]_0$ family. Concretely, the strategy translate to the following SDP that depends on the parameters $\{\ds,\de,\f_{\e\e\e}/\f_{\s\s\e}\}$ :
\begin{align}
	&objective=\vec{\alpha}\cdot(\Vp_{\Delta=0,\ell=0}+(\lambda_{\s\s\e},\lambda_{\e\e\e}).\Vth.(\lambda_{\s\s\e},\lambda_{\e\e\e})\nonumber\\
	&\qquad\qquad\qquad+\lambda_T^2(\Ds,\De).\Vp_{\Delta=3,\ell=2}.(\Ds,\De)) \nonumber\\
	&conditions:\nonumber\\
	&\vec{\alpha}\cdot(\Vp_{\D,\ell \ge 2})\ge 0 \text{ for } \D^{(+)}_{upper} \ge \D \ge \D^{(+)}_{lower} \text{ or } \D \ge \D^{(+)}_{gap}, \nonumber\\
	&\vec{\alpha}\cdot(\vec{V}^-_{\D,\ell \ge 1})\ge 0 \text{ for } \D^{(-)}_{upper} \ge \D \ge \D^{(-)}_{lower} \text{ or } \D \ge \D^{(-)}_{gap}, \nonumber\\
	&\textit{and the conditions of (\ref{cond:spin012}).}
	\label{cond:soliderr}
\end{align}
\begin{align}
	&\textit{conditions for $\dsZ_2$ even spin 0, 2 and $\dsZ_2$ odd spin 0, 1}:\nonumber\\
	&\vec{\alpha}\cdot(\Vp_{\D,\ell=0})\ge 0 \text{ for } \D \ge 3.7,\nonumber\\
	&\vec{\alpha}\cdot(\Vp_{\D,\ell=2})\ge 0 \text{ for } \D \ge 5,\nonumber\\
	&\vec{\alpha}\cdot(\Vm_{\D,\ell=0})\ge 0 \text{ for } \D \ge 5,\nonumber\\
	&\vec{\alpha}\cdot(\Vm_{\D,\ell=1})\ge 0 \text{ for } \D \ge 5.
	\label{cond:spin012}
\end{align}
The normalization of the SDP is the same as Section \ref{sec:numerics}. For $\dsZ_2$ even spin 0, 2 and $\dsZ_2$ odd spin 0, 1, we manually set up some gaps. Those conditions will be used repeatedly in different setups in later sections, so we label them separately. The full setup of the problem is
\begin{hybrid}\label{hybrid:mild_gaps}
The framework of the SDP is given by (\ref{cond:soliderr}). In the \textit{conditions}, we set
\begin{align*}
&\D^{(+)}_{\text{upper}/\text{lower}}=\D_{\text{GFF}}+\delta^{(analytic)}_{[\s\s]_{0,\ell}}(1+\text{Err}^{(upper/lower)}_{[\s\s]_0}),\\
&\D^{(+)}_\text{gap}=\D_{[\s\s]_{0,\ell}}^{(analytic)}+1,\\
&\D^{(-)}_{\text{upper}}=\D^{(-)}_{\text{lower}}=\D^{(-)}_{\text{gap}}=\D_{\text{unitary}}.
\end{align*}
where $\text{Err}^{(upper/lower)}_{[\s\s]_0}$ is the two dashed line of Figure~\ref{fig:soliderr}. We maximize/minimize the parameters (such as $\ds$) while maintaining $objective\le0$.
\end{hybrid}
The result for maximizing/minimizing $\ds$ at $\Lambda=11$ is presented in Figure \ref{fig:soliderr}. To benchmark the result, we also run another pure numerics (without using the analytic information). The setup are
\begin{numerics}\label{numerics:pure}
Minimize the navigator (\textit{i.e.} objective) or maximize/minimize the parameters (such as $\ds$) of the SDP of (\ref{cond:soliderr}) while maintaining $objective\le0$. For both $\bbZ_2$-even and $\bbZ_2$-odd channel, we set $\D^{(-)}_{upper}=\D^{(-)}_{lower}=\D^{(-)}_{gap}=\D_{unitary}$.
\end{numerics}
The result for $\Lambda=11$ is presented in Figure~\ref{fig:soliderr}. Comparing the Numerics Setup \ref{numerics:pure}, the error bar of Hybrid Setup \ref{hybrid:mild_gaps} shrinks but not very much. At higher $\Lambda$, the difference would be even less (the $\Lambda=19$ results are given in the Appendix \ref{app:results}). The effectiveness of the approach will really rely on how small the analytic error is in the intermediate region. One might make some compromise between this approach and our main approach in Section \ref{sec:main_strategy}, \textit{i.e.} demand the numerics to be fairly close to the analytics but still leave it with some finite size feasible region and use that as an estimation of error bar. The confidence level of the error bar will be the same as the confidence level of the error bar of the analytics.

If our analytics is very accurate compared to the numerics, both of the alternatives can actually work.



\subsection{Bootstrapping the 3D Ising CFT}\label{sec:fullsetup}

In this section, we use the 3D Ising $\{\s,\e\}$ system as an example to implement our hybrid bootstrap strategy. 

To demand the leading twist is as close as possible to the analytics, we use the following prescription. For $\ell<L_0$, we require $\D \ge \D_{unitary}$. For $\ell\ge L_0$, we sum up operators at $\D_{analytic}$ with OPE coefficient $\f=\f_{analytic} (1-e^{a \ell})$ where $a$ is a negative number. Meanwhile, we assume an operator can exist in the interval $[\D_{GFF}+\delta_{lower},\D_{GFF}+\delta_{upper}]$, where $\D_{GFF}$ is the GFF dimension and $\delta_{lower/upper}$ is given by minimum/maximum of $\delta_{analytic} (1\pm e^{a \ell})$. Then we minimize the parameter $a$ until the navigator function is zero. Note that in general the error bar of a CFT data from lightcone analytics does not decay exponentially. Instead, most like the decay behavior is a linear combination of the power function. Here we choose the exponential function because after $L_0$ the numerics is exponentially less sensitive, so we can just trust analytical prediction. The $1\pm e^{a \ell}$ form is just a model independent choice to glue the numerics with the analytics. There are more quantitative arguments in Section \ref{sec:how_to_determine_L0} to show why the exponential function is a good choice.

Technically the above strategy translates to the following SDP problem that smoothly depending on the parameters $\{\ds,\de,\fT,\fsse,\feee,a\}$:
\begin{align}
	&objective=\vec{\alpha}.(\Vp_{\Delta=0,\ell=0}+(\lambda_{\s\s\e},\lambda_{\e\e\e}).\Vth.(\lambda_{\s\s\e},\lambda_{\e\e\e})+
	\lambda_T^2(\Ds,\De).\Vp_{\Delta=3,\ell=2}.(\Ds,\De)+\nonumber\\
	&\sum_{\ell_{max}\ge \ell_\calO \ge L_0, \calO=[\s\s]_{0,\ell}} (\f^*_{\s\s\calO},\f^*_{\e\e\calO})\cdot \Vp_{\Delta_\calO,\ell_\calO} \cdot(\f^*_{\s\s\calO},\f^*_{\e\e\calO})+
	\sum_{\ell_{max}\ge \ell_\calO \ge L_0, \calO=[\s\e]_{0,\ell}} {\f^*_{\s\e\calO}}^2 \Vm_{\Delta_\calO,\ell_\calO}
	; \nonumber\\
	&conditions:\nonumber\\
	&\vec{\alpha}\cdot(\Vp_{\D,\ell \ge 2})\ge 0 \text{ for } \D^{(+)}_{\text{upper}} \ge \D \ge \D^{(+)}_{\text{lower}} \text{ or } \D \ge \D^{(+)}_{\text{gap}}, \nonumber\\
	&\vec{\alpha}\cdot(\Vm_{\D,\ell \ge 1})\ge 0 \text{ for } \D^{(-)}_{\text{upper}} \ge \D \ge \D^{(-)}_{\text{lower}} \text{ or } \D \ge \D^{(-)}_{\text{gap}}, \nonumber\\
	&\textit{and the conditions of (\ref{cond:spin012}).}
	\label{cond:experr}
\end{align}
The normalization of the SDP is defined similarly as in Section \ref{sec:numerics}. The full setup of the hybrid bootstrap is \footnote{Due to the proof in \cite{Kundu:2020gkz}, the leading twist in the singlet channel actually enjoys the convexity starting from spin 2, therefore we can demand twist gaps increase linearly between spin 2 and $L_0$, which is a stronger condition. The effect of this condition on the result is small though. In this paper, we will not use this condition. We thank Johan Henriksson for discussion on the convexity of in the singlet channel and suggest this stronger condition.}
\begin{hybrid}\label{hybrid:fullsetup}
The framework of the SDP is given by (\ref{cond:experr}). In the \textit{conditions}, we take
\begin{align*}
&\D^{(+/-)}_{\text{upper}/\text{lower}}=\D^{(+/-)}_{\text{gap}}=\D_{\text{unitary}} \quad\text{for}\quad \ell<L_{0},\\
&\D^{(+)}_{\text{upper}/\text{lower}}=\D_{\text{GFF}}+\delta^{(analytic)}_{[\s\s]_{0,\ell}}(1\pm e^{a \ell}) \quad\text{for}\quad \ell \ge L_{0},\\
&\D^{(-)}_{\text{upper}/\text{lower}}=\D_{\text{GFF}}+\delta^{(analytic)}_{[\s\e]_{0,\ell}}(1\pm e^{a \ell}) \quad\text{for}\quad \ell \ge L_{0},\\
&\D^{(+)}_\text{gap}=\D_{[\s\s]_{0,\ell}}^{(analytic)}+1 \quad\text{for}\quad \ell \ge L_{0},\\
&\D^{(-)}_\text{gap}=\D_{[\s\e]_{0,\ell}}^{(analytic)}+1 \quad\text{for}\quad \ell \ge L_{0}.
\end{align*}
In the \textit{objective}, we take $\f^*=\f_{analytic} (1-e^{a \ell})$ and $\ell_{max}=100$\footnote{Note $\ell_{max}$ only affects the summation in the \textit{objective}, while in the \textit{conditions} the spin set is given in Appendix~\ref{app:parameters}}. All analytics are computed using Analytics Setup \ref{analytics:2}. We minimize the $a$ parameter until the navigator (\itie \textit{objective}) is zero.
\end{hybrid}

The different analytic information ($\D^{(analytic)}_{[\s\s]_{0,\ell}}$, $\f^{(analytic)}_{\s\s[\s\s]_{0,\ell}}$, $\f^{(analytic)}_{\e\e[\s\s]_{0,\ell}}$, $\D^{(analytic)}_{[\s\e]_{0,\ell}}$, $\f^{(analytic)}_{\s\e[\s\e]_{0,\ell}}$) we used in the hybrid bootstrap may come with different accuracy. In a situation where one analytic information is much more accurate than another, we might have to choose the $a$ parameter differently in different places (say if $\D^{(analytic)}_{[\s\s]_{0,\ell}}$ is much more accurate than $\D^{(analytic)}_{[\s\e]_{0,\ell}}$, then the magnitude of the $a$ parameter for $\dsZ_2$-even channel should be larger than that of $\dsZ_2$-odd channel, \itie in intermediate region we demand the matching for $\D^{(analytic)}_{[\s\s]_{0,\ell}}$ should be better than for $\D^{(analytic)}_{[\s\e]_{0,\ell}}$). This is a complicated situation and a preliminary approach to determine different bounding bands will be discussed in Section \ref{sec:liebig}. In this section let's simply use same $a$ for all analytics. We also empirically fix some $L_0$. The justification for those choices will be discussed in the next section.
\begin{figure}[!htpb]
\centering
\includegraphics[width = 0.6\textwidth]{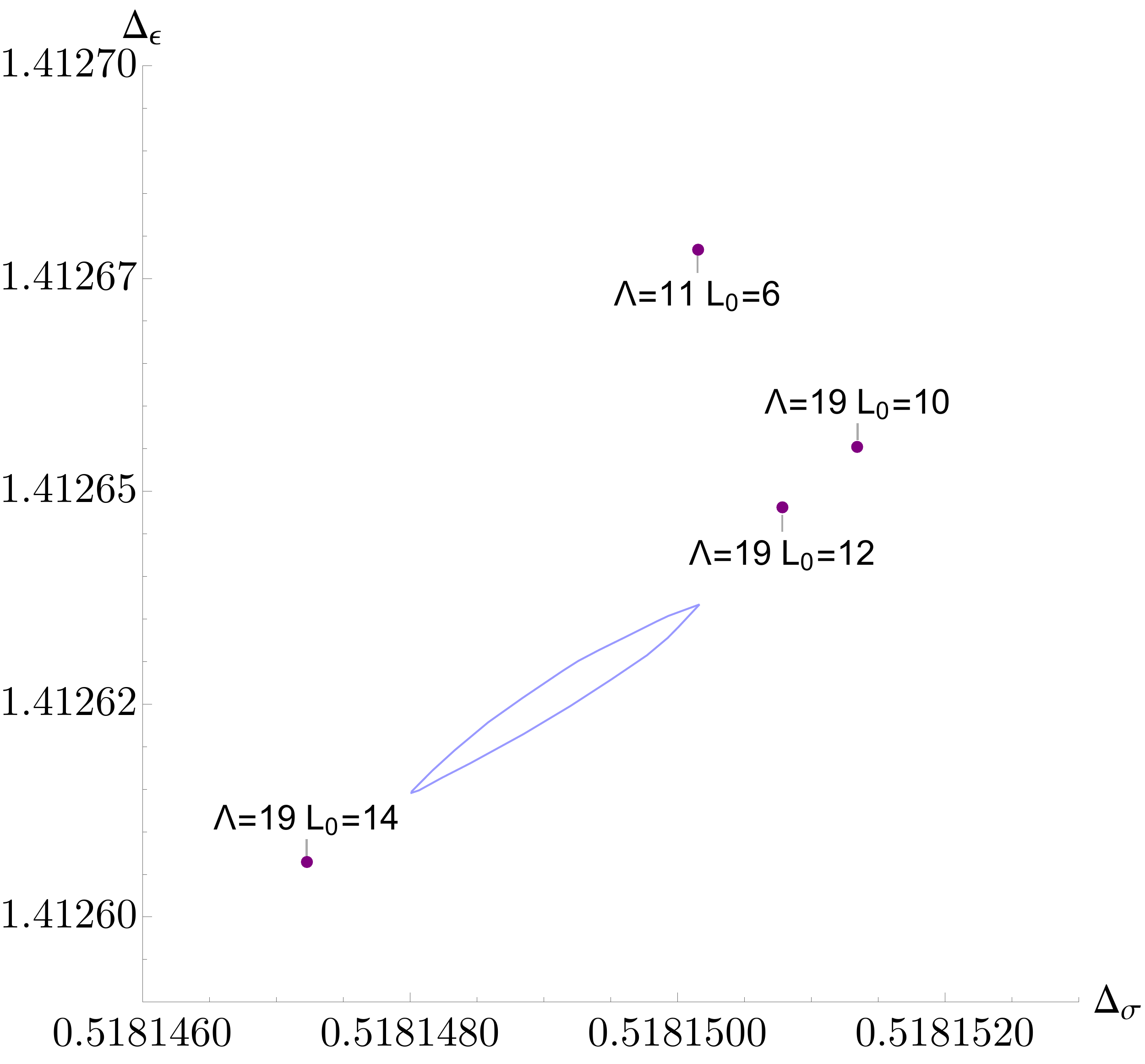}
\caption{\label{fig:hybrid_fullsetup}Results of the computations for the Hybrid Setup \ref{hybrid:fullsetup} at various $\Lambda$ and $L_0$. The light blue curve is the boundary of the $\Lambda=43$ Ising island of \cite{Kos:2016ysd}}
\end{figure} 
\begin{figure}[!htpb]
\centering
\includegraphics[width = 0.85\textwidth]{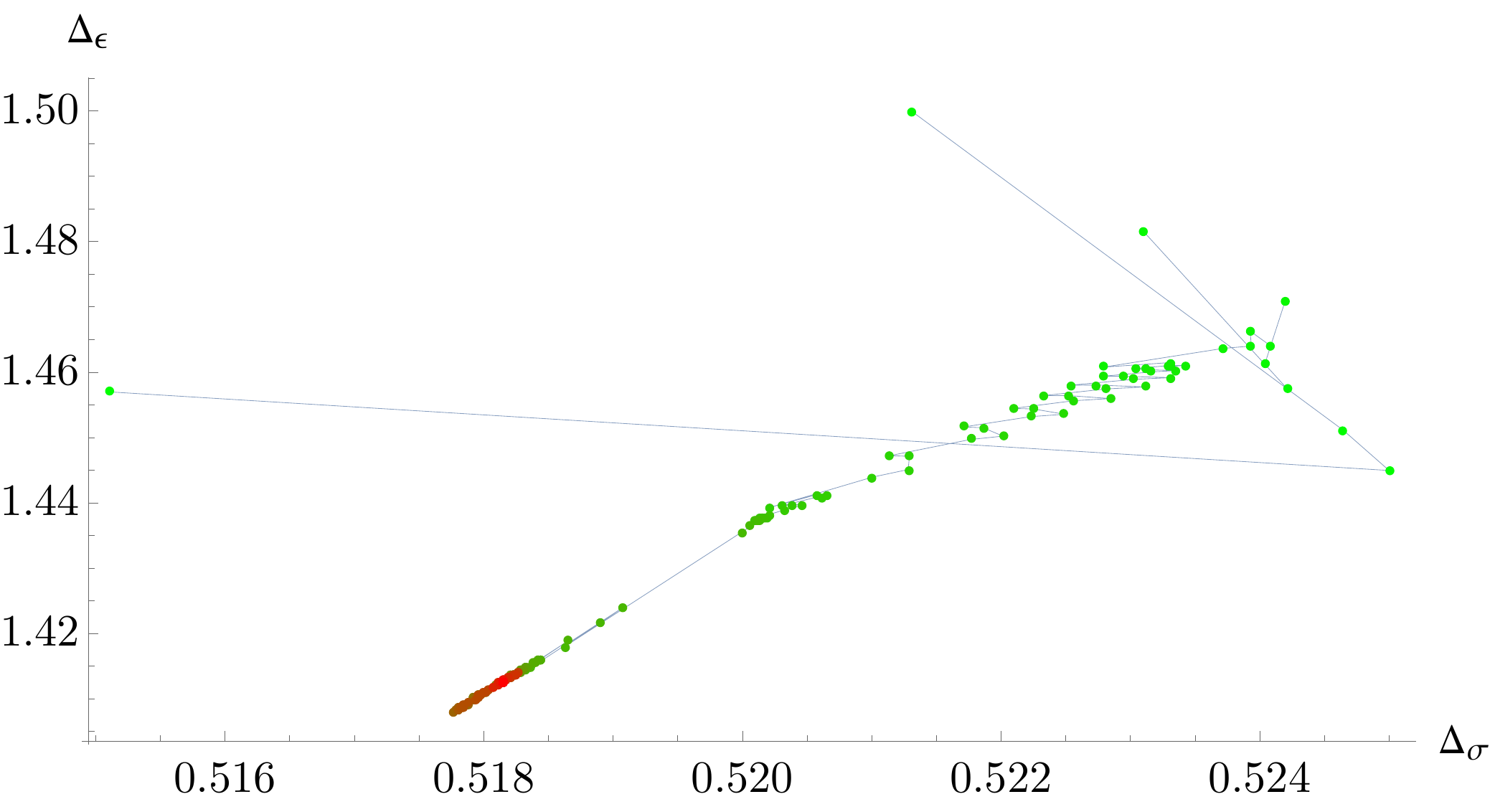}
\caption{\label{fig:minerr_fullpath}The path of the computation of Hybrid Setup \ref{hybrid:fullsetup} with $\Lambda=11, L_0=6$. The greenish points are points from early steps while the redish points are from later steps.}
\end{figure}

The result of Hybrid Setup \ref{hybrid:fullsetup} for various $\Lambda$ and $L_0$ in $(\D_\s,\D_\e)$ plane is plotted in Figure~\ref{fig:hybrid_fullsetup}, while the full results are in the Appendix \ref{app:results}. We see even with relatively small $\Lambda=11, 19$, the result is already quite accurate, comparing to the Numerics Setup \ref{numerics:1} and \ref{numerics:pure}. In Figure~\ref{fig:minerr_fullpath}, we show a path for the minimization $a$ in $\Lambda=11, L_0=6$ case, starting from $(\ds,\de,\fsse,\feee,\fT,a)=(0.5151, 1.457, 1, 1.5, 1.25, -0.1)$, which is quite far from the actual CFT. The algorithm is a modification of the Algorithm 2 of \cite{Reehorst:2021ykw} (see Appendix \ref{app:parameters} for more details). After 256 function calls, it eventually reached $(\ds,\de)=(0.51815015, 1.4126784)$.

In the next two subsections, we will use only one analytic information (either $\D_{[\s\e]_0}$ or $\D_{[\s\s]_0}$) for the hybrid bootstrap. The advantage is that with only one analytic information it's easier to analyze why and how the strategy works.

\subsection{Why it works? How to determine $L_0$?}\label{sec:how_to_determine_L0}
In this subsection, we will try to understand why the strategy works and analyze $L_0$ more carefully. To make the situation simpler, we will only use the analytics for $\D_{[\s\e]_0}$. The full setup is:

\begin{hybrid}\label{hybrid:se0}
The framework of the SDP is given by (\ref{cond:experr}). In the \textit{conditions}, we take
\begin{align*}
&\D^{(-)}_{\text{upper}/\text{lower}}=\D^{(-)}_{\text{gap}}=\D_{\text{unitary}} \quad\text{for}\quad \ell<L_{0},\\
&\D^{(-)}_{\text{upper}/\text{lower}}=\D_{\text{GFF}}+\delta^{(analytic)}_{[\s\e]_{0,\ell}}(1\pm e^{a \ell}) \quad\text{for}\quad \ell \ge L_{0},\\
&\D^{(-)}_\text{gap}=\D_{[\s\e]_{0,\ell}}^{(analytic)}+1 \quad\text{for}\quad \ell \ge L_{0},\\
&\D^{(+)}_{\text{upper}}=\D^{(+)}_{\text{lower}}=\D^{(+)}_{\text{gap}}=\D_{\text{unitary}} \quad\text{for all $\ell$}.
\end{align*}
In the \textit{objective}, we take all $\f^*=0$. All analytics are computed using Analytics Setup \ref{analytics:2}. We minimize the $a$ parameter until the navigator (\itie \textit{objective}) is zero.
\end{hybrid}

\begin{figure}[!htpb]
\centering
\includegraphics[width = 0.6\textwidth]{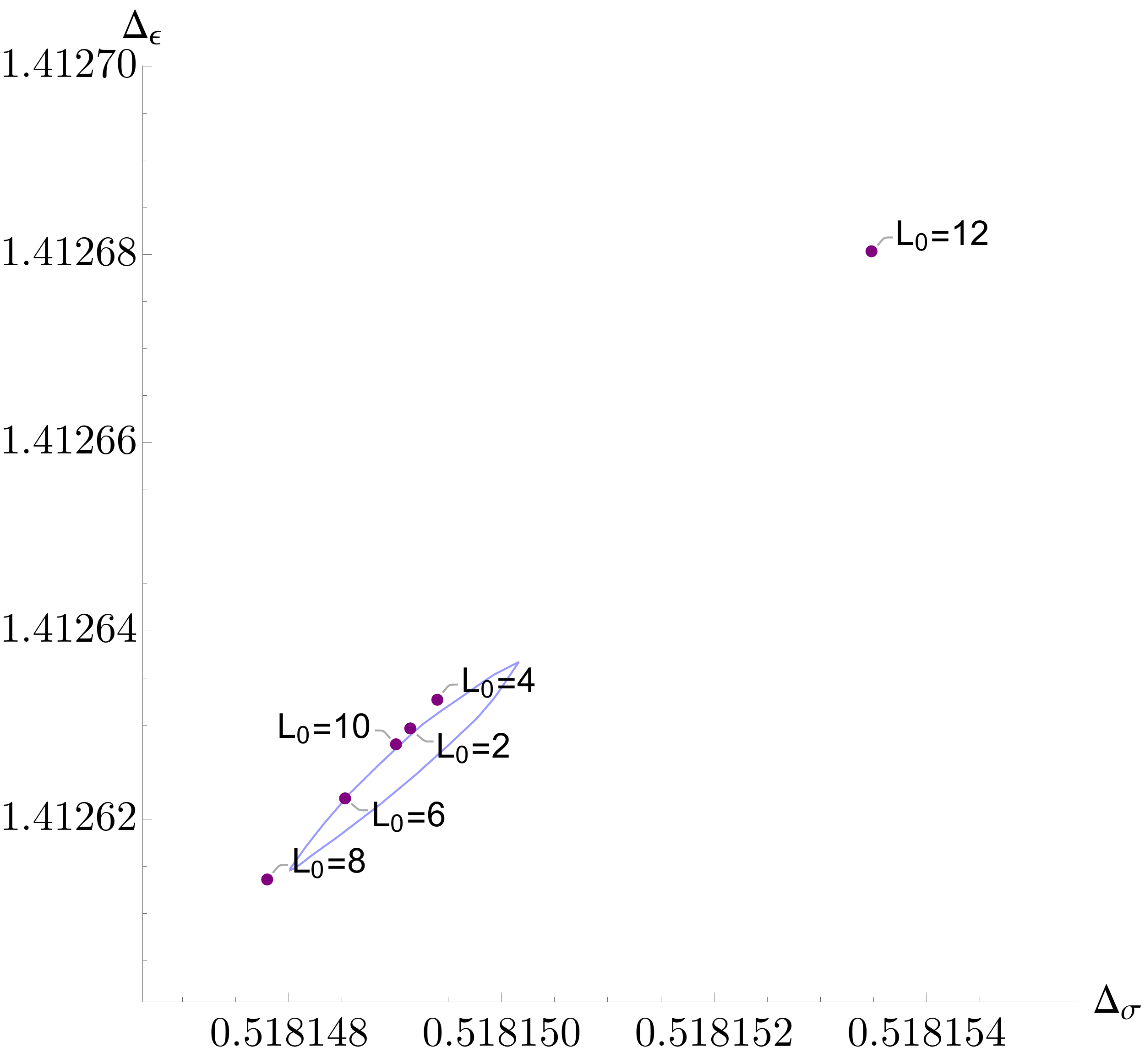}
\caption{\label{fig:hybrid_Dse0}Results of the computations for the Hybrid Setup \ref{hybrid:se0} at $\Lambda=19$ and various $L_0$. The light blue curve is the boundary of the $\Lambda=43$ Ising island of \cite{Kos:2016ysd}}
\end{figure}

\begin{figure}[!htpb]
\centering
\includegraphics[width = 0.6\textwidth]{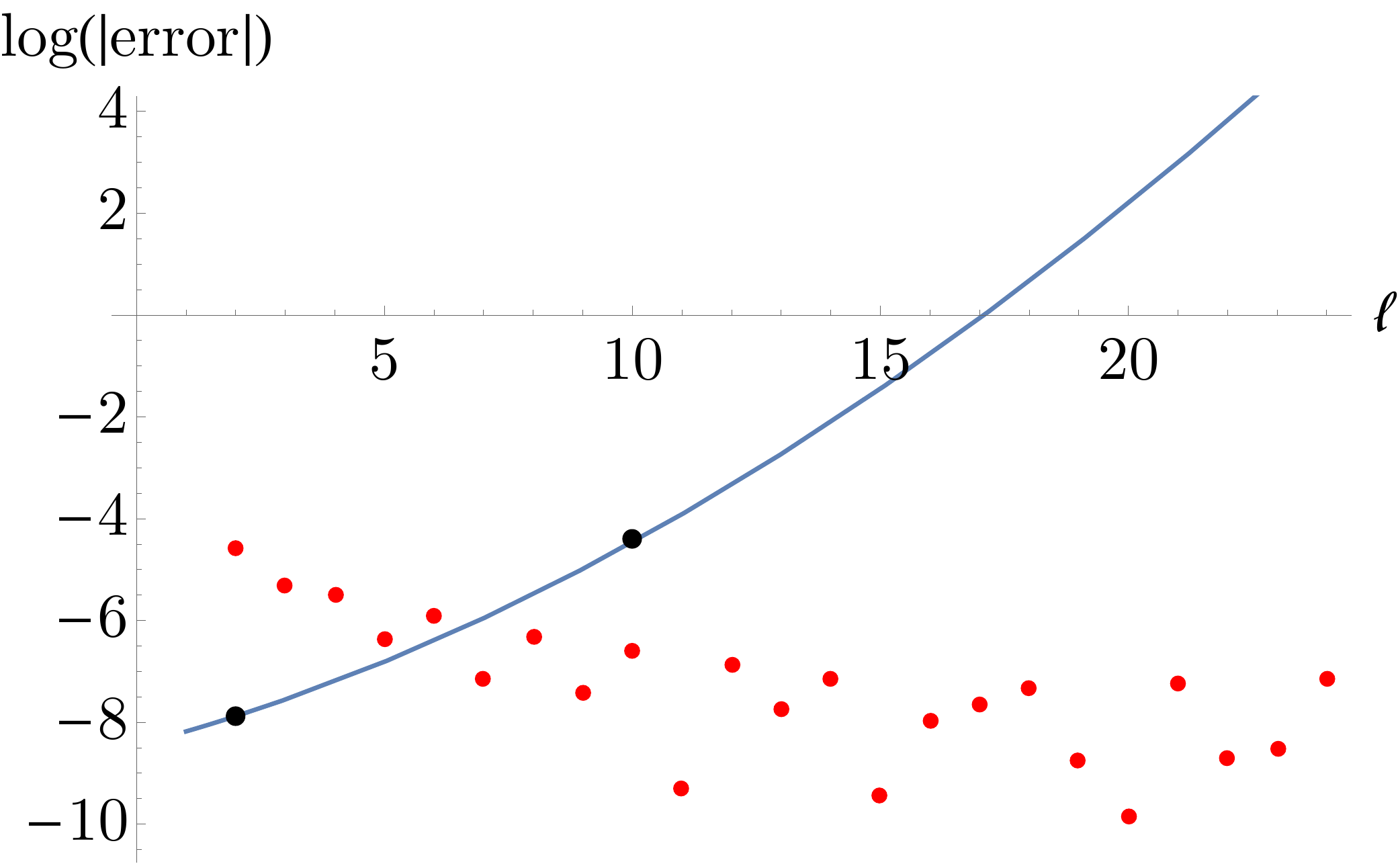}
\caption{\label{fig:NAcurve_se0}$\text{log}(\text{error}(\D_{[\s\e]_0}))$ v.s. spin. Red dots : estimated error of the analytics. Blue curve : estimated error of the numerics. Black dot: rigorous error bar for the numerics.
}
\end{figure}

The result for $\Lambda=19$ with various $L_0$ is plotted in Figure~\ref{fig:hybrid_Dse0}.
To understand how the numerics and analytics glued together, let's make a quantitative version of the Figure \ref{fig:NAcurve_schematic} for Hybrid Setup \ref{hybrid:se0}. In Figure~\ref{fig:NAcurve_se0}, the red dots are estimated errors of the analytics, obtained by comparing with the $\Lambda=43$ EFM spectrum. The blue curve is the estimated error of the pure numerics, obtained in the following way. We first set up a computation as (\ref{cond:experr}), but we take $L_0$ to be large enough so we can exactly plug in the scaling dimension from the analytics, \itie $\D^{(+/-)}_{\text{upper}}=\D^{(+/-)}_{\text{lower}}=\D^{(analytic)}$. We minimize the navigator function and find minimal navigator = $3.74\times10^{-12}$ (this order of magnitude doesn't sensitive to $L_0$). At the minimal navigator point, we compute the derivative of the navigator function respect to $\D^{(analytic)}$ at each spin, \itie $\partial_{\D_\ell}\cN$ (the order of magnitude is also insensitive to $L_0$ except when $\ell$ is close to $L_0$). Then $\partial_{\D_\ell}\cN/(3.74\times10^{-12})$ should be the estimation of the error bar at large $\ell$. For small spin, we can rigorously bootstrap the error bar. The computation is the same as (\ref{cond:nvg3par}) except for a certain spin $\ell_0$ we plug in $\D_{\ell_0}$ and assume next operator has dimension $\ge \D_{\ell_0}+1$. We then maximize/minimize $\D_{\ell_0}$ and the rigorous error bar is given by (max-min)/2. The black dots are the results of such computations for $\ell=2,10$. With those data, we fit the large spin log(error) with a degree 4 polynomial and demand the fit to match with $\ell=2$ rigorous data. The result is the blue curve : $-10.9349 + 0.242082 \ell + 0.0193981 \ell^2 - 0.000138822 \ell^3 + 4.02564\times10^{-7} \ell^4$. 

From Figure~\ref{fig:NAcurve_se0}, we see the errors of analytics and numerics intersecting around $\ell=7$. However this error of the numerics is for the pure numeric bootstrap (without injecting analytics), not the hybrid bootstrap. Once we injecting the information of analytics at large spin, the numerical error for the small spins will also be improved (as we have seen in Figure \ref{fig:hybrid_fullsetup} and Figure \ref{fig:hybrid_Dse0}), and the intermediate region should shift to the right. Therefore we may say the small spin region is $L_0\le7$. 

\begin{figure}[!htpb]%
\centering
\includegraphics[width = 0.47\textwidth]{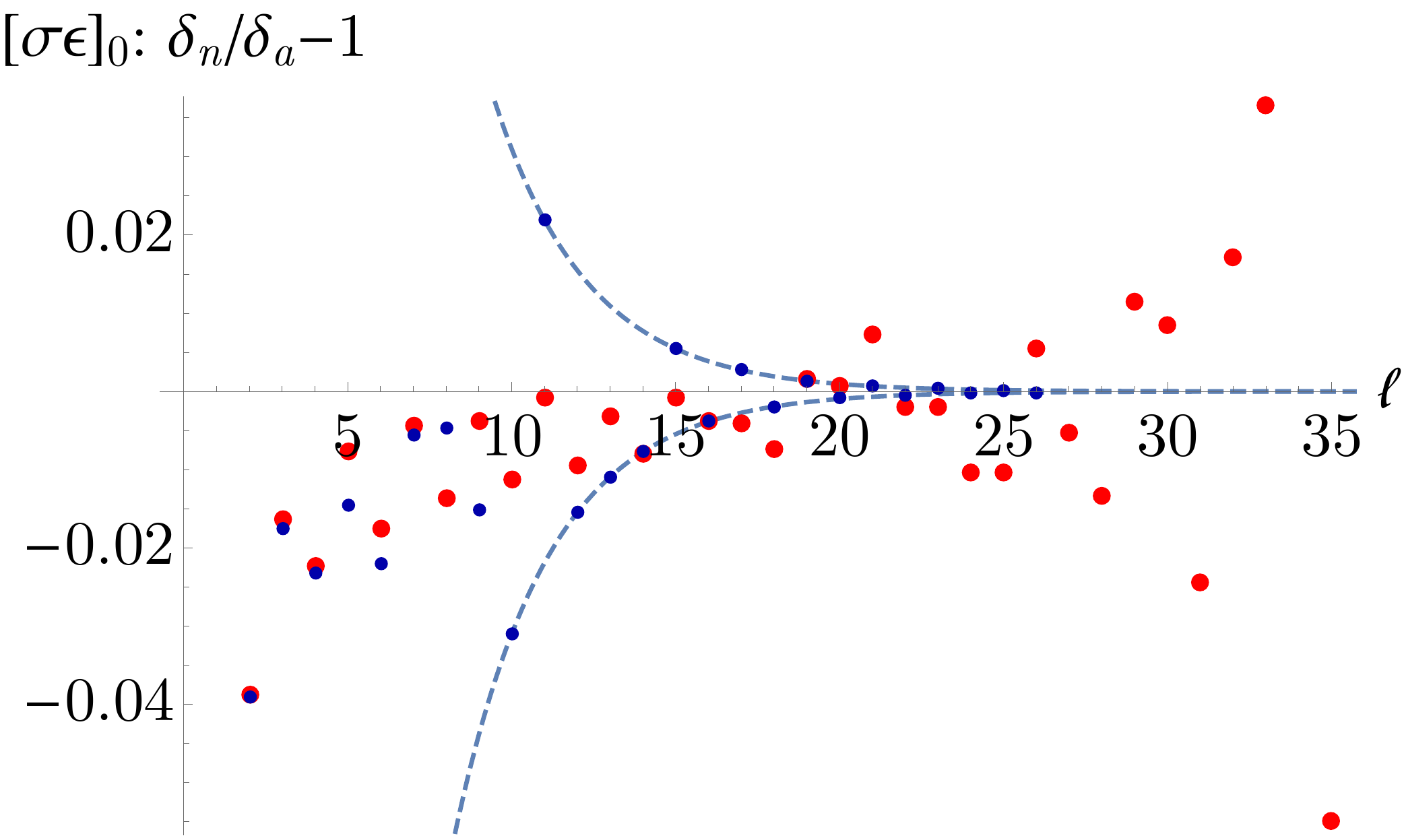}
\includegraphics[width = 0.47\textwidth]{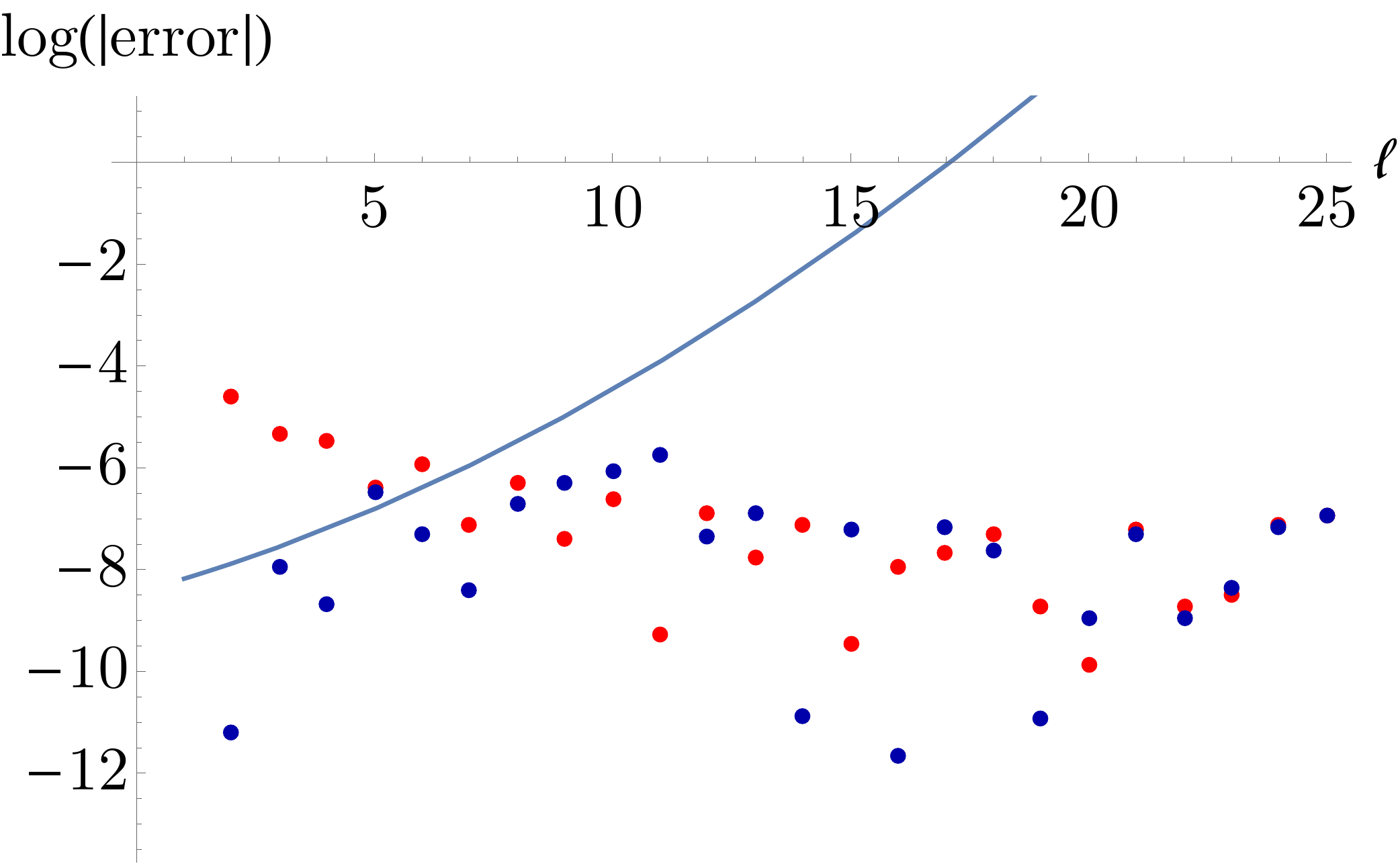}
\caption{\label{fig:hybrid_Dse0_err} In both figures, the blue dots are related to the EFM spectrum of $\D_{[\s\e]_0}$ from the Hybrid Setup \ref{hybrid:se0} at $\Lambda=19, L_0=10$. On the left: comparing the spectrum from the hybrid setup with the $\D^{(analytic)}_{[\s\e]_0}$ of the Analytics Setup \ref{analytics:2} computed at the values of (\ref{data:pd43}). Horizontal axes : spin. Vertical axes : $\delta^{(numerics)}/\delta^{(analytic)}-1$, where $\delta$ is the correction to the GFF dimension. Blue dots : $\delta^{(numerics)}$ is from the hybrid setup. Red dots:  $\delta^{(numerics)}$ is from the $\Lambda=43$ EFM spectrum (the same as the right of Figure \ref{fig:analyticerr}).  Dashed lines: $\pm e^{a \ell}$ with $a=-0.347263$ from the result of the hybrid computation. On the right: The red dots and blue line is the same as Figure \ref{fig:NAcurve_se0}. The blue dots is log(error) for the EFM spectrum $\D_{[\s\e]_0}$ of the hybrid setup, where the error is computed by comparing with the $\Lambda=43$ EFM spectrum of Section \ref{sec:numerics}.
}
\end{figure}

We expect the error in the hybrid bootstrap should schematically look like the dashed curve of Figure \ref{fig:NAcurve_hybrid_vs_pure} in Section \ref{sec:introduction}. Let's try to quantitatively confirm this scenario in the example of the Hybrid Setup \ref{hybrid:se0}. We take the EFM spectrum of the $L_0=10$ case and plot it on top of the $\Lambda=43$ EFM spectrum for $\D_{[\s\e]_0}$. The relative difference by comparing with the Analytics Setup \ref{analytics:2} is shown in the left of Figure \ref{fig:hybrid_Dse0_err}\footnote{We have to plot the relative difference, because had we plot the absolute $\D_{[\s\e]_0}$ or $\tau_{[\s\e]_0}$, we won't be able to see the tiny difference with bare eyes.}. One may notice that indeed for larger spin, the $\D_{[\s\e]_0}$ is forced to be sit on the analytic prediction $\D^{analytic}_{[\s\e]_0}$. In fact for $\ell=25, 26$, our $\Lambda=19$ EFM spectrum is likely to be even more precise than the $\Lambda=43$ EFM spectrum, because at $\ell=25, 26$ the $\Lambda=43$ EFM spectrum already start to fluctuating away from the analytic prediction\footnote{In this paper, for $\Lambda=19$ we used the spin set in (\ref{tab:spinsets}), so we only see operators with spin $\ell=...25, 26, 49, 50$ appear in the EFM. The EFM spectrum for the last two operators are $\D_{[\s\e]_{0,\ell=49}}=50.8677486, \D_{[\s\e]_{0,\ell=50}}=51.9847355$, sitting inside the narrow band around the analytic prediction. They are certainly much more precise than the $\Lambda=43$ EFM spectrum.}. The logarithemic of the error (estimated by comparing with $\Lambda=43$ EFM spectrum) is shown on the right of Figure \ref{fig:hybrid_Dse0_err}. Indeed we see the small spin accuracy is improved while the large spin operators match with the analytics, confirming our intuition in the schematic Figure \ref{fig:NAcurve_hybrid_vs_pure}.

Now let's try to determine the large spin region. We do a simple test: if we exactly plug in the information from analytics, how small we can go in $\ell$. For the case in this section, the setup is
\begin{hybrid}\label{hybrid:se0_Lmin}
The SDP is the same as Hybrid Setup \ref{hybrid:se0} with $a=-\infty$. We decrease $L_0$ until we can't decrease it anymore without losing the feasible region. We call it $L_\text{min}=\text{min}(L_0)$ subject to $objective\le0$.
\end{hybrid}
For the Hybrid Setup \ref{hybrid:se0}, $L_\text{min}=14$. It means up to $\ell=14$, the numerics still couldn't tell if the analytics is right or wrong. So we may say $\ell\ge 14$ is the large spin and $8\le\ell\le13$ is the intermediate region. In practice, we found a simple choice for $L_0$ is a few spins smaller than $L_\text{min}$. In the $\Lambda=11$ run of the Hybrid Setup \ref{hybrid:fullsetup}, we find $L_\text{min}=7$; and $L_\text{min}=15$ for $\Lambda=19$, so we made those choices of $L_0$ in Figure~\ref{fig:hybrid_fullsetup}.

\begin{figure}[!htpb]
\centering
\includegraphics[width = 0.6\textwidth]{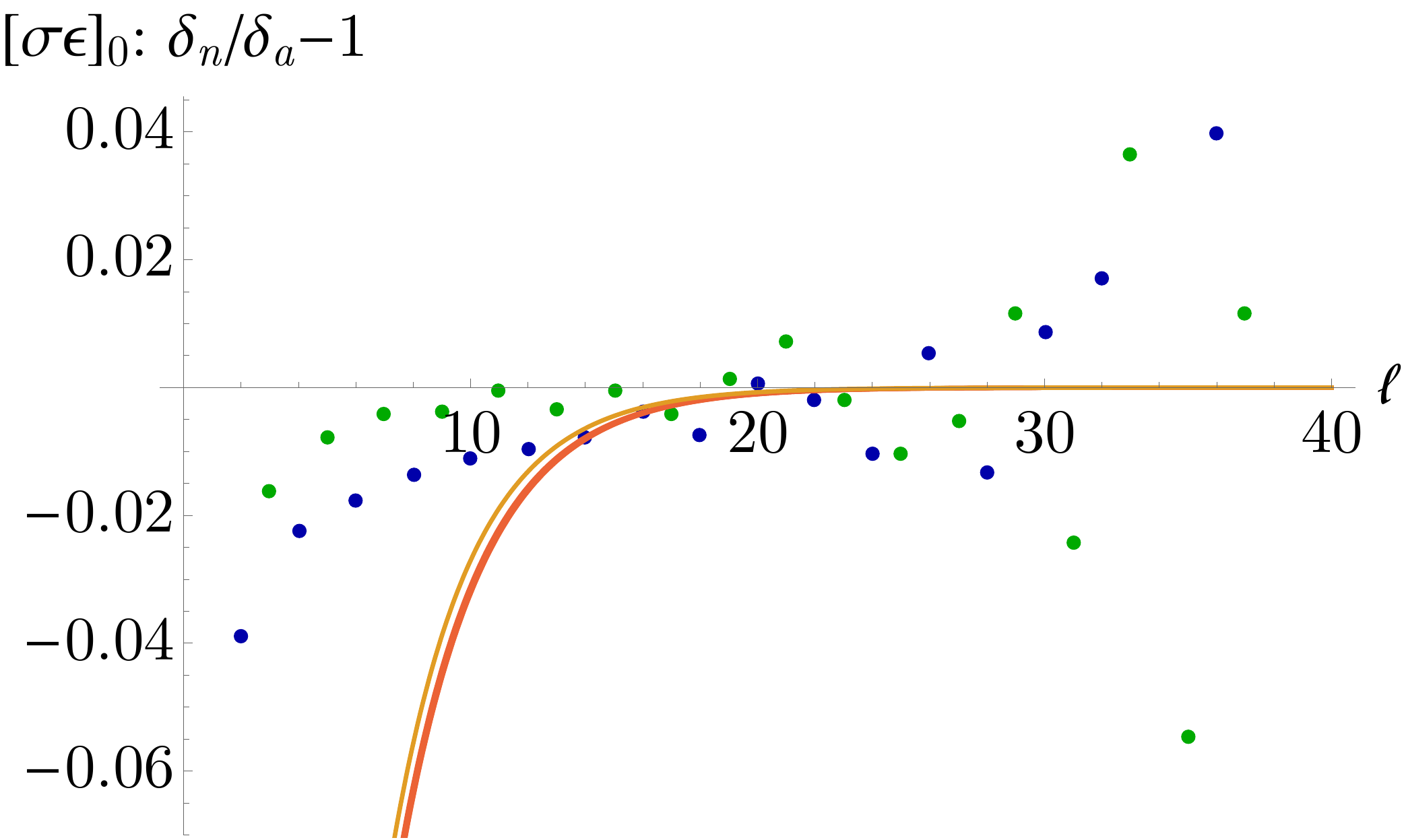}
\caption{\label{fig:err_and_exp}The blue and green dots are the same as the right of Figure \ref{fig:analyticerr}. Orange curve : $-e^{a \ell}$ with $a=-0.361$ for $L_0=12$. Red curves : $-e^{a \ell}$ with $a$ between -0.343 and -0.347 for $L_0=2, 4, ...10$.
}
\end{figure}

One may wonder why $L_0=2, 4$ seems to also work in Figure~\ref{fig:hybrid_Dse0}. It will be clear by looking at the $a$ parameter at those points. In Figure \ref{fig:err_and_exp}, we plot the $-e^{a \ell}$ on top of the left of Figure~\ref{fig:analyticerr}. We see that for $L_0=2, ...12$, the $a$ parameter is almost the same (roughly around $-0.35$). Indeed, since for small spin, the physical operator is well inside the $\pm e^{a \ell}$ bound, so even if we impose the interval posivity from $\ell=2$, it will do no harm. Of course it may do harm if we also plug in OPE coefficients, because such an operator will sit at a dimension that the numerics can say it’s wrong. So we shouldn't choose $L_0$ too small for the runs of Figure~\ref{fig:hybrid_fullsetup}. We see indeed the exponential decay function is a good choice for gluing the numerics and analytics, although other choices may also work, as long as they decay faster enough\footnote{If the function doesn't decay fast enough, the risk is that when we shrink the size of the bounding band, it may get stuck at the small spins, so the accurate information for the larger spins won't used efficiently.}.

\subsection{How to improve further?} \label{sec:improve_error}

\begin{figure}[!htpb]
\centering
\includegraphics[width = 0.5\textwidth]{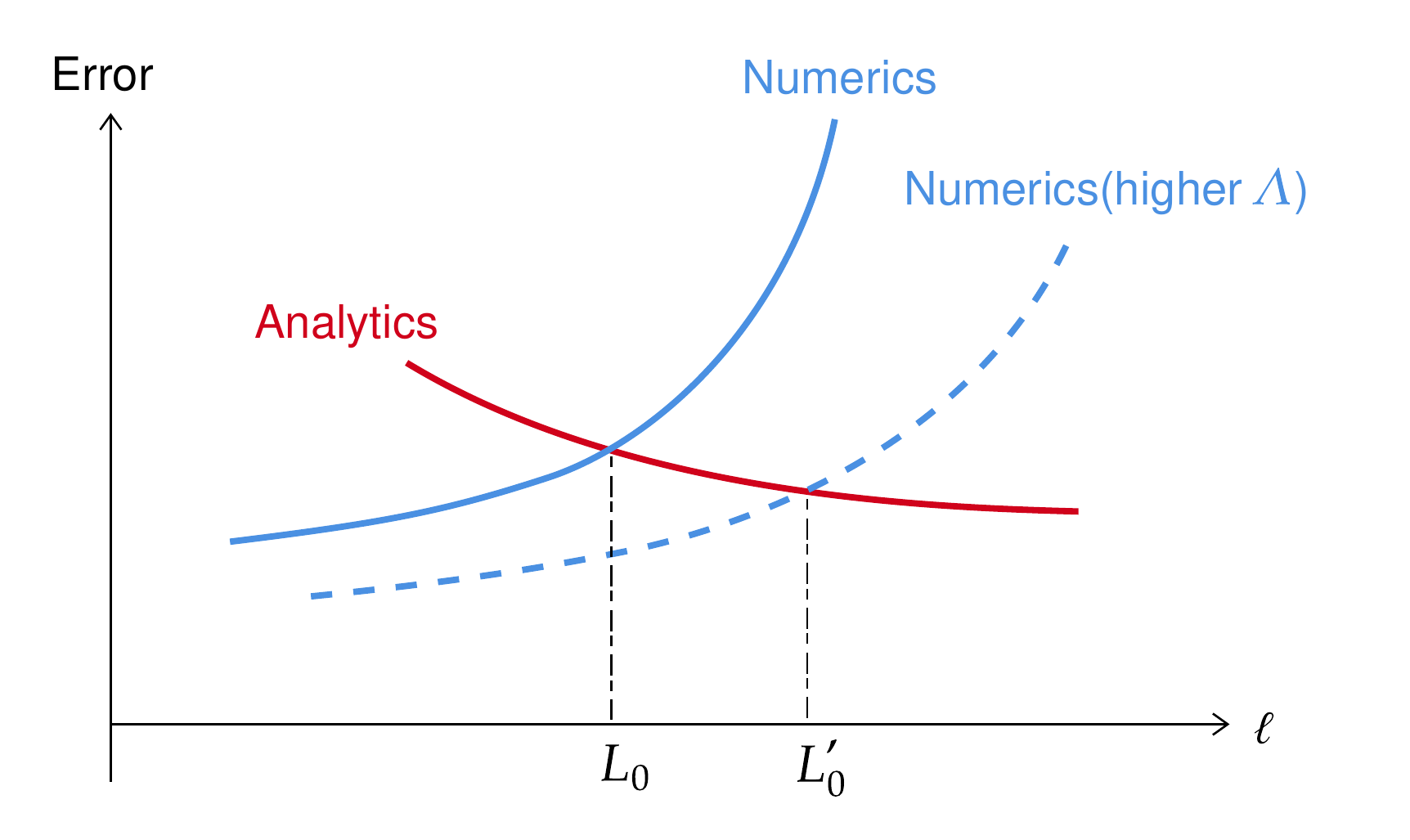}%
\includegraphics[width = 0.5\textwidth]{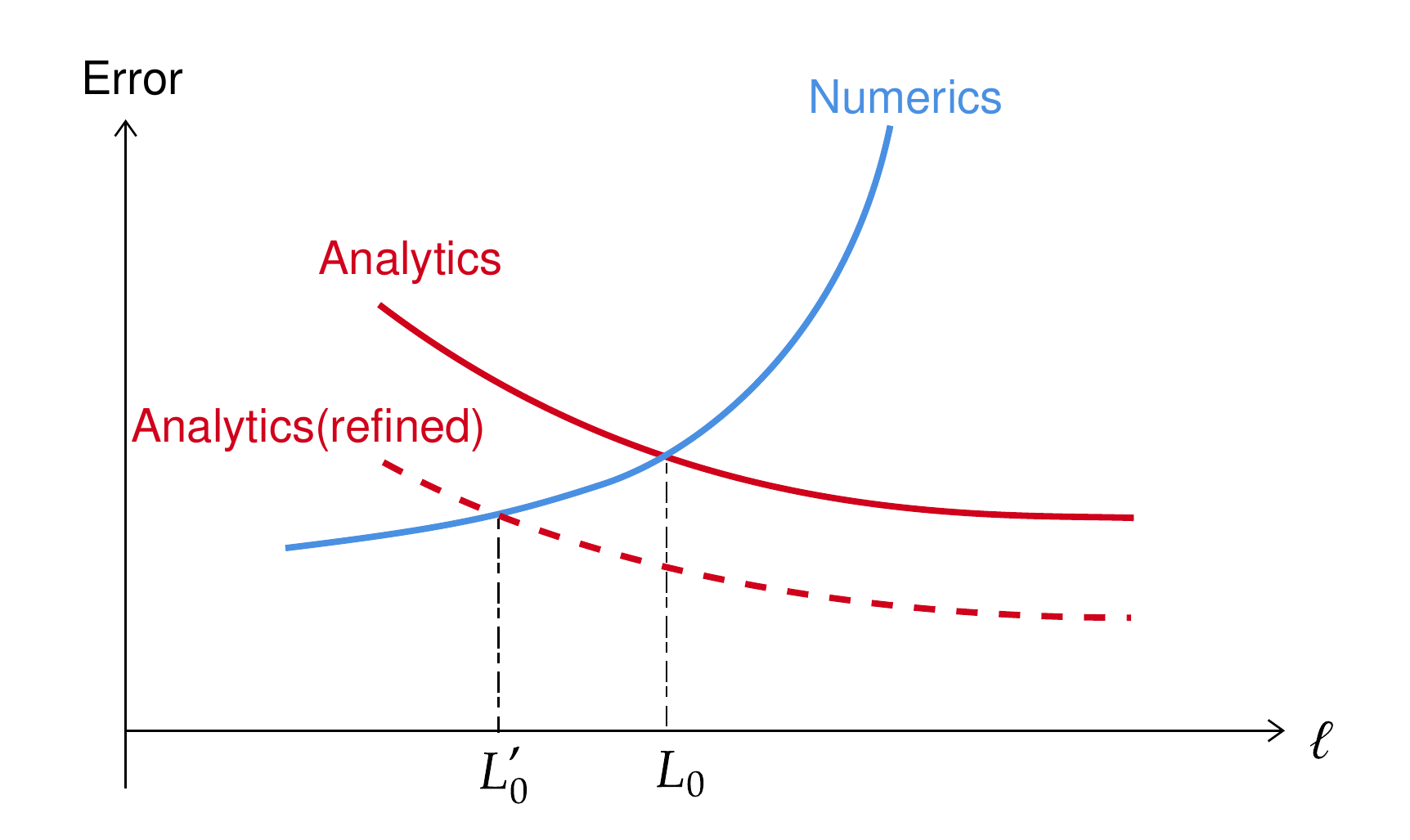}%
\caption{\label{fig:NAcurve_improve_NA}If we improve the numerics, the intermediate region will shift to the right; while if we improve the analytics, the intermediate region will shift to the left.} 
\end{figure}

Let's think about how to systematically improve the hybrid bootstrap result. We can compute the numerics at higher $\Lambda$, so the error curve for the numerics will move downward and the intermediate region shift to the right (see Figure~\ref{fig:NAcurve_improve_NA}). We have already seen such an example in Figure \ref{fig:hybrid_fullsetup}. In this section, we will try to push in the other direction : after improving the analytics, we expect the intermediate region will shift to the left and the result should be more accurate. Let's use the analytics of $[\s\s]_{0,\ell}$ as an example. The full setups for two different versions of the analytics are

\begin{hybrid}\label{hybrid:ss0_v2}
The framework of the SDP is given by (\ref{cond:experr}). In the \textit{conditions}, we take
\begin{align*}
&\D^{(+)}_{\text{upper}/\text{lower}}=\D^{(+)}_{\text{gap}}=\D_{\text{unitary}} \quad\text{for}\quad \ell<L_{0},\\
&\D^{(+)}_{\text{upper}/\text{lower}}=\D_{\text{GFF}}+\delta^{(analytic)}_{[\s\s]_{0,\ell}}(1\pm e^{a \ell}) \quad\text{for}\quad \ell \ge L_{0},\\
&\D^{(+)}_\text{gap}=\D_{[\s\s]_{0,\ell}}^{(analytic)}+1 \quad\text{for}\quad \ell \ge L_{0},\\
&\D^{(-)}_{\text{upper}}=\D^{(-)}_{\text{lower}}=\D^{(-)}_{\text{gap}}=\D_{\text{unitary}} \quad\text{for all $\ell$}.
\end{align*}
In the \textit{objective}, we take all $\f^*=0$. All analytics are computed using Analytics Setup \ref{analytics:2}. We minimize the $a$ parameter until the navigator (\itie \textit{objective}) is zero.
\end{hybrid}

\begin{hybrid}\label{hybrid:ss0_v3}
The setup is the same as Hybrid Setup \ref{hybrid:ss0_v2}, except we use Analytics Setup \ref{analytics:1}.
\end{hybrid}

\begin{figure}[!htpb]
\centering
\includegraphics[width = 0.5\textwidth]{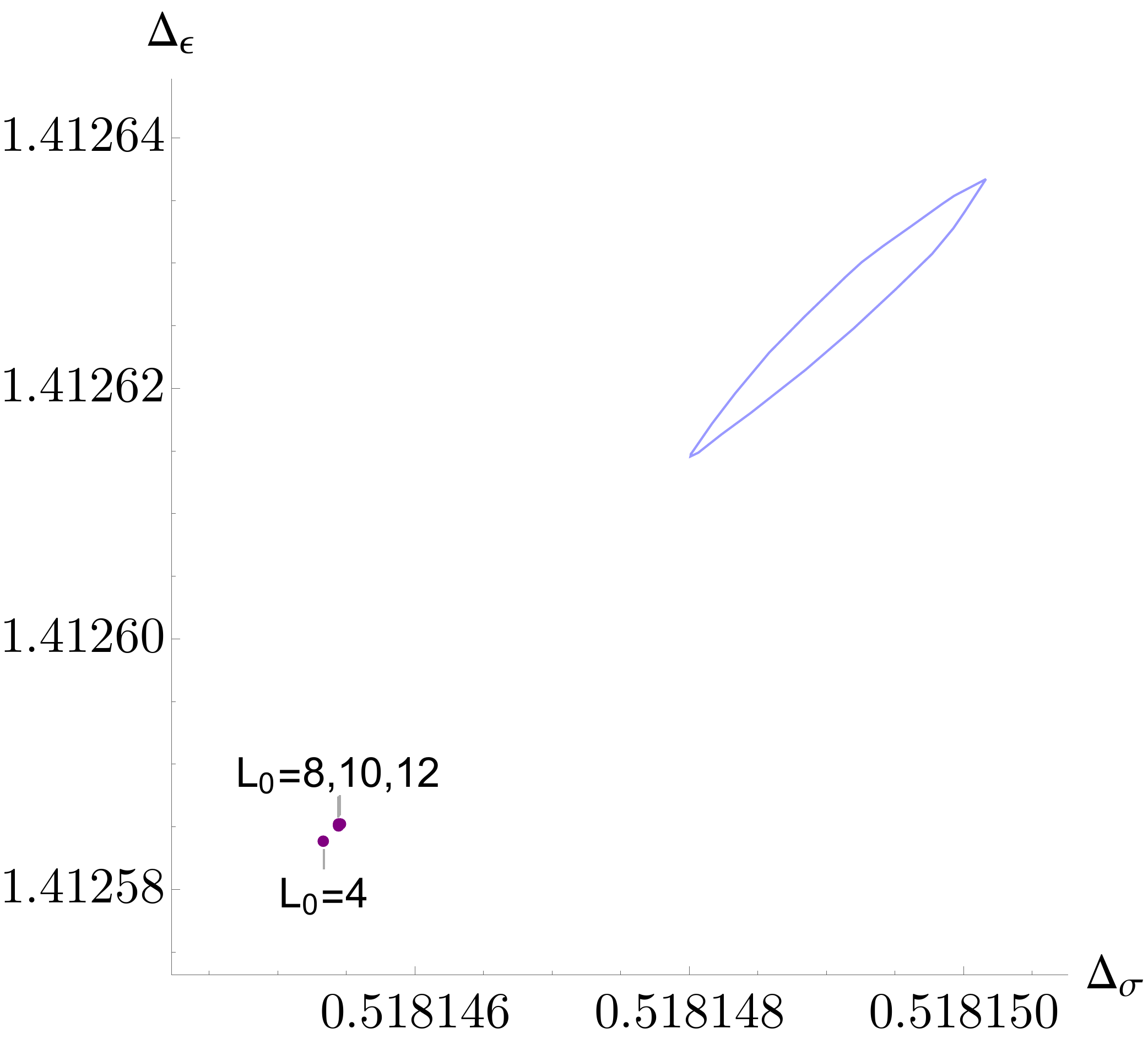}%
\includegraphics[width = 0.5\textwidth]{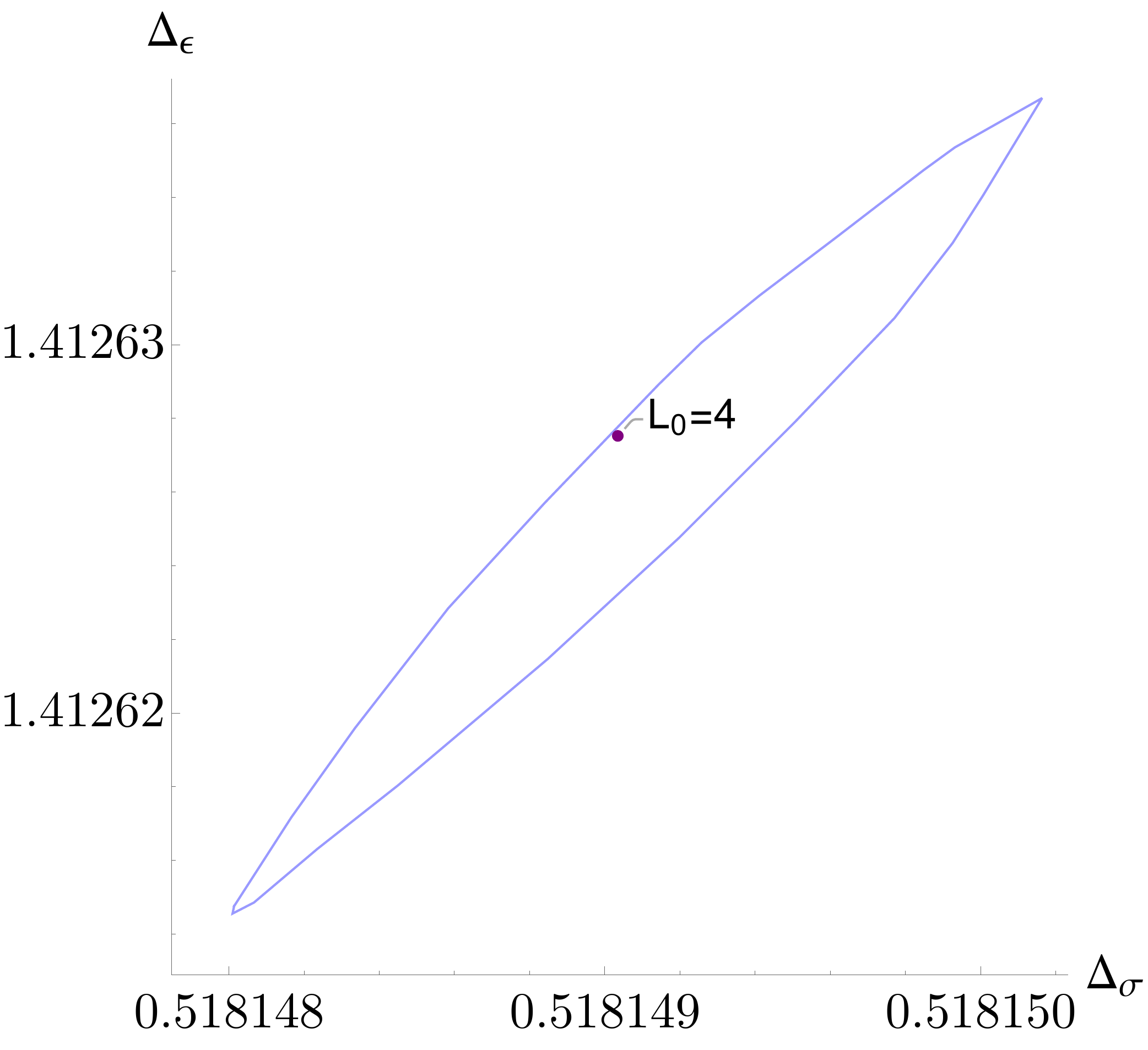}%
\caption{\label{fig:hybrid_ss0}The results of the computations for the Hybrid Setup \ref{hybrid:ss0_v2} and \ref{hybrid:ss0_v3} at $\Lambda=19$ are shown on the left and the right respectively. The light blue curve is the boundary of the $\Lambda=43$ Ising island of \cite{Kos:2016ysd}.} 
\end{figure}

We first follow the logic of Section \ref{sec:how_to_determine_L0} to determine $L_\text{min}$. Set $a=-\infty$ in the Hybrid Setup \ref{hybrid:ss0_v2}, we find $L_\text{min}=14$. For the Hybrid Setup \ref{hybrid:ss0_v3}, our analytics is so accurate that we can actually exactly plug in $\D^{(analytic)}_{[\s\s]_{0,\ell}}$ up to $L_\text{min}=6$! The result of both setups for some $L_0<L_\text{min}$ are shown in Figure \ref{fig:hybrid_ss0}. One can see indeed the Analytics Setup \ref{analytics:1} significantly improve the accuracy of the hybrid bootstrap result.
\begin{figure}[!htpb]
\centering
\includegraphics[width = 0.7\textwidth]{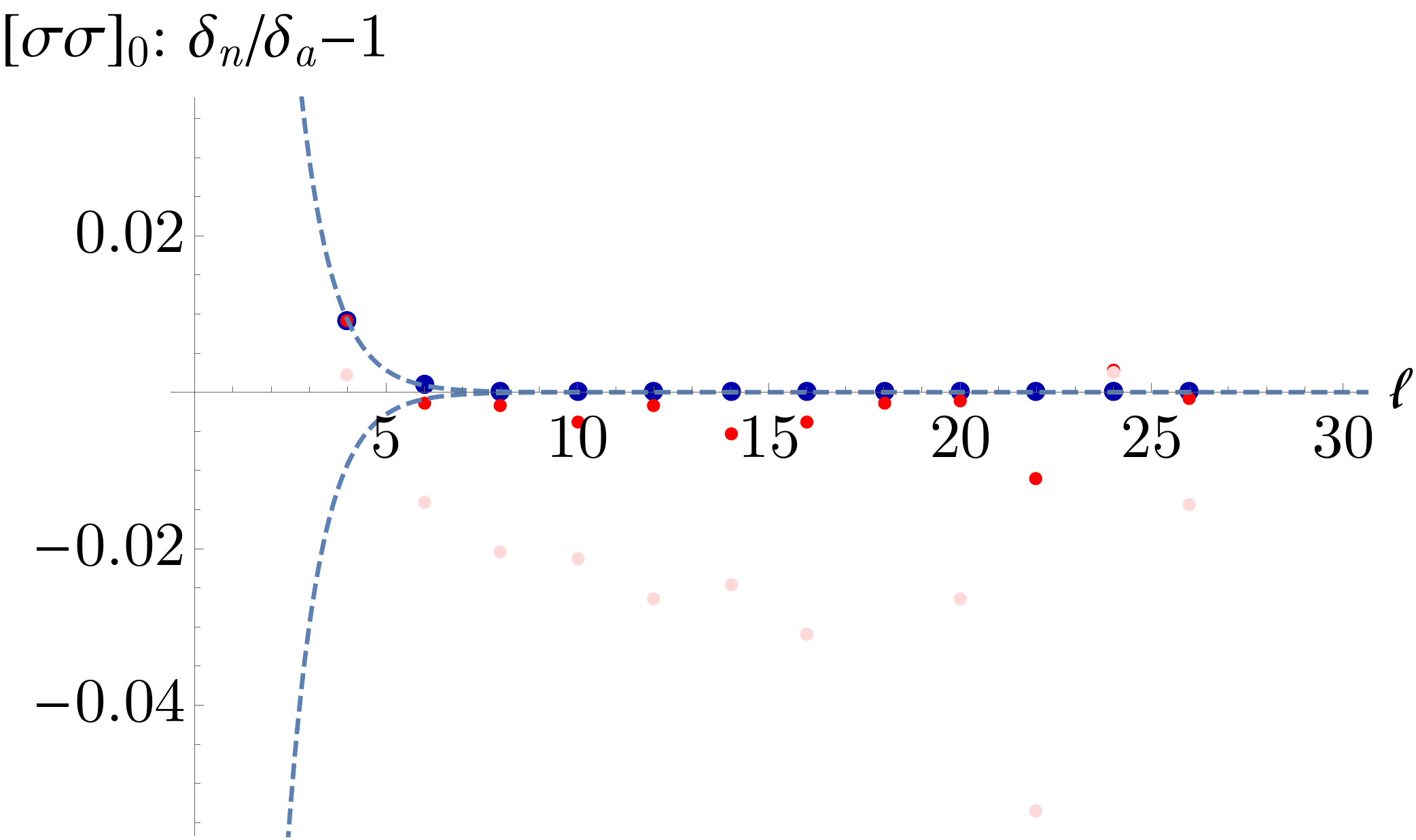}%
\caption{\label{fig:ss0_err_and_exp}Comparing the EFM spectrum $\D_{[\s\s]_0}$ from the computation of the Hybrid Setup \ref{hybrid:ss0_v3} at $\Lambda=19, L_0=4$ with the $\D^{(analytic)}_{[\s\s]_0}$ of the Analytics Setup \ref{analytics:1} (computed at the values of $(\ds,\de,\fsse,\fT)$ from the hybrid setup). Horizontal axes : spin. Vertical axes : $\delta^{(numerics)}/\delta^{(analytic)}-1$, where $\delta$ is the correction to the GFF dimension. Blut dots: $\delta^{(numerics)}$ is from the hybrid computation. Red dots : $\delta^{(numerics)}$ is from the $\Lambda=43$ EFM spectrum (after compensating the sharing effect). Light red dots : $\delta^{(numerics)}$ is from the raw $\Lambda=43$ EFM spectrum (without compensating the sharing effect). Dashed lines: $\pm e^{a \ell}$ with $a=-1.17306876127$ from the result of the hybrid computation.
}
\end{figure}
Next we will look at the EFM spectrum of $\D_{[\s\s]_{0,\ell}}$ extracted at $L_0=4$ point of the Hybrid Setup \ref{hybrid:ss0_v3}. The result (relative to Analytics Setup \ref{analytics:1}) is shown in Figure \ref{fig:ss0_err_and_exp}.
We see that $\Lambda=19$ EFM spectrum from Hybrid Setup \ref{hybrid:ss0_v3} is even more accurate than the $\Lambda=43$ EFM spectrum raw data (without averaging over both real operator and fake operator)\footnote{In fact, the evidence seems suggests blue dots of Figure \ref{fig:ss0_err_and_exp} is also more accurate than the red dots, because (1), if the red dots are more close to the real operators, we expected the blue dots should sit at the lower bound of the band, instead of the upper bound; (2),  the red dots is obtained through the averaging method aiming to compensate the sharing effect (see Section \ref{sec:numerics}). Since it's just an empirical method, it's hard to believe after such a process the data could precisely located on the actual operators. The data from Table 3 of \cite{Simmons-Duffin:2016wlq} is close to the red dots.}. Another interesting feature is that the $\Lambda=19$ spectrum only have one operator near $\D_{[\s\s]_{0,\ell}}$ in the bounding band\footnote{In Figure \ref{fig:ss0_err_and_exp}, the analytics is computed using the values from Table \ref{tab:hybrid:ss0_v3} and the blue dot is precisely on the upper dashed line. Had we use the values of (\ref{data:pd43}), the difference is tiny and invisible in the figure.}, \itie it doesn't suffer from the sharing effect\footnote{It would be interesting to see how the pure bootstrap numerics changes if we are able to remove the sharing effect, \itie demand there is only one operator near the $\D_{[\s\s]_{0,\ell}}$, although such condition might be hard to implement.}. On the other hand, the spectrum for $\D_{[\s\e]_{0,\ell}}$ doesn't perfectly match the analytics, as expected. This suggests we should throw more analytic information into the numerics.

Lastly, we would like to comment on the gaps of (\ref{cond:spin012}). If we relax those gaps to 
\begin{align}
	&\vec{\alpha}\cdot(\Vp_{\D,\ell=0})\ge 0 \text{ for } \D \ge 3.1,\nonumber\\
	&\vec{\alpha}\cdot(\Vp_{\D,\ell=2})\ge 0 \text{ for } \D \ge 4,\nonumber\\
	&\vec{\alpha}\cdot(\Vm_{\D,\ell=0})\ge 0 \text{ for } \D \ge 3.1,\nonumber\\
	&\vec{\alpha}\cdot(\Vm_{\D,\ell=1})\ge 0 \text{ for } \D \ge 3.1,
	\label{cond:mild_spin012}
\end{align}
we find surprisingly the computations of setups only use $\D^{(analytic)}_{[\s\s]_0}$ will yield less accurate results, but the computations of setups only use $\D^{(analytic)}_{[\s\e]_0}$ won't change much. In particularly, we checked that the Hybrid Setup \ref{hybrid:ss0_v3} with condition (\ref{cond:mild_spin012}) at $\Lambda=19, L_0=4$ gives result $(\ds,\de)=(0.51816320, 1.41276289)$, while the Hybrid Setup \ref{hybrid:se0} with condition (\ref{cond:mild_spin012}) at $\Lambda=19, L_0=4$ gives result $(\ds,\de)=(0.51814895, 1.41262720)$. We don't have a good explanation of the difference. But it does suggest again that we should use both analytic information at the same time. We will discuss some preliminary attempts in Section \ref{sec:liebig}. Before doing that, we need to have a tool to estimate the error.

\subsection{A preliminary attempt on estimation of the error bar} \label{sec:estimate_error}
A missing piece in our method is a satisfactory method to estimate the error bar of the result. We don't have a perfect method yet, but in this subsection we will make a preliminary attempt.

The result we obtained in various hybrid setups in this paper will have some errors. Let's think about the source of errors. Taking the Hybrid Setup \ref{hybrid:ss0_v3} as an example, the result we obtained has errors, because in Figure \ref{fig:ss0_err_and_exp} we pushed the numerics too close to the analytics, which is not exact. By the moment the feasible region shrinking to a point, we definitely have already ``over pushed" it and over passed the spin 4 operator ($[\s\s]_{0,\ell=4}$) of the actual CFT. We should expect the actual operator for spin 4 in Figure \ref{fig:ss0_err_and_exp} is slightly above the blue dot
\footnote{One may reason in this way: by the moment the feasible region shrinking to a point, the numerics can almost tell the analytics is wrong, because had the $a$ decrease even a tiny bit more, the navigator function would be positive. We expect the main power that the numerics can tell the analytics is marginally wrong comes from the spin 4 operators, because for spin 6 even if we exactly plug in the analytic value the numerics won't complain (\itie $L_\text{min}=6$).}. 

But it's not obvious how to estimate how much had we overpassed the dimension of the actual operator. It's also not obvious what is the consequence of this over-push. It may not necessarily damage the prediction of $(\ds,\de)$. Let's consider the example of $\D_{\s'}$ bound, where $\s'$ is the next spin 0 operator (after $\s$) appearing in the $\s \times \e$ OPE. If we assume $\D_\epsilon$ saturates the $\<\s\s\s\s\>$ single correlator bootstrap bound and the first operator in the $\dsZ_2$-odd spin 1 channel has dimension $\ge 2.3$, the plot of the bound on $\D_{\s'}$ v.s. $\ds$ looks like dark blue region of the Figure 6 of \cite{Karateev:2019pvw}. If we maximize $\D_{\s'}$ in the feasible region, although it would overpass the real operator, the $\ds$ actually get more accurate. But it doesn't necessarily mean the spectrum at the optimal point of the maximization is overall a more accurate spectrum. Therefore there is a risk that the accurate result we obtain in this paper might partially due to similar effect. With this in mind, let's now try to estimate the errors.

We may expect, if the numerics is more sensitive to the bounding band, the scale of $\partial_a \cN$ should be larger. In fact, if the navigator function around the feasible region of $\{\ds,\de,\fT,\fsse,\feee,\D_{[\s\s]_{0,\ell=4}}\}$ is a quadratic function, we can simply use the gradient and Hessian at a local point to reconstruct an approximate navigator function and estimate the size of the feasible region based on that. In reality, the navigator function is not a quadratic function. But let's still try to estimate the error based on the approximate quadratic navigator function. Hopefully the estimation may scale with the actual error.

After each hybrid bootstrap computation, we use the procedure in Appendix C of \cite{Reehorst:2021ykw} to compute the Hessian \footnote{As a rough estimation, one may use the the approximate Hessian at the end of Algorithm 2 of \cite{Reehorst:2021ykw}. However we found the precision of the approximate Hessian is not good enough for our purpose. For example, the approximate Hessian is always positive definite, but in reality the Hessian could be negative.}, and construct the approximate navigator function as
\be
\cN^{(approx)}(x) = g(x_f) (x-x_f) + \frac{1}{2} (x-x_f)^T B (x-x_f).
\ee
where $g$ is the gradient, $B$ is the Hessian, and $x=(\ds,\de,\fT,\fsse,\feee,a)$. Now we can maximize/minimize various parameters subject to $\cN^{(approx)}(x)\le 0$ and use it as an estimation of the error bars.

Let's take the computation in Hybrid Setup \ref{hybrid:ss0_v3} with $\Lambda=19, L_0=4$ as an example. The gradient at the optimal point is
\be\nonumber
(4.59\times 10^{-17}, 7.70\times 10^{-17}, 2.81\times 10^{-16}, 7.97\times 10^{-18}, 7.53\times 10^{-17}, -8.05231\times 10^{-9})
\ee
while the Hessian is given by
\be\nonumber
\begin{pmatrix}
$28828.7$ & $-6254.82$ & $-8598.62$ & $427.614$ & $-1345.41$ & $-0.0429191$ \\
$-6254.82$ & $1914.45$ & $3752.10$ & $-51.2511$ & $905.168$ & $0.0741709$ \\
$-8598.62$ & $3752.10$ & $8956.93$ & $14.2040$ & $2481.62$ & $0.231068$ \\
$427.614$ & $-51.2511$ & $14.2040$ & $9.71850$ & $24.8149$ & $0.00365034$ \\
$-1345.41$ & $905.168$ & $2481.62$ & $24.8149$ & $776.322$ & $0.0795515$ \\
$-0.0429191$ & $0.0741709$ & $0.231068$ & $0.00365034$ & $0.0795515$ & $0.0000548614$ 
\end{pmatrix}.
\ee
Maximizing/minimizing various parameters in the approximate navigator function, we obtain the data in Table \ref{tab:error_of_ss0v3_pd19}\footnote{The error in Table \ref{tab:error_of_ss0v3_pd19} is defined by $(\text{max}-\text{min})$. For the last column, we took the data from Table 2 of \cite{Simmons-Duffin:2016wlq}. For $\fT$ in the last column, we estimate it by $\fT=(\feeT/\ds+\fssT/\de)/2$ and convert to our convention, while the error is derived from the error of $\feeT$.}. We found all the error bars are compatible with the data in \cite{Simmons-Duffin:2016wlq}.

\begin{table}[!h]
\begin{center}
\begin{tabular}{|| c | *{4}{c} |l ||}\hline
$\text{Parameter}$ & $\text{value}$ & $\text{min}$ & $\text{max}$ & $\text{error}$ & $\text{value of \cite{Simmons-Duffin:2016wlq}}$ \\
\hline
$\Delta _{\sigma }$ & $0.51814904$ & $0.51814721$ & $0.51815085$ & $0.00000363$ & $\text{0.5181489(10)}$ \\
$\Delta _{\epsilon }$ & $1.41262750$ & $1.41260625$ & $1.41264878$ & $0.00004253$ & $\text{1.412625(10)}$ \\
$f_{\sigma \sigma \epsilon }$ & $1.05185297$ & $1.05186070$ & $1.05184522$ & $0.00001548$ & $\text{1.0518537(41)}$ \\
$f_{\epsilon \epsilon \epsilon }$ & $1.53244587$ & $1.53240724$ & $1.53248517$ & $0.00007793$ & $\text{1.532435(19)}$ \\
$f_T$ & $1.25885744$ & $1.25885555$ & $1.25885928$ & $0.00000373$ & $\text{1.2588576(90)}$ \\
\hline
\end{tabular}
\caption{\label{tab:error_of_ss0v3_pd19}Estimation of the errors in Hybrid Setup \ref{hybrid:ss0_v3} with $\Lambda=19, L_0=4$.}
\end{center}
\end{table}

For various setups in this paper, the estimated errors $\Delta _{\sigma ,\max }-\Delta _{\sigma ,\min }$, $\Delta _{\e ,\max }-\Delta _{\e ,\min }$ and the total error $\prod^6_{i=1} (x_{i,\text{max}}-x_{i,\text{min}})$ is presented in Table \ref{tab:errors}\footnote{For some of the computations, the Hessian is not positive, therefore the error would be $\infty$ based on the approximate navigator function.}. We find all error boxes are compatible with the data in \cite{Simmons-Duffin:2016wlq}. We also observe that this estimation can tell qualitatively : (1), Hybrid Setup \ref{hybrid:ss0_v3} is better than other setups; (2), in Hybrid Setup \ref{hybrid:fullsetup}, $\Lambda=19$ computations are better than the $\Lambda=11$ computation; (3), the results from Hybrid Setup \ref{hybrid:ss0_v2} are the least accurate ones; (4), When $L_0$ is too large, usually the result is not good. Indeed this method is still helpful. On the other hand, we are not sure how robust the method is in other scenarios and other models.

\begin{table}
\begin{center}
\begin{tabular}{|| l | c|c|c ||}\hline
$\text{Setup}$ & $\Delta _{\sigma ,\max }-\Delta _{\sigma ,\min }$ & $\Delta _{\epsilon ,\max }-\Delta _{\epsilon ,\min }$ & $\prod _{i=0}^6\left(x_{i,\max }-x_{i,\min }\right)$ \\
\hline
$\text{Hybrid Setup \ref{hybrid:fullsetup} $\Lambda $=11 }L_0\text{= 6}$ & $0.00016363$ & $0.00198749$ & $4.225\times 10^{-17}$ \\
$\text{Hybrid Setup \ref{hybrid:fullsetup} $\Lambda $=19 }L_0\text{=10}$ & $0.00001632$ & $0.00019559$ & $6.392\times 10^{-24}$ \\
$\text{Hybrid Setup \ref{hybrid:fullsetup} $\Lambda $=19 }L_0\text{=12}$ & $0.00001679$ & $0.00020124$ & $1.095\times 10^{-23}$ \\
$\text{Hybrid Setup \ref{hybrid:fullsetup} $\Lambda $=19 }L_0\text{=14}$ & $\infty$ & $\infty$ & $\infty$ \\
$\text{Hybrid Setup \ref{hybrid:se0} $\Lambda $=19 }L_0\text{= 2}$ & $0.00001939$ & $0.00023415$ & $2.872\times 10^{-23}$ \\
$\text{Hybrid Setup \ref{hybrid:se0} $\Lambda $=19 }L_0\text{= 4}$ & $0.00001620$ & $0.00019514$ & $1.125\times 10^{-23}$ \\
$\text{Hybrid Setup \ref{hybrid:se0} $\Lambda $=19 }L_0\text{= 6}$ & $0.00003387$ & $0.00040541$ & $4.811\times 10^{-22}$ \\
$\text{Hybrid Setup \ref{hybrid:se0} $\Lambda $=19 }L_0\text{= 8}$ & $0.00001720$ & $0.00020704$ & $1.612\times 10^{-23}$ \\
$\text{Hybrid Setup \ref{hybrid:se0} $\Lambda $=19 }L_0\text{=10}$ & $0.00002485$ & $0.00029873$ & $1.204\times 10^{-22}$ \\
$\text{Hybrid Setup \ref{hybrid:se0} $\Lambda $=19 }L_0\text{=12}$ & $0.00003137$ & $0.00037567$ & $1.735\times 10^{-21}$ \\
$\text{Hybrid Setup \ref{hybrid:ss0_v2} $\Lambda $=19 }L_0\text{= 4}$ & $\infty$ & $\infty$ & $\infty$ \\
$\text{Hybrid Setup \ref{hybrid:ss0_v2} $\Lambda $=19 }L_0\text{= 8}$ & $\infty$ & $\infty$ & $\infty$ \\
$\text{Hybrid Setup \ref{hybrid:ss0_v2} $\Lambda $=19 }L_0\text{=10}$ & $\infty$ & $\infty$ & $\infty$ \\
$\text{Hybrid Setup \ref{hybrid:ss0_v2} $\Lambda $=19 }L_0\text{=12}$ & $\infty$ & $\infty$ & $\infty$ \\
$\text{Hybrid Setup \ref{hybrid:ss0_v3} $\Lambda $=19 }L_0\text{= 4}$ & $0.00000363$ & $0.00004253$ & $2.453\times 10^{-28}$ \\
\hline
\end{tabular}
\caption{\label{tab:errors}Estimation of errors for various setups in this paper.}
\end{center}
\end{table}

It might be possible to find other methods to estimate the error. For example, instead of using $\pm e^{a \ell}$ for the form of the bounding band, we can use $\pm A e^{a \ell}$. By varying $A$, we glue the numerics and the analytics slightly differently and the result will be slightly different. The range where the parameters fluctuate may be used as estimation of the error. This is just like in the loop computation, different Padé approximation schemes give slightly different results and one may use the range as an estimation of the error. Similarly, we may change the value of $z_0$ in the analytics setup, observe how much the result of the hybrid bootstrap fluctuate with $z_0$, and use that as an error estimation. However we haven't thoroughly investigated those approaches.

\subsection{Balance out multiple analytics with different accuracy}\label{sec:liebig}
Although the Hybrid Setup \ref{hybrid:fullsetup} used all the information from Analytics Setup \ref{analytics:2}, it seems the result isn't as good as the Hybrid Setup \ref{hybrid:se0}, which only used $\D^{analytic}_{[\s\e]_0}$. After our analysis in Section \ref{sec:how_to_determine_L0} and \ref{sec:improve_error}, this shouldn't be too surprising. Imagine in a hybrid bootstrap setup, we have two analytic information with very different accuracy, say $\D^{analytic}_{[\s\e]_0}$ from Analytics Setup \ref{analytics:2} and $\D^{analytic}_{[\s\s]_0}$ from Analytics Setup \ref{analytics:1}. If we use the same $a$ for the bounding bands for the both analytics, $a$ certainly can't decrease below the optimal value in Hybrid Setup \ref{hybrid:se0} (around $-0.343$). Therefore the very accurate information from $\D^{analytic}_{[\s\s]_0}$, which allows $a\sim-1.17$, won't be used effectively. This situation remind us the Liebig's barrel, or the law of the minimum : a barrel's capacity is limited by its shortest stave.

To overcome this issue, the natural idea is that we set different decay functions for the bands that bound analytics with different accuracy. A main difficulty is that we have to choose how they decay relative to each other as we shrink the feasible region to a point. In Section \ref{sec:estimate_error} we found a method to estimate the error (or an indicator that more or less scales with the error), therefore we may choose a ratio between different decay factors such that the estimated error is minimized. Since our method for the error is not precise, the ratio may not be perfect. But let's still try it as a preliminary attempt. In this subsection, we will use both $\D^{analytic}_{[\s\s]_0}, \D^{analytic}_{[\s\e]_0}$ from the Analytics Setup \ref{analytics:2}. One may notice from Figure \ref{fig:hybrid_ss0} and \ref{fig:hybrid_Dse0}, the result from using $\D^{analytic}_{[\s\s]_0}$ only is less accurate than the result from using $\D^{analytic}_{[\s\e]_0}$ only and both predict a $\ds$ that is too small. We want to see if we can get a better result by using both analytics. The full setup is
\begin{hybrid}\label{hybrid:liebig_ratio}
The framework of the SDP is given by (\ref{cond:experr}). In the \textit{conditions}, we take
\begin{align*}
&\D^{(+)}_{\text{upper}/\text{lower}}=\D^{(+)}_{\text{gap}}=\D_{\text{unitary}} \quad\text{for}\quad \ell<L_{0},\\
&\D^{(+)}_{\text{upper}/\text{lower}}=\D_{\text{GFF}}+\delta^{(analytic)}_{[\s\s]_{0,\ell}}(1\pm e^{a_e \ell}) \quad\text{for}\quad \ell \ge L_{0},\\
&\D^{(+)}_\text{gap}=\D_{[\s\s]_{0,\ell}}^{(analytic)}+1 \quad\text{for}\quad \ell \ge L_{0},\\
&\D^{(-)}_{\text{upper}/\text{lower}}=\D^{(-)}_{\text{gap}}=\D_{\text{unitary}} \quad\text{for}\quad \ell<L_{0},\\
&\D^{(-)}_{\text{upper}/\text{lower}}=\D_{\text{GFF}}+\delta^{(analytic)}_{[\s\e]_{0,\ell}}(1\pm e^{a_o \ell}) \quad\text{for}\quad \ell \ge L_{0},\\
&\D^{(-)}_\text{gap}=\D_{[\s\e]_{0,\ell}}^{(analytic)}+1 \quad\text{for}\quad \ell \ge L_{0}.
\end{align*}
In the \textit{objective}, we take all $\f^*=0$. All analytics are computed using Analytics Setup \ref{analytics:2}. We fix the ratio $r=a_e/a_o$ and minimize the $a$ parameter until the navigator (\itie \textit{objective}) is zero. We change the $r$ until the estimated total error (using the method of Section \ref{sec:estimate_error}) is minimized.
\end{hybrid}

\begin{figure}[!htpb]
\centering
\includegraphics[width = 0.7\textwidth]{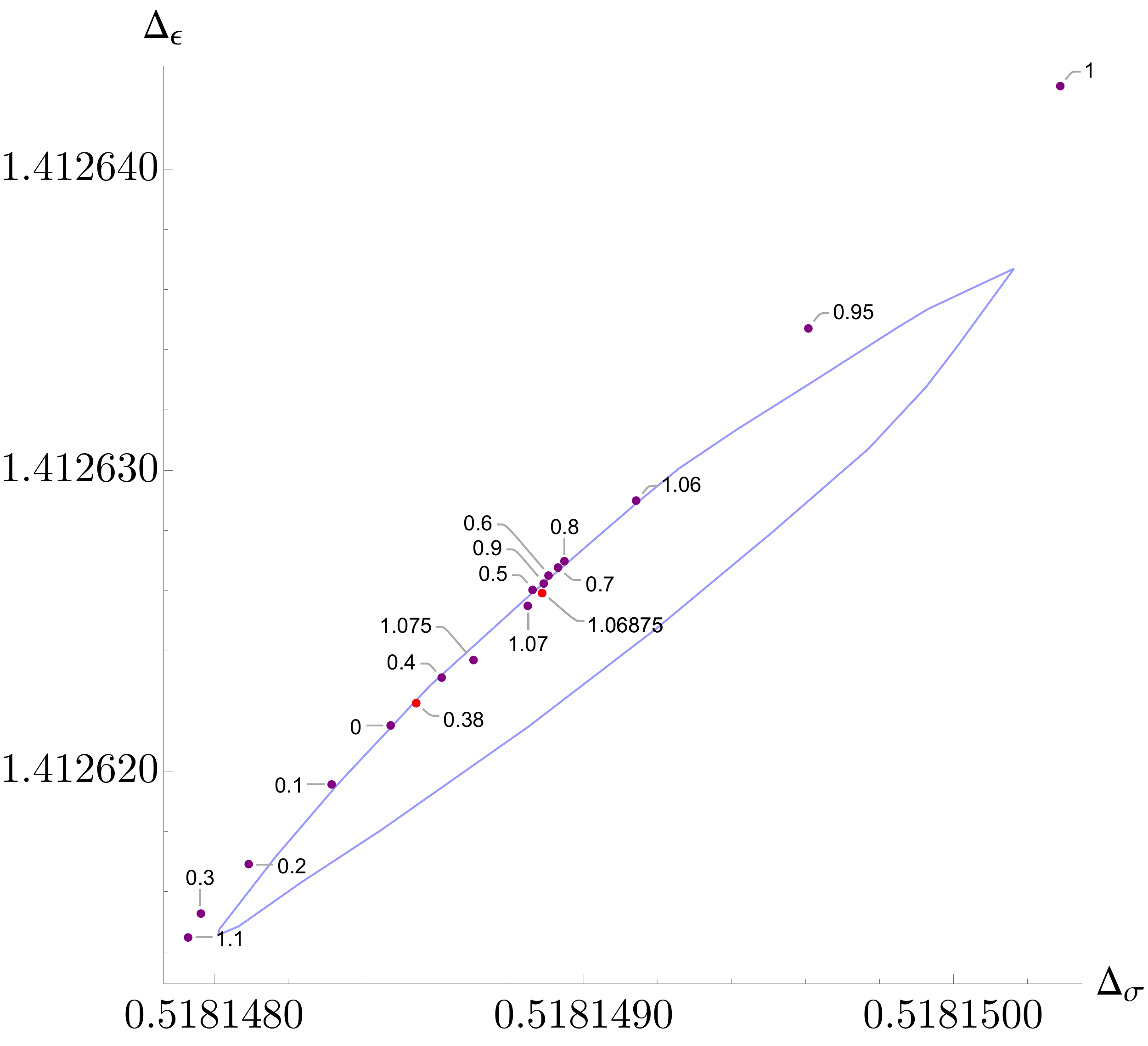}%
\caption{\label{fig:hybrid_liebig}Results of the computations for the Hybrid Setup \ref{hybrid:liebig_ratio} at $\Lambda=19$, $L_0=8$ and various $r=a_e/a_o$ ratios. The red dots correspond to ratios where the estimated errors are small. The light blue curve is the boundary of the $\Lambda=43$ Ising island of \cite{Kos:2016ysd}.
} 
\end{figure}

\begin{figure}[!htpb]
\centering
\includegraphics[width = 0.7\textwidth]{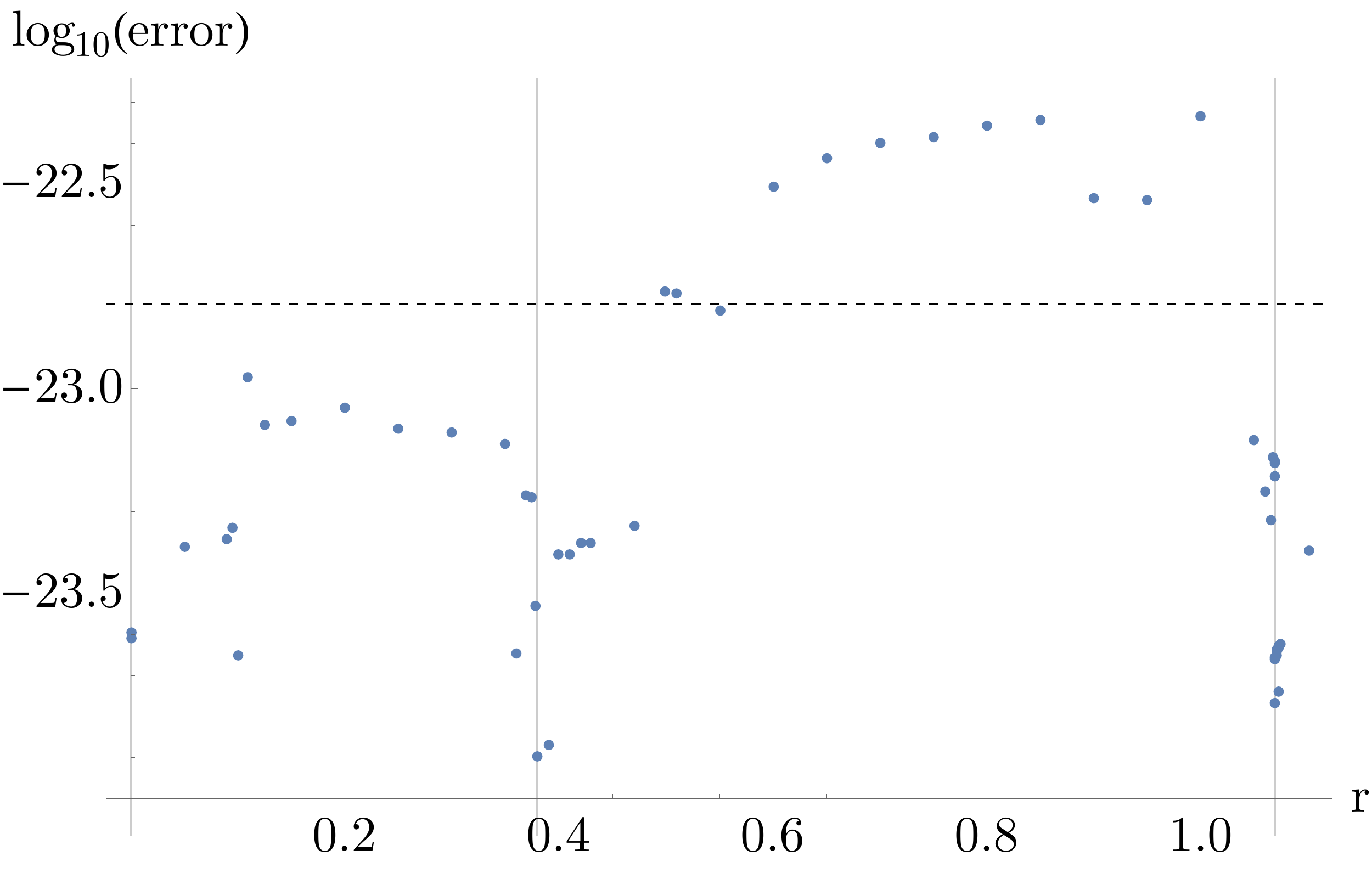}%
\caption{\label{fig:hybrid_liebig_error}The estimated error for the Hybrid Setup \ref{hybrid:liebig_ratio} at $\Lambda=19$, $L_0=8$. Horizontal : $r=a_e/a_o$. Vertical : $\text{log}_{10}(\text{error})$, where $\text{error}=\prod _{i=0}^6(x_{i,\max }-x_{i,\min })$ is estimated using the method of Section \ref{sec:estimate_error}. The two vertical grids correspond to two local minimums with $r=0.38, 1.06875$. The horizontal grid is the $\text{log}(\text{error})$ for the Hybrid Setup \ref{hybrid:se0}, while the Hybrid Setup \ref{hybrid:ss0_v2} has $\text{error}=\infty$.
} 
\end{figure}

The results for some $r$ are shown in Figure \ref{fig:hybrid_liebig}. We found surprisingly the prediction fluctuate around the $\Lambda=43$ island as $r$ varies. The estimated total error is also not smooth on $r$. So in practice, we scanned many values of $r$. The estimated errors are shown in Figure \ref{fig:hybrid_liebig_error}. The logic to choose this range of $r$ is the following. From Table \ref{tab:hybrid:ss0_v2} and \ref{tab:hybrid:se0}, we have seen $a_e=-0.3951$ from the Hybrid Setup \ref{hybrid:ss0_v2} and $a_o=-0.3459$ from the Hybrid Setup \ref{hybrid:se0} when we only use one analytics. So a natural choice could be $r=(-0.3951)/(-0.3459)=1.1422$\footnote{The result for $r=1.1422$ is $(0.518146278, 0.51814627)$, outside the range of Figure \ref{fig:hybrid_liebig}. The estimated error is $\infty$.}. But since the Hybrid Setup \ref{hybrid:ss0_v2} has estimated error $\infty$, we should choose $r<1.1422$ to favor the analytics in $\dsZ_2$-odd channel. Thus we scanned those values of $r$ in Figure \ref{fig:hybrid_liebig} and \ref{fig:hybrid_liebig_error}. From those result, we observe the prediction from various $r$ come cross a certain region of the island several times, where the estimated errors are small. And for most ratios, the result is better than both Hybrid Setup \ref{hybrid:ss0_v2} and \ref{hybrid:se0}. So this preliminary method seems to have worked for our setup. If we want to incorporate more analytics, such as the OPE coefficients, it might be difficult to find global minimum due to the non-smoothness of error function on the ratios. On the other hand, as we seen in Figure \ref{fig:hybrid_liebig} and \ref{fig:hybrid_liebig_error}, local minimums of the error function also give good predictions. The validity and robustness of this method is subject to future studies in other scenarios and models. It would also be interesting to find other methods to balance out multiple analytics\footnote{We may naively think, why not just let $a_e, a_o$ be free and minimize the total area of both bounding bands? Unfortunately this won't work. The reason is that the numerics is more sensitive to the band of the more accurate analytics. By minimizing the total area, we more or less ``equalized" the sensitivity. But this will favor the less accurate analytics instead of the more accurate one. In other words, the area of different bounding bands can't be compared directly. To carefully compensate for the difference, it seems we have to use the information of the 2nd order derivative of the navigator function. Our method of estimating error is of this type.}.

\section{Conclusion and outlook}

In this work, we have found a hybrid bootstrap method to combine numerical conformal bootstrap with the analytical lightcone bootstrap. We have explained the logic of the hybrid bootstrap and the general strategy. Partial implementation of the strategy in the examples in 3D Ising $\{\s,\e\}$ system have been worked out. We have observed that once injecting the analytical information into the numerics, the prediction for the actual CFT in the hybrid setups is significantly improved. We also have made a preliminary attempt to estimate the errors. We studied a method to use multiple analytic information at the same time, where we have found the result is better than the result of just using one.

\begin{figure}[!h]
\centering
\includegraphics[width = 0.6\textwidth]{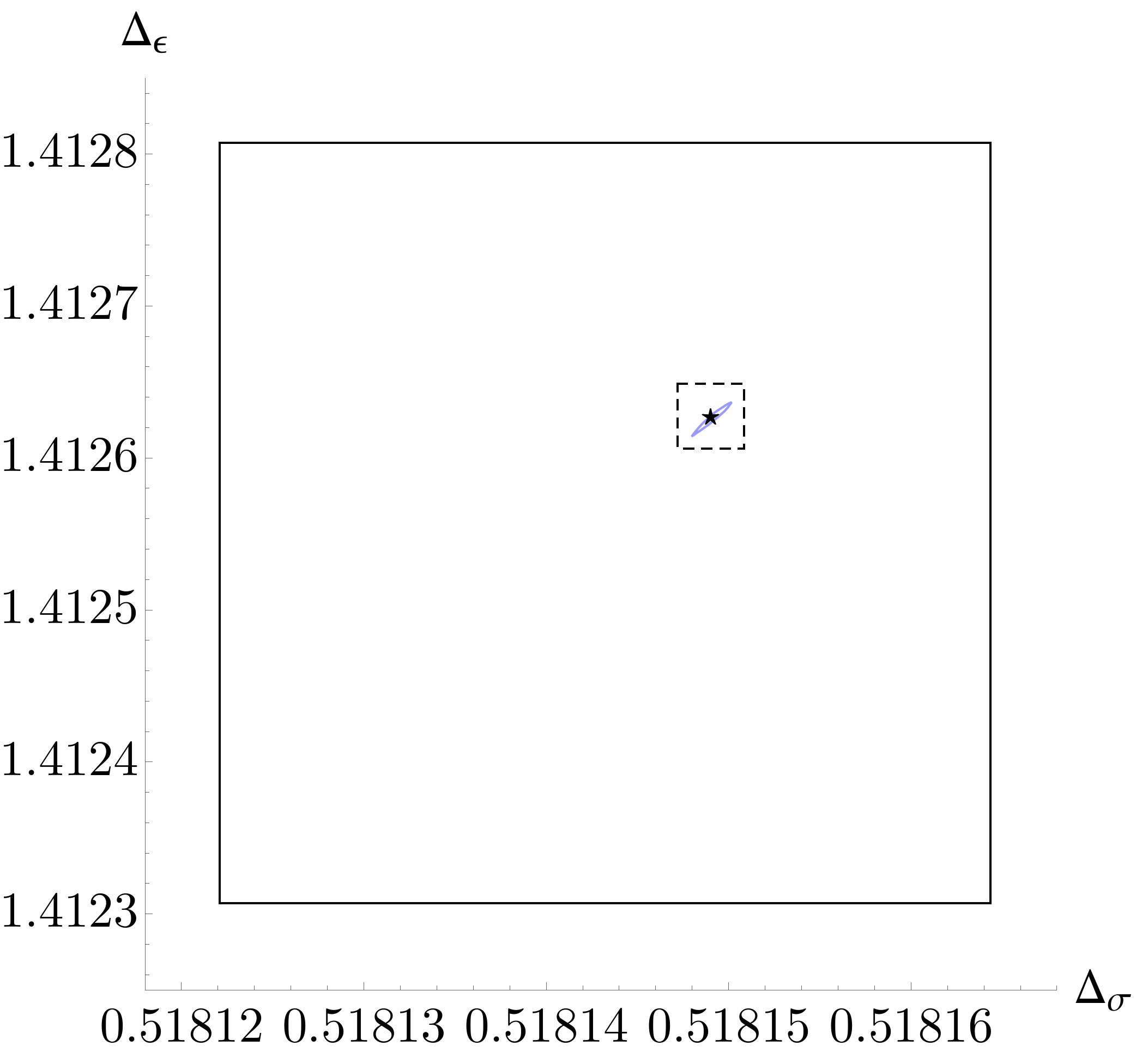}%
\caption{\label{fig:errorbar_Dss0v3} The star : the result of the Hybrid Setup \ref{hybrid:ss0_v3} at $\Lambda=19, L_0=4$. The dashed box : an estimation of the error for the star (the data is from Table \ref{tab:error_of_ss0v3_pd19}). The black box : the error box of the Numerics Setup \ref{numerics:pure} at $\Lambda=19$ (the data is from Table \ref{tab:numerics_pure}).
}
\end{figure}

The best computation in this paper comes from combining a refined version of $\D^{(analytic)}_{[\s\s]_{0}}$ with the $\Lambda=19$ numerics. The estimated error box of our best hybrid run, as well as the error box of the pure numerics (non-hybrid) at $\Lambda=19$ are presented in Figure \ref{fig:errorbar_Dss0v3}. Even in relatively low derivative order $\Lambda=19$ with some partial analytic information of the leading twist operators, the prediction of hybrid bootstrap is very accurate, clearly showing the potential of this method to advance the conformal bootstrap computation.

More work is needed to really bring this method to its full fledged form. For 3D Ising CFT, we need to systematically improve the efficiency and accuracy of the analytics, carefully balance the accuracy of various analytic information, and do the computation at higher $\Lambda$. We also need to further investigate methods for the error estimation and test their validity in more scenarios. For most computations in this paper, we used the $\Lambda=43$ island of \cite{Kos:2016ysd} as a benchmark. But as we go beyond $\Lambda=19$ in the hybrid bootstrap, we will be in an uncharted territory and the validity of the error estimation will be essential.

This hybrid bootstrap won't give the rigorous error box as the traditional numerical bootstrap does, so it can’t completely replace the rigorous method. But by sacrificing some rigorousness, we might gain a lot in the accuracy of the prediction for the actual CFT. Such a trade-off is certainly favorable in many scenarios. If we can obtain a non-rigorous but still very trustable spectrum, we will still be able to answer important questions in physics.

Our method can be applied to many other models. As we have shown, even some partial analytic information could significantly improve the result. When applying the hybrid bootstrap to other CFTs, one may start with the most reliable analytics that one has. Quick applications could be super-Ising and the O(N) CFTs, where some analytics were worked out in \cite{Liu:2020tpf,Atanasov:2022bpi}\footnote{It's likely that one will have to further improve the analytics for large $\hb$ before applying it to the hybrid bootstrap, similar to what we did in Section \ref{sec:analytics}.}. 

Besides the lightcone limit, the analytic information from other limits could be useful as well. The numerical bootstrap is not sensitive to large $\D$. If we have some estimation of the effective contribution from larger $\D$, we probably can inject that information into the numerical bootstrap as well. Another interesting limit is the large charge expansion for the $O(N)$ CFTs \cite{Gaume:2020bmp}. To access operators with larger $O(N)$ charge in the numerical bootstrap, we have to include more and more external operators with large charges in the bootstrap equation. This looks like a very bad situation because the number of bootstrap equations grows very fast with the number of external operators. But as we have shown in this paper, if we inject the analytics of the leading twist into the numerics, we probably don't need very high $\Lambda$ for every equation. Without increasing the total number of components of the functional $\alpha$, we may redistribute those components to other equations that have access to higher charge. In this way, we might see a higher charge spectrum and hopefully could connect the large charge analytics with the numerics. Beside the conformal bootstrap, it would be interesting to see if a similar hybrid method works for other bootstrap methods. For example, in the modular bootstrap there is an analogy of the lightcone analytics in terms of the Cardy formula.



Another interesting direction to explore in the future is to see if the hybrid bootstrap can help with bootstrapping a four-point function whose external operator has a large dimension. CFTs from gauge theories are of this type. It is known that in general numerical bootstrap works less well if the external operator has a larger dimension. For example, the single correlator bound \cite{El-Showk:2012cjh} converges less well for larger $\ds$. We also see an example in gauge theory \cite{He:2021xvg}, where the numerics for 2d scalar QED converge much better than the 3d case. We don't fully understand this phenomenon. But one contributing factor might be, as the dimension of the external operator gets larger, the leading twist family usually also has a larger twist that is further away from the unitary bound, so the unitary bound we put in each channel may not be sufficient. A relevant hint is that the EFM spectrum of the super-Ising CFT \cite{Atanasov:2022bpi} also suffers from the sharing effect, \itie it has fake operators on the unitary bound. Since the $\ds$ in the super-Ising CFT is larger than $\ds$ in the Ising CFT, those fake operators will further away from the true operators, thus could do more damage to the convergence. On the other hand, in the hybrid bootstrap the bounding band is imposed close to the true operator and the EFM spectrum in the band doesn't suffer from the sharing effect. If this fact (free from the sharing effect) really contributes to the nice convergence of the hybrid bootstrap computations in this work, we expect the hybrid bootstrap with leading twist analytics is important for many bootstrap setups involving large external operators. To confirm this scenario, we should implement the hybrid bootstrap for the super-Ising CFT.

\section*{Acknowledgements}
We thank Johan Henriksson, Yinchen He, Brian McPeak, David Poland, Marten Reehorst, Jongchen Rong, Slava Rychkov, David Simmons-Duffin, Benoit Sirois, Alessandro Vichi, Balt van Rees for discussions. We thank Shai Chester, Johan Henriksson, Jongchen Rong, Alessandro Vichi for their critical opinions on this manuscript. We thank Walter Landry for very helpful advice on C++ parallel scientific programming. This project has received funding from the European Research Council (ERC) under the European Union’s Horizon 2020 research and innovation programme (grant agreement no. 758903). We thank Yinchen He and David Simmons-Duffin for support on computational resources. The computations in this paper were partially run on the Symmetry cluster of Perimeter institute. The computations presented here were partially conducted in the Resnick High Performance Computing Center, a facility supported by Resnick Sustainability Institute at the California Institute of Technology.

\appendix

\section{Code availability}
\label{app:code}
During this project, we developed several programs to facilitate the implementation. All the programs are open-source software and available online.

To compute the conformal block we used a fork version of \texttt{scalar\_blocks} with the temporary name \texttt{scalar\_blocks\_mod}(\href{https://gitlab.com/suning-git/scalar-blocks-mod}{\tt https://gitlab.com/suning-git/scalar-blocks-mod}). This program computes the scalar block, convolves it and saves the block polynomials to a binary format or the Mathematica format. One difference between \texttt{scalar\_blocks\_mod} and original version is that for correlators of the form $\<\s\s\e\e\>$, certain poles have zero residue and they are explicitly removed in \texttt{scalar\_blocks\_mod}.

To generate the SDP for \texttt{sdpb}, we used a fork version of \texttt{sdp2input} with the temporary name  \texttt{sdp2input\_mod}(\href{https://github.com/suning-git/sdpb/tree/sdp2input_mod}{\tt https://github.com/suning-git/sdpb/tree/sdp2input\_mod}). This program depends on three inputs. The first one is the block data from the \texttt{scalar\_blocks\_mod}. The second one is called a parameter file, which contains the values for a set of user-defined variables. The third one is called a SDP template file. This is a Mathematica file similar to the input for original \texttt{sdp2input}, except that all the polynomials from the blocks are not plugged in but just stubs, and it also allows user-defined variables. This program parses the SDP template file, plugs in the actual values for the block polynomials and user-defined variables, evaluates elementary Mathematica expressions, and generates the final SDP for \texttt{sdpb}. The typical runtime of \texttt{sdp2input\_mod} for generating a SDP at $\Lambda=19, 43$ in Section \ref{sec:numerics} is about 7 seconds and 110 seconds respectively on a computer with 64 AMD EPYC 7532 32-Core Processor.

The workflow is managed using \texttt{simpleboot 3.0} (\href{https://gitlab.com/bootstrapcollaboration/simpleboot}{\tt https://gitlab.com/bootstrap\\collaboration/simpleboot}) as follows. Given a bootstrap problem, the SDP might depend on various variables (such as the dimensions of the external operators and the gaps in each channel). The \texttt{simpleboot} package generates a template file for \texttt{sdp2input\_mod} that contains those variables, but the value is not plugged in. Such a template file is generated once for all in a setup (with fixed $\Lambda$). In a specific \texttt{sdpb} run, simpleboot will compute the actual values for the variables in the template and save it as the parameter file, then call \texttt{scalar\_blocks\_mod} and \texttt{sdp2input\_mod} to generate the SDP. This workflow minimizes the operations in Mathematica and the most heavy computation is done in C++. For the computations in this work, the evaluation of the gradient of the navigator function requires generating the SDP multiple times, thus the efficient workflow is essential. For users on a cluster without Mathematica, one possible way to proceed is to use \texttt{simpleboot} to generate a template file locally, then upload it to the cluster and write a script to replace above management role of \texttt{simpleboot} (\itie decide the variables for the template file).

For the computation of the gradient of the navigator function, we used an experimental version of \texttt{approx\_objective}. The difference from the original version is that this program will take the change of the SDP variable ``$A$" (defined in (2.14) of \cite{Simmons-Duffin:2015qma}) into consideration. However we later found this is not really necessary for the computations in this paper, because if we fix the non-polynomial part of the rational approximation of the conformal block ($\chi_\ell(\Delta)$ in (3.7) of \cite{Simmons-Duffin:2015qma}), the $A$ variable won't change. 

By the time this paper is published, all the programs are still in the testing stage and there is very little documentation support. But we are committed to make the method in this paper more accessible to the reader. Interested users may come back to those links in a few months for the documentation and example files.

\section{Technical details}
\label{app:parameters}

For all computations, we used the following choices for the set of spins at each value of $\Lambda$:
\begin{align}\label{tab:spinsets}
S_{11} &= \{0,\dots,20\}\cup \{49, 52\}\,,\nonumber\\
S_{19} &= \{0,\dots,26\}\cup \{49, 52\}\,,\nonumber\\
S_{27} &= \{0,\dots,31\}\cup \{33, 34, 37, 38, 41 ,42 ,45 ,46 ,49 ,50\}\,,\nonumber\\
S_{35} &= \{0,\dots,44\}\cup \{47, 48, 51, 52, 55, 56, 59, 60, 63, 64, 67, 68\}\,,\nonumber\\
S_{43} &= \{0,\dots,64\}\cup \{67, 68, 71, 72, 75, 76, 79, 80, 83, 84, 87, 88\}\,. 
\end{align}
For $\Lambda=11, 19, 27, 35, 43$, the ``--keptPoleOrder" parameter of \texttt{scalar\_blocks\_mod} are $12, 14, 20, 30, 40$ and the ``--order" parameter are 4 times ``--keptPoleOrder". The \texttt{sdpb} parameters are given in table~\ref{tab:params}.

\begin{table}[!h]
\begin{center}
\begin{tabular}{@{}|c|c|c|c|c@{}}
	\toprule
$\Lambda$ &  11,19,27,35  & 43 \\
{\small\texttt{precision}} & 765 & 1024 \\
{\small\texttt{dualityGapThreshold}} &  $10^{-40}$ & $10^{-50}$ \\
{\small\texttt{primalErrorThreshold}}&  $10^{-40}$ & $10^{-50}$ \\
{\small\texttt{dualErrorThreshold}} & $10^{-40}$ & $10^{-50}$\\ 
{\small\texttt{initialMatrixScalePrimal}} & $10^{20}$& $10^{20}$\\
{\small\texttt{initialMatrixScaleDual}} &  $10^{20}$ &  $10^{20}$\\
{\small\texttt{feasibleCenteringParameter}} &  0.1 & 0.1\\
{\small\texttt{infeasibleCenteringParameter}} &  0.3 & 0.3\\
{\small\texttt{stepLengthReduction}} &  0.7 & 0.7\\
{\small\texttt{maxComplementarity}} & $10^{70}$ & $10^{70}$\\
 \bottomrule
\end{tabular}
\caption{\label{tab:params}Parameters used for the \texttt{sdpb}.}
\end{center}
\end{table}

To compute the gradient of the navigator function, we used a finite difference of $10^{-40}$ in each argument of the navigator function. To maximize/minimize a parameter in the navigator method, we used Algorithm 2 in \cite{Reehorst:2021ykw} with $g_{\text{tol}}=10^{-15}$. To minimum navigator, we used Algorithm 1 in \cite{Reehorst:2021ykw} with $g_{\text{tol}}=10^{-15}$. We made a small modification to Algorithm 1. The standard BFGS line search is good for the most of the runs except the $\Lambda=43$ run of Figure~\ref{fig:3par}. The issue is that the gradient of the navigator function for points outside the $\Lambda=43$ island is typically in the order of $10^{-3}$, while inside the island it's $10^{-16}$. If we use the standard Broyden–Fletcher–Goldfarb–Shanno (BFGS) line search, at the first feasible point, the line search step length will be very small because the gradient is very small while the scale of the Hessian is still inferred based on the points outside the island. The standard BFGS line search simply multiple the step length by 4 in subsequent trials, but it will take too many steps ($\sim \text{log}_4(10^{16})=26.5$) to have the correct scale for the step length. We fix this issue in the following way : if the first step with $\alpha=1$ in the line search failed, we use the information at $\alpha=0,1$ to fit a quadratic polynomial in $\alpha$ and use this polynomial to do extrapolation to determine the next step. This fix effectively reduce the $\sim 26$ steps to 3 steps. We also made a small modification to Algorithm 2. We add an option that if one start from an infeasible initial point, certain parameters will be fixed until the first feasible point is found. The behavior of algorithm before finding first feasible point is very similar to Algorithm 1 of \cite{Reehorst:2021ykw}. For the run of Figure \ref{fig:minerr_fullpath}, the fixed parameter is $a$. The complete code of both algorithms can be found in \texttt{simpleboot}.

\section{Collection of results}\label{app:results}

In the tables of this section, we listed the results of computations in this paper.

\begin{table}[h!]
\begin{center}
\begin{tabular}{|| c | *{3}{c} ||}
\hline
$\Lambda$ & $\Delta _{\sigma }$ & $\Delta _{\epsilon }$ & $f_{\sigma \sigma \epsilon }/f_{\epsilon \epsilon \epsilon }$ \\
\hline
$11$ & $0.518258316$ & $1.41269841$ & $1.45655643$ \\
$19$ & $0.518134913$ & $1.41244118$ & $1.45641973$ \\
$27$ & $0.518149340$ & $1.41261849$ & $1.45684755$ \\
$35$ & $0.518148980$ & $1.41262326$ & $1.45688261$ \\
$43$ & $0.518148886$ & $1.41262383$ & $1.45688799$ \\
\hline
\end{tabular}
\caption{\label{tab:numerics1}Computations of the Numerics Setup \ref{numerics:1}.}
\end{center}
\end{table}
\begin{table}
\begin{center}
\begin{tabular}{|| c | *{5}{c} ||}
\hline
$\text{Setup}$ & $\Delta _{\sigma }$ & $\Delta _{\epsilon }$ & $f_{\sigma \sigma \epsilon }$ & $f_{\epsilon \epsilon \epsilon }$ & $f_T$ \\
\hline
$\left.\text{$\Lambda $=11 max(}\Delta _{\sigma }\right)$ & $0.51833135$ & $1.4147102$ & $1.0510383$ & $1.5356715$ & $1.2592846$ \\
$\left.\text{$\Lambda $=11 min(}\Delta _{\sigma }\right)$ & $0.51775600$ & $1.4079531$ & $1.0536891$ & $1.5244978$ & $1.2579070$ \\
$\left.\text{$\Lambda $=19 max(}\Delta _{\sigma }\right)$ & $0.51816152$ & $1.4127727$ & $1.0518006$ & $1.5327088$ & $1.2588684$ \\
$\left.\text{$\Lambda $=19 min(}\Delta _{\sigma }\right)$ & $0.51812509$ & $1.4123433$ & $1.0519570$ & $1.5319139$ & $1.2588299$ \\
\hline
\end{tabular}
\caption{\label{tab:numerics1}Computations of the Hybrid Setup \ref{hybrid:mild_gaps}.}
\end{center}
\end{table}
\begin{table}
\begin{center}
\begin{tabular}{|| c | *{5}{c} ||}
\hline
$\text{Setup}$ & $\Delta _{\sigma }$ & $\Delta _{\epsilon }$ & $f_{\sigma \sigma \epsilon }$ & $f_{\epsilon \epsilon \epsilon }$ & $f_T$ \\
\hline
$\left.\text{$\Lambda $=11 max(}\Delta _{\sigma }\right)$ & $0.51853855$ & $1.4172801$ & $1.0501657$ & $1.5413379$ & $1.2591992$ \\
$\left.\text{$\Lambda $=11 min(}\Delta _{\sigma }\right)$ & $0.51772067$ & $1.4075014$ & $1.0538377$ & $1.5235323$ & $1.2579841$ \\
$\text{$\Lambda $=11 min(nvg)}$ & $0.51798601$ & $1.4106308$ & $1.0525454$ & $1.5281088$ & $1.2588351$ \\
$\left.\text{$\Lambda $=19 max(}\Delta _{\sigma }\right)$ & $0.51816437$ & $1.4128073$ & $1.0517879$ & $1.5327764$ & $1.2588716$ \\
$\left.\text{$\Lambda $=19 min(}\Delta _{\sigma }\right)$ & $0.51812208$ & $1.4123071$ & $1.0519696$ & $1.5318388$ & $1.2588280$ \\
$\text{$\Lambda $=19 min(nvg)}$ & $0.51813719$ & $1.4124867$ & $1.0519042$ & $1.5321763$ & $1.2588426$ \\
\hline
\end{tabular}
\caption{\label{tab:numerics_pure}Computations of the Numerics Setup \ref{numerics:pure}. ``nvg" means the navigator function.}
\end{center}
\end{table}

\begin{table}
\begin{center}
\begin{tabular}{|| c | *{6}{c} ||}
\hline
$\text{Setups}$ & $\Delta _{\sigma }$ & $\Delta _{\epsilon }$ & $f_{\sigma \sigma \epsilon }$ & $f_{\epsilon \epsilon \epsilon }$ & $f_T$ & $a$ \\
\hline
$\Lambda =11,L_0=6$ & $0.51815015$ & $1.4126784$ & $1.0518300$ & $1.5327085$ & $1.2588618$ & $-0.55899947$ \\
$\Lambda =19,L_0=10$ & $0.51815135$ & $1.4126552$ & $1.0518428$ & $1.5324985$ & $1.2588596$ & $-0.19533902$ \\
$\Lambda =19,L_0=12$ & $0.51815078$ & $1.4126480$ & $1.0518455$ & $1.5324834$ & $1.2588589$ & $-0.19681209$ \\
$\Lambda =19,L_0=14$ & $0.51814723$ & $1.4126063$ & $1.0518600$ & $1.5324020$ & $1.2588556$ & $-0.75148583$ \\
\hline
\end{tabular}
\caption{\label{tab:numerics1}Computations of the Hybrid Setup \ref{hybrid:fullsetup}.}
\end{center}
\end{table}

\begin{table}
\begin{center}
\begin{tabular}{|| c | *{6}{c} ||}
\hline
$L_0$ & $\Delta _{\sigma }$ & $\Delta _{\epsilon }$ & $f_{\sigma \sigma \epsilon }$ & $f_{\epsilon \epsilon \epsilon }$ & $f_T$ & $a$ \\
\hline
$2$ & $0.51814914$ & $1.4126296$ & $1.0518523$ & $1.5324528$ & $1.2588570$ & $-0.34342837$ \\
$4$ & $0.51814939$ & $1.4126327$ & $1.0518510$ & $1.5324596$ & $1.2588574$ & $-0.34375960$ \\
$6$ & $0.51814853$ & $1.4126222$ & $1.0518549$ & $1.5324393$ & $1.2588564$ & $-0.34511252$ \\
$8$ & $0.51814780$ & $1.4126136$ & $1.0518578$ & $1.5324206$ & $1.2588557$ & $-0.34595671$ \\
$10$ & $0.51814901$ & $1.4126279$ & $1.0518529$ & $1.5324495$ & $1.2588568$ & $-0.34726307$ \\
$12$ & $0.51815349$ & $1.4126802$ & $1.0518339$ & $1.5325445$ & $1.2588612$ & $-0.36100753$ \\
\hline
\end{tabular}
\caption{\label{tab:hybrid:se0}Computations of the Hybrid Setup \ref{hybrid:se0}.}
\end{center}
\end{table}

\begin{table}
\begin{center}
\begin{tabular}{|| c | *{6}{c} ||}
\hline
$L_0$ & $\Delta _{\sigma }$ & $\Delta _{\epsilon }$ & $f_{\sigma \sigma \epsilon }$ & $f_{\epsilon \epsilon \epsilon }$ & $f_T$ & $a$ \\
\hline
$4$ & $0.51814534$ & $1.4125839$ & $1.0518688$ & $1.5323629$ & $1.2588538$ & $-0.39513721$ \\
$8$ & $0.51814545$ & $1.4125852$ & $1.0518683$ & $1.5323654$ & $1.2588539$ & $-0.39519309$ \\
$10$ & $0.51814544$ & $1.4125851$ & $1.0518683$ & $1.5323651$ & $1.2588539$ & $-0.39525413$ \\
$12$ & $0.51814545$ & $1.4125852$ & $1.0518683$ & $1.5323653$ & $1.2588539$ & $-0.39525418$ \\
\hline
\end{tabular}
\caption{\label{tab:hybrid:ss0_v2}Computations of the Hybrid Setup \ref{hybrid:ss0_v2}.}
\end{center}
\end{table}

\begin{table}
\begin{center}
\begin{tabular}{|| c | *{6}{c} ||}
\hline
$\text{Setup}$ & $\Delta _{\sigma }$ & $\Delta _{\epsilon }$ & $f_{\sigma \sigma \epsilon }$ & $f_{\epsilon \epsilon \epsilon }$ & $f_T$ & $a$ \\
\hline
$\text{$\Lambda $=19 }L_0\text{=4}$ & $0.51814904$ & $1.4126275$ & $1.0518530$ & $1.5324459$ & $1.2588574$ & $-1.1730688$ \\
\hline
\end{tabular}
\caption{\label{tab:hybrid:ss0_v3}Computations of the Hybrid Setup \ref{hybrid:ss0_v3}.}
\end{center}
\end{table}

\begin{table}
\begin{center}
\begin{tabular}{|| c | *{6}{c} ||}
\hline
$r$ & $\Delta _{\sigma }$ & $\Delta _{\epsilon }$ & $f_{\sigma \sigma \epsilon }$ & $f_{\epsilon \epsilon \epsilon }$ & $f_T$ & $a_o$ \\
\hline
$0$ & $0.51814848$ & $1.4126215$ & $1.0518548$ & $1.5324348$ & $1.2588571$ & $-0.34387402$ \\
$0.2$ & $0.51814809$ & $1.4126169$ & $1.0518568$ & $1.5324276$ & $1.2588560$ & $-0.34326748$ \\
$0.38$ & $0.51814855$ & $1.4126223$ & $1.0518546$ & $1.5324367$ & $1.2588571$ & $-0.34297320$ \\
$0.4$ & $0.51814861$ & $1.4126231$ & $1.0518542$ & $1.5324375$ & $1.2588572$ & $-0.34281567$ \\
$0.6$ & $0.51814891$ & $1.4126265$ & $1.0518528$ & $1.5324416$ & $1.2588574$ & $-0.34109519$ \\
$0.8$ & $0.51814895$ & $1.4126270$ & $1.0518525$ & $1.5324415$ & $1.2588574$ & $-0.34014029$ \\
$0.9$ & $0.51814889$ & $1.4126263$ & $1.0518528$ & $1.5324397$ & $1.2588574$ & $-0.33986723$ \\
$0.975$ & $0.51815014$ & $1.4126410$ & $1.0518478$ & $1.5324710$ & $1.2588587$ & $-0.33875725$ \\
$1$ & $0.51815029$ & $1.4126428$ & $1.0518472$ & $1.5324750$ & $1.2588589$ & $-0.33811048$ \\
$1.06875$ & $0.51814889$ & $1.4126259$ & $1.0518534$ & $1.5324424$ & $1.2588575$ & $-0.33557516$ \\
$1.1$ & $0.51814793$ & $1.4126145$ & $1.0518576$ & $1.5324205$ & $1.2588565$ & $-0.33393662$ \\
$1.14216$ & $0.51814628$ & $1.4125946$ & $1.0518649$ & $1.5323818$ & $1.2588547$ & $-0.33053125$ \\
\hline
\end{tabular}
\caption{\label{tab:numerics1}Computations of the Hybrid Setup \ref{hybrid:liebig_ratio}.}
\end{center}
\end{table}

\clearpage
\bibliography{Biblio}
\bibliographystyle{utphys}
\end{document}